\pdfoutput=1
\documentclass[10pt]{article}
\usepackage[left=.75in,right=.75in,top=.75in,bottom=.75in]{geometry}
\usepackage[utf8]{inputenc}
\usepackage{amssymb,amsmath,latexsym,amsthm,amsfonts,mathtools}

\usepackage{mathrsfs}
\usepackage{multirow}
\usepackage{graphicx}
\usepackage{color}
\usepackage{textcomp,siunitx}
\usepackage{comment}
\usepackage{float}
\usepackage{subcaption}
\usepackage{booktabs}
\usepackage{enumitem}
\usepackage{comment}
\usepackage[dvipsnames]{xcolor}
\definecolor{ao(english)}{rgb}{0.0, 0.5, 0.0}
\definecolor{Tan}{rgb}{0.82, 0.71, 0.55}
\usepackage{hyperref}
\hypersetup{colorlinks, breaklinks, citecolor=blue, linkcolor=ao(english), urlcolor=blue}
\usepackage[nameinlink,noabbrev,sort&compress]{cleveref}
\crefname{assumption}{Assumption}{Assumptions}
\usepackage{setspace}
\usepackage{cite}
\theoremstyle{plain}
\newtheorem{theorem}{Theorem}
\newtheorem{lemma}{Lemma}

\newtheorem{assumption}{Assumption}

\theoremstyle{remark}
\newtheorem{remark}{Remark}

\theoremstyle{definition}
\newtheorem{definition}{Definition}

\newtheorem{problem}{Problem}
\newcommand{\sign}{\mathrm{sign}}

\usepackage{newtxmath}

\usepackage{anyfontsize}
\usepackage{accents}

\usepackage{scalerel}
\usepackage{tikz}
\usepackage{pgfplots}
\usetikzlibrary{automata, shapes, arrows, calc, arrows.meta, fit, positioning}
\usetikzlibrary{svg.path}

\definecolor{orcidlogocol}{HTML}{A6CE39}
\tikzset{
	orcidlogo/.pic={
		\fill[orcidlogocol] svg{M256,128c0,70.7-57.3,128-128,128C57.3,256,0,198.7,0,128C0,57.3,57.3,0,128,0C198.7,0,256,57.3,256,128z};
		\fill[white] svg{M86.3,186.2H70.9V79.1h15.4v48.4V186.2z}
		svg{M108.9,79.1h41.6c39.6,0,57,28.3,57,53.6c0,27.5-21.5,53.6-56.8,53.6h-41.8V79.1z M124.3,172.4h24.5c34.9,0,42.9-26.5,42.9-39.7c0-21.5-13.7-39.7-43.7-39.7h-23.7V172.4z}
		svg{M88.7,56.8c0,5.5-4.5,10.1-10.1,10.1c-5.6,0-10.1-4.6-10.1-10.1c0-5.6,4.5-10.1,10.1-10.1C84.2,46.7,88.7,51.3,88.7,56.8z};
	}
}

\newcommand\orcidicon[1]{\href{https://orcid.org/#1}{\mbox{\scalerel*{
				\begin{tikzpicture}[yscale=-1,transform shape]
					\pic{orcidlogo};
				\end{tikzpicture}
			}{|}}}}

\newcommand\ith{$i$\textsuperscript{th} interceptor}

\title{Predefined-time One-Shot Cooperative Estimation, Guidance, and Control for Simultaneous Target Interception}

\author{Lohitvel Gopikannan\and Shashi Ranjan Kumar\thanks{L. Gopikannan and S. R. Kumar are with the Intelligent Systems and Control (ISaC) Lab, Department of Aerospace Engineering, Indian Institute of Technology Bombay, Powai, MH, 4000076,  India.
 e-mails: 24m0023@iitb.ac.in, srk@aero.iitb.ac.in.}
	\and Abhinav~Sinha       	
        \thanks{A. Sinha is with the Guidance, Autonomy, Learning, and Control for Intelligent Systems (GALACxIS) Lab, Department of Aerospace Engineering and Engineering Mechanics, University of Cincinnati, Cincinnati, OH, 45221,  USA.
 e-mail: abhinav.sinha@uc.edu.}
}

\date{}

\begin{document}

\maketitle
\doublespacing

\begin{abstract}
This work develops a unified nonlinear estimation-guidance-control framework for the cooperative simultaneous interception of a stationary target under a heterogeneous sensing topology, where sensing capabilities are non-uniform across interceptors. Specifically, only a subset of agents is instrumented with onboard seekers (informed/seeker-equipped agents), whereas the rest of them (seeker-less agents) acquire the information about the target indirectly via the informed agents and execute a distributed cooperative guidance for simultaneous target interception. To address the resulting partial observability, a predefined-time distributed observer is leveraged, guaranteeing convergence of the target state estimates for seeker-less agents through information exchange with seeker-equipped neighbors over a directed communication graph. Thereafter, an improved time-to-go estimate accounting for wide launch envelopes is utilized to design the distributed cooperative guidance commands. This estimate is coupled with a predefined-time consensus protocol, ensuring consensus in the agents' time-to-go values. The temporal upper bounds within which both observer error and time-to-go consensus error converge to zero can be prescribed as design parameters. Furthermore, the cooperative guidance commands are realized by means of an autopilot, wherein the interceptor is steered by canard actuation. The corresponding fin deflection commands are generated using a predefined-time convergent sliding mode control law. This enables the autopilot to precisely track the commanded lateral acceleration within a design-specified time, while maintaining non-singularity of the overall design. Theoretical guarantees are supported by numerical simulations across diverse engagement geometries, verifying the estimation accuracy, the cooperative interception performance, and the autopilot response using the proposed scheme. 
\medskip
	
\noindent \emph{\textbf{Keywords}}--- Cooperative estimation-guidance-control, predefined-time convergence, impact time, salvo guidance.
\end{abstract}
\section{Introduction}\label{sec:intro}
The primary objective of modern guidance systems is to achieve precise target interception while satisfying stringent terminal requirements such as prescribed impact time \cite{sp31,sp32,sp33 }, and impact angle \cite{sp30,sp34,sp35}. Although single-interceptor engagements have been extensively investigated, their practical effectiveness is often compromised by inherent limitations, including restricted seeker field-of-view, actuator saturation, and system lags. At the same time, advanced defense architectures employ layered protection and countermeasures such as close-in weapon systems, making penetration by a lone attacker highly improbable. These challenges highlight the need for coordinated multi-interceptor engagements, commonly referred to as salvo guidance, to address the limitations of traditional one-to-one interception.

Contemporary salvo strategies are broadly categorized into non-cooperative and cooperative approaches. Non-cooperative schemes rely on preassigned impact times without information exchange, making them susceptible to errors during flight and reducing the likelihood of simultaneous interception. Cooperative schemes, in contrast, exploit real-time information sharing and coordination among interceptors, enabling superior synchronization and robustness against uncertainties.  For instance, the work in \cite{sp11} introduced a distributed cooperative guidance law based on biased proportional navigation (PN). Building on this, the authors of \cite{sp12} reformulated impact-time control as a range-tracking problem. The work in \cite{sp6} aimed to minimize time-to-go variance by varying navigation gains, but relied on all-to-all communication. This approach was extended in \cite{sp13} under acceleration constraints, and the authors of \cite{sp14} incorporated finite-time consensus with terminal constraints, all considering undirected networks. 

The work in \cite{sp7} proposed finite-time distributed laws requiring only local time-to-go estimates, reducing communication, whereas in \cite{sp15}, a receding-horizon cooperative guidance law was proposed where interceptors exchanged only neighbor data and solved local finite-horizon optimizations. The work in \cite{sp16,sp17} avoided explicit time-to-go estimation using a two-stage scheme relying on decentralized consensus on range/heading followed by PN guidance. In \cite{sp18}, coordination with local guidance in both centralized and distributed forms was integrated. Together, the strategies \cite{sp16,sp17,sp18,sp7,sp15} ensured consensus over undirected networks, though only asymptotically. This could lead to performance degradation in short-duration engagements because strict control over it may be lacking.

Leader–follower design combined with PN-based time-to-go estimation and super-twisting sliding mode control was proposed in \cite{sp1} to guarantee finite-time convergence even under large heading errors for simultaneous interception of stationary targets. However, it should be taken into consideration that failure of the leader will compromise the mission. The authors in \cite{sp8} developed time-to-go–dependent guidance laws based on the suboptimal finite-time state-dependent Riccati equation (FT-SDRE) for simultaneous target interception; however, the inherent sub-optimality of the FT-SDRE may constrain performance in highly dynamic engagement scenarios. In \cite{sp19}, a cooperative guidance was developed that explicitly addressed lateral acceleration constraints within a nonlinear framework, ensuring stability even under large heading errors, and extended applicability to both directed and undirected networks. However, the consensus over the network is reached only asymptotically. Fixed-time cooperative guidance laws, for example, \cite{sp20,sp21,sp22,sp23,sp24}, may address the shortcomings of asymptotic and finite-time schemes used in aforementioned works by guaranteeing consensus independent of initial engagement conditions, thereby ensuring reliable guidance precision. Nevertheless, these methods often rely on conservatively large controller gains, and their design parameters must be re-tuned whenever a different consensus time is desired, which can limit practical flexibility. To overcome this, leaderless cooperative guidance based on predefined-time consensus was proposed by \cite{sp25} for interceptor swarms, guaranteeing arbitrarily set time-to-go consensus and simultaneous interception under stationary or maneuvering targets with undirected dynamic topologies. The work in \cite{sp26} proposed a leader–follower cooperative salvo guidance law with arbitrary-time consensus ensuring predefined-time agreement on time-to-go and simultaneous interception under nonlinear kinematics and autopilot lag. 

Most of the aforementioned guidance strategies have been developed based solely on the kinematics of interceptor–target engagement, assuming that the interceptor can instantaneously realize any commanded acceleration. This assumption effectively neglects the dynamic characteristics of the interceptor’s control loop, which can lead to significant performance degradation under non-ideal or rapidly changing flight conditions, particularly during the terminal phase of engagement. To address this limitation, it is essential to account for the interceptor’s dynamics while formulating practical guidance strategies to ensure that the proposed guidance law remains effective and implementable in real-world scenarios. In this regard, authors in \cite{sp27} proposed an integrated interceptor guidance–autopilot loop employing a sliding mode control framework to achieve seamless coordination between guidance and flight control for a canard-controlled interceptor. The effectiveness of the integrated design was demonstrated through multiple endgame interception scenarios. Furthermore, the work in \cite{sp28} utilized this decoupled guidance–autopilot design, where canard and tail surfaces were coordinated to track the commanded lateral acceleration using predefined-time convergent sliding mode control. This ensured precise target interception within a specified time, as demonstrated through simulations. Such a framework gives an accurate assessment of real-time performance, including actuator limits, control–guidance coupling, and endgame response characteristics.

To the authors’ knowledge, cooperative salvo guidance strategies reported in contemporary literature (for example, the aforementioned works and their cited references) generally assume that all interceptors have access to complete real-time target position information to synchronize their interception times. In practice, however, equipping every interceptor with an onboard seeker is often infeasible in large-scale engagements due to constraints on cost, payload capacity, and energy resources. Moreover, maintaining full bidirectional communication is often impractical due to limited bandwidth, computational load, and onboard power. Therefore, directed interaction topologies, where information flows asymmetrically, provide an effective compromise, ensuring both coordination efficiency and practical feasibility in real-world salvo engagements. In time-critical scenarios with short engagement durations, achieving consensus in the interception time among interceptors within this short duration becomes essential, thereby motivating the use of predefined-time consensus methods that guarantee synchronization within the available window. Similarly, obtaining an instantaneous autopilot response is crucial to the implementation of the cooperative guidance commands, thereby further necessitating accurate control over the tracking error convergence (commanded versus actual input). 

In this work, we present a new perspective on cooperative simultaneous target interception by developing a one-shot nonlinear cooperative estimation-guidance-control framework to enhance the guidance precision in the endgame. The objective of this joint framework is to simultaneously address the three critical components of the interception problem within a unified design. By embedding the estimator, guidance law, and autopilot into a single coherent architecture, all within predefined-time convergence, the proposed framework explicitly accounts for the performance limitations of contemporary designs, which treat these aspects separately without a clear indication of their mutual interactions, closed-loop stability guarantees, or the resulting impact on terminal accuracy. We summarize our main contributions below.

In our treatment, only a subset of interceptors is equipped with seekers, while the remaining agents operate without direct target measurements. These seeker-less interceptors reconstruct the required engagement variables through a predefined-time distributed observer under a heterogeneous sensing architecture that exploits information fusion and local coordination over directed sensing networks. 

The estimated states are then utilized within a cooperative simultaneous interception scheme that incorporates an improved time-to-go estimate, free from small heading angle assumptions and thus effective under large initial heading angle errors, together with a predefined-time consensus protocol to guarantee simultaneous interception despite heterogeneous target observability. This opens up the full circular range of the interceptors' bearing angles for interception, thereby providing greater flexibility in realizing a simultaneous interception.

Thereafter, a predefined-time convergent sliding mode control law is developed for generating the interceptor’s canard deflection commands. This guarantees that the autopilot tracks the commanded lateral acceleration or the cooperative guidance commands within a designer-specified settling time, independent of initial conditions. By eliminating lag in the inner-loop dynamics, the proposed scheme ensures that the autopilot does not degrade the overall system response, thereby enabling the interceptor to realize the desired lateral acceleration from the guidance law in an almost instantaneous manner.

The proposed one-shot nonlinear cooperative estimation-guidance-control framework eliminates the need for full sensing redundancy, supports decentralized decision making, and ensures strict control on the convergence of necessary variables. It is worth noting that all convergence times (whether associated with estimation error dynamics, consensus error dynamics, or autopilot tracking error) are predefined-time stable, in the sense that the settling time can be explicitly assigned a priori as a design parameter. 

The remainder of this paper is organized as follows. \Cref{sec:problem} introduces the preliminaries, including the nonlinear engagement kinematics, interceptor dynamics, and the main objectives. \Cref{sec:main} develops the proposed one-shot estimation–guidance–control framework in detail. \Cref{sec:simulation} demonstrates the effectiveness of the proposed framework through various engagement scenarios. Finally, \Cref{sec:conclusions} concludes the paper.

\section{Background and Problem Formulation}\label{sec:problem}
\begin{figure}[h!]
        \centering
        \includegraphics[width=0.48\linewidth]{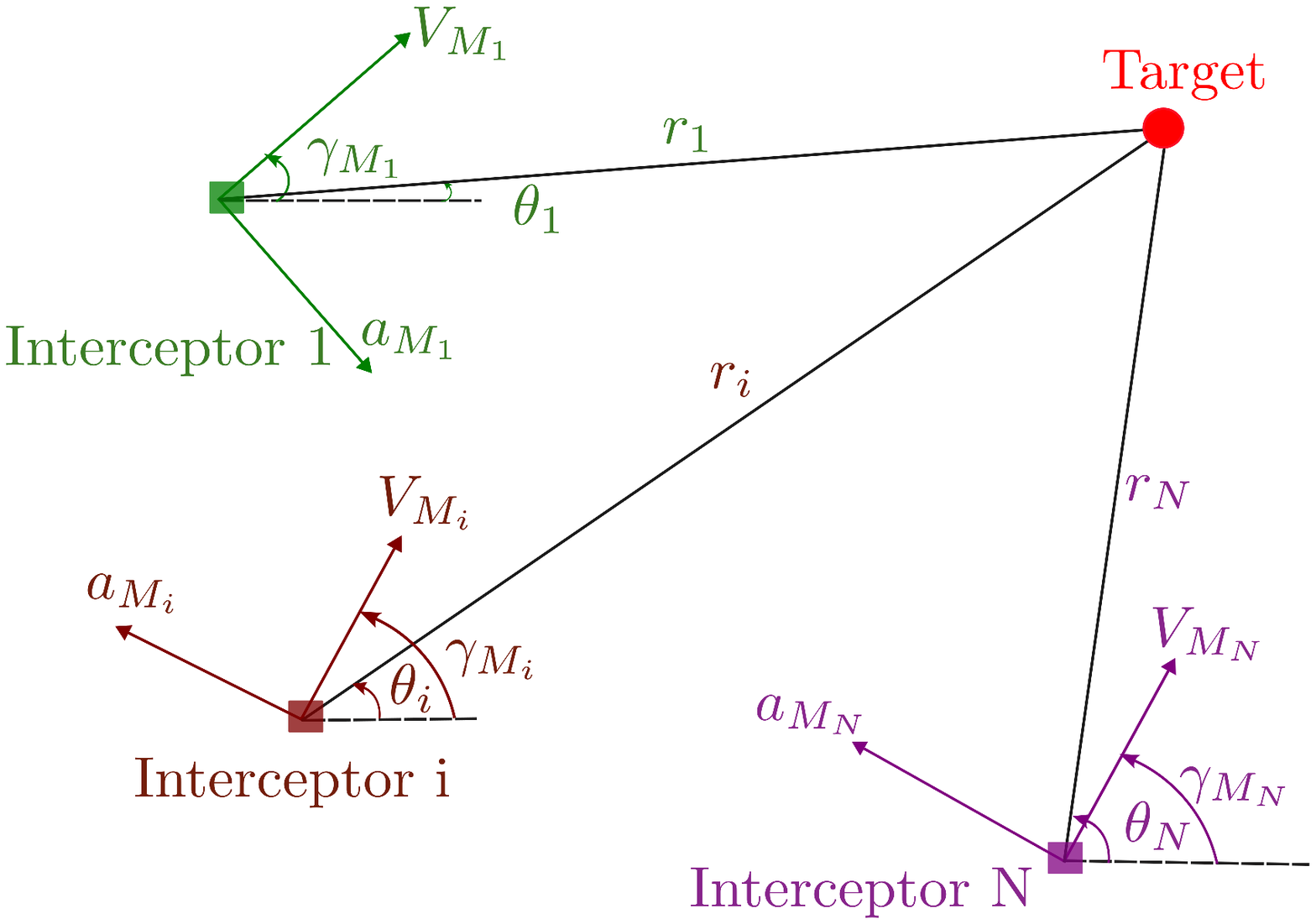}
        \caption{Cooperative multi-interceptor engagement}
        \label{fig:enggeo}
\end{figure}
Consider a multi-interceptor target engagement scenario involving $N$ interceptors and a stationary target, as shown in \Cref{fig:enggeo}. The speed of the $i$\textsuperscript{th} interceptor is given by $V_{M_i}$, whereas the relative range and line-of-sight (LOS) angle with respect to the target are $r_i$ and $\theta_i$, respectively. The flight path angle for the $i$\textsuperscript{th} interceptor is $\gamma_{M_i}$ and it is steered using its lateral acceleration $a_{M_i}$. The kinematic engagement equations in a relative coordinate system are given by
\begin{subequations}\label{eq:enggeo}
\begin{align}
\dot{r}_{i} &= - V_{M_{i}} \cos\left(\gamma_{M_{i}} - \theta_{i}\right) = V_{r_{i}}, \\
r_{i} \, \dot{\theta}_{i} &= - V_{M_{i}} \sin\left(\gamma_{M_{i}} - \theta_{i}\right) = V_{\theta_{i}}, \\
\dot{\gamma}_{M_{i}} &= \dfrac{a_{M_{i}}}{V_{M_{i}}},\label{eq:aMd}
\end{align}
\end{subequations}
where $V_{r_{i}}$ and $V_{\theta_{i}}$ denote the components of the relative velocity along and perpendicular to the LOS, respectively. The multi-interceptor swarm is assumed to exchange necessary information over a shared network, which can be modeled via graphs. It should be noted that the multi-interceptor swarm may have different sensing and actuation architectures.

We first model the \emph{sensing topology} by a directed graph $\mathscr{S} = \{\mathcal{V}, \mathcal{E}\}$, where the node set $\mathcal{V} = \{s_0, s_1, \ldots, s_N\}$ consists of a single target agent $s_0$ and $N$ interceptor agents $s_1$ through $s_N$. The directed edge set $\mathcal{E} \subseteq \mathcal{V} \times \mathcal{V}$ encodes the \emph{communication topology}, such that an edge $\varepsilon_{ij} \in \mathcal{E}$ denotes a directed information link from agent $s_i$ to agent $s_j$. For each node $s_i \in \mathcal{V}$, the \emph{in-neighbor set} $\mathcal{M}_i$ is the set of all agents $s_j$ such that $\varepsilon_{ji} \in \mathcal{E}$, i.e., all agents from which $s_i$ receives information. An agent is classified as a leader if $\mathcal{M}_i = \emptyset$ and as a follower otherwise. In this setting, the target agent $s_0$ serves as the unique leader (transmitting but not receiving information), while all interceptor agents act as followers. The communication structure is captured by the adjacency matrix $\mathcal{A} = [a_{ij}] \in \mathbb{R}^{(N+1) \times (N+1)}$, where
\begin{equation*}
    a_{ij}=\begin{cases}
        1,&\text{if}~ e_{ji}\in \mathcal{E}\\
        0, & \text{otherwise}
    \end{cases}\quad\text{and}~a_{ii}=0.
\end{equation*}
The in-degree matrix is given by $\mathcal{D} = \mathrm{diag}\{d_0, d_1, \ldots, d_N\}$, where each element $d_i = \sum_{j=0}^{N} a_{ij}$ denotes the total number of incoming edges to node $v_i$. The corresponding graph Laplacian is defined as $\mathcal{L} = \mathcal{D} - \mathcal{A}$, while the Kronecker product $\otimes$ is utilized to extend scalar graph representations to higher-dimensional system dynamics.
\begin{assumption}\label{asm:seeker}
The interceptor team comprises two distinct classes of autonomous agents-- those equipped with onboard seekers capable of directly measuring the target’s position (informed/seeker-equipped agents), and those without such sensing capability who rely on local information exchanged among neighboring agents over a directed sensing graph $\mathscr{S}$ to estimate the target's position (seeker-less agents).
\end{assumption}
\Cref{asm:seeker} introduces a multi-interceptor sensing framework for the proposed cooperative integrated estimation and guidance approach. The proposed framework explicitly accounts for heterogeneous information availability within the interceptor network to facilitate cooperative guidance among interceptors and also forms the mathematical basis for the cooperative interception problem addressed in this study.
\begin{assumption}[\cite{sp2}]\label{asm:asm2}
The directed graph $\mathscr{S}$ is assumed to possess a directed spanning tree with its root at the target node $s_0$. Then, the associated Laplacian matrix is given by
\begin{equation}
    \mathcal{L} = 
    \begin{bmatrix}
        0 & \mathbf{0}_{1\times N} \\
        \mathcal{L}_{TI} & \mathcal{L}_{II}
    \end{bmatrix},
\end{equation}
where $\mathcal{L}_{TI} \in \mathbb{R}^{N \times 1}$ captures the influence of leader node (target) on each follower (interceptors), and $\mathcal{L}_{II} \in \mathbb{R}^{N \times N}$ is the reduced Laplacian for the follower subgraph. 
\end{assumption}
\begin{lemma}
All eigenvalues of $ \mathcal{L}_{II} $ have strictly positive real parts, satisfying $0 < \text{Re}\{\lambda_1\} \leq \text{Re}\{\lambda_2\} \leq \cdots \leq \text{Re}\{\lambda_N\}$. 
\end{lemma}
\begin{lemma}[\cite{sp3}]\label{lem:ineq}
Under \Cref{asm:asm2}, the matrix $\mathcal{L}_{II}$ is invertible. 
Define $\mathbf{M} \coloneqq \mathrm{diag}(\boldsymbol{\mu})$, where $\boldsymbol{\mu} = (\mathcal{L}_{II}^\top)^{-1} \mathbf{1}_N = [\mu_1, \ldots, \mu_N]^\top$. 
Then, we have $\mathbf{M} \succ 0$  and $
\mathbf{C}(\mathcal{L}_{II}) = \mathbf{M} \mathcal{L}_{II} +\mathcal{L}_{II}^\top \mathbf{M} \succ 0.
$

\end{lemma}
Let $\mathscr{A} = (\mathscr{V}, \mathscr{E}) $ denote the \emph{actuation graph}, representing the interaction topology used for coordinating guidance commands among interceptors. Here, $\mathscr{V} = \{v_{1_a}, \dots, v_{N_a}\}$ consists only of the interceptor nodes, and $\mathscr{E} \subseteq \mathscr{V} \times \mathscr{V}$ is the set of directed edges that need not coincide with $\mathscr{S}$. Unlike the sensing graph  $\mathscr{S}$, the actuation graph $\mathscr{A}$ is assumed to be \emph{leaderless} and a strongly connected digraph. The associated Laplacian matrix with $\mathscr{A}$ is given by $\mathcal{L_I}$. 
\begin{remark}
The choice of different sensing and actuation graphs introduces significant design flexibility, which allows independent tuning of the estimation and control layers without compromising overall coordination and stability. This modular architecture allows sensing strategies to be tailored to information availability while the guidance layer simultaneously enforces time-to-go consensus and cooperative interception.
\end{remark}

\begin{lemma}[\cite{sp4}]\label{lem:L}
Consider a network of agents with a fixed topology represented by a strongly connected digraph $\mathcal{G}$ with Laplacian matrix $\mathcal{L}$. 
Let $\mathcal{\hat L}$ denote the Laplacian matrix of the underlying undirected graph $\mathcal{\hat G}$ (that is, the mirror of $\mathcal{G }$). 
Then, the network states $\mathbf{x}(t)$ satisfy $
\mathbf{x}^\top \mathcal{L} \mathbf{x} \ge \lambda_2(\mathcal{\hat G}) \|\mathbf{x}\|^2,
$
where $\lambda_2(\mathcal{\hat G})$ is the Fiedler eigenvalue of the mirror graph induced by $\mathcal{G}$.

\end{lemma}
\begin{figure}[h!]
    \centering
        \includegraphics[width=.5\linewidth]{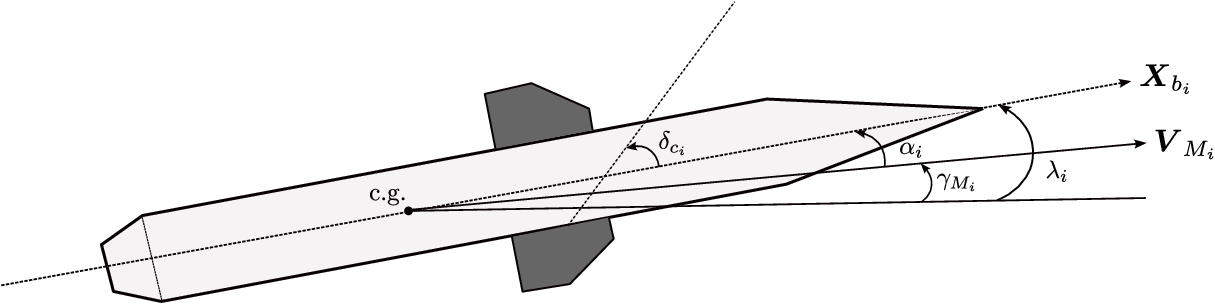}
        \caption{Planar dynamics of the \(i\)\textsuperscript{th} interceptor.}
        \label{planardynamics}
\end{figure}
In this work, we consider the $i$\textsuperscript{th} interceptor to be a roll-stabilized, skid-to-turn, cruciform type, which is steered exclusively by its canard aerodynamic surface, as illustrated in \Cref{planardynamics}. The planar (pitch plane) dynamics of such an interceptor can be described using
\begin{subequations}\label{eq:inter_dynamics}
     \begin{align}
    \dot \alpha_i&= q_i-\dfrac{a_{M_i}^{ac}}{V_{M_i}}=q_i-\dfrac{L_{{\alpha_i}}^\beta k_{1_i}(\alpha_i)+L_{\delta_{c_i}}k_{2_i}(\alpha_i+\delta_{c_i})}{V_{M_i}} ,\label{eq:alphai_dot}\\
    \dot \lambda_i&=q_i,\\
    q_i&= M_{q_i}q_i+M_{{\alpha_i}}^\beta k_{3_i}(\alpha_i)+M_{\delta_{c_i}}k_{4_i}(\alpha_i+\delta_{c_i}),\\
    \dot \delta_{c_i}&= \dfrac{\delta_{c_i}^c-\delta_{c_i}}{\tau_{c_i}},\label{eq:deltaci_dot}
     \end{align}
 \end{subequations}
 where $L_{{\alpha_i}}^\beta,\;L_{\delta_{c_i}}$ denotes the lift force coefficients whereas $M_{q_i},\;M_{{\alpha_i}}^\beta,\;M_{\delta_{c_i}}$ represents the pitching moment coefficients. In this formulation,  $\alpha_i$ and $\lambda_i$ denote the angle of attack and the pitch angle, respectively, while $\delta_{c_i}$ represents the canard deflection. The pitch rate is given by $q_i$. To achieve guidance objectives, the achieved lateral acceleration $a_{M_i}^{ac}$, generated by aerodynamic lift, must accurately track the desired command $a_{M_i}$ from \eqref{eq:aMd}.
 
The canard aerodynamic surface is modeled as a first-order actuator with a time-constant $\tau_{c_i}$ and the commanded canard deflection is given by $\delta_{c_i}^c$. The $i$\textsuperscript{th} interceptor's aerodynamic lift and moment characteristics are described by nonlinear functions $k_{m_i}(\cdot), \; m = 1,\ldots,4$ which are assumed to be smooth and bounded, thereby providing a realistic yet analytically tractable representation of the aerodynamic behavior. Although the exact forms of these functions are unknown, their rates are assumed to be bounded to ensure physically realistic changes in the aerodynamic forces and moments.
\begin{remark}
    In time-critical engagements, interceptors must achieve consensus in estimated target states and in the common interception time within a short duration. Moreover, it is also desirable to have an almost instantaneous autopilot response to minimize any lags. This motivates the use of a predefined-time approach, which guarantees necessary error convergence within a user-specified bound, irrespective of initial conditions, for precise and simultaneous interception.
\end{remark}

\begin{definition}[\cite{sp3}]\label{def:assignabletime}
Consider the dynamical system 
\begin{equation}\label{system}
    \dot{\mathbf{x}}(t) = g(t, \mathbf{x}(t)), \quad t \in \mathbb{R}_+, \quad \mathbf{x}(0) = \mathbf{x}_0,
\end{equation}
where \( \mathbf{x} \in \mathbb{R}^n \) is the state vector and 
\( g : \mathbb{R}_+ \times \mathbb{R}^n \to \mathbb{R}^n \) is a nonlinear vector field that is locally bounded uniformly in time.  The origin of \eqref{system} is said to be \emph{globally prescribed-time stable} if it is globally uniformly finite-time stable and the settling time $T$ can be chosen as a user-specified finite constant. In particular, for any $0 < T_p \leq T_{\max} < \infty$, there exists a prescribed settling time $T$ such that
$T_p \leq T \leq T_{\max},  \forall\, \mathbf{x}_0 \in \mathbb{R}^n$, where $T_p$ denotes the minimal physically realizable convergence time.
\end{definition}

Before proceeding further, we introduce a time-varying scaling function based on \cite{sp5} as 
\begin{align}\label{timevarying}
f(t,t_c) =
\begin{cases}
\displaystyle \frac{t_c}{\pi} \sin\left(\frac{\pi}{t_c} t \right) + t - t_c, & 0 \le t < t_c, \\
1, & t \ge t_c,
\end{cases}
\end{align} 
whose derivative is obtained as 
\begin{align}
\dot{f}(t,t_c) =
\begin{cases}
\displaystyle \cos\left(\frac{\pi}{t_c} t \right) + 1, & 0 \le t < t_c, \\[8pt]
0, & t \ge t_c.
\end{cases}
\end{align}
where $t_c$ is a user-specified time. 
\begin{remark}\label{remark:scaling_function}
   In this work, we adopt the trigonometric shaping functions employed in \eqref{timevarying} to ensure smooth and singularity-free convergence of the closed-loop trajectories. Specifically, since \( f(t,t_c) <0 \) and \(\dot{f}(t,t_c) > 0\) with \( |f(t,t_c)|  \) being monotonically decreasing for all $t\in [0,t_c)$,  and satisfying \(\lim_{t \to t_c^-}|\dot{f}(t,t_c)|= 0\). In subsequent analysis, it will be established that the corresponding control laws remain continuous, bounded, and well-defined throughout the entire finite convergence horizon $[0,t_c)$.
\end{remark}
\begin{lemma}[\cite{sp3}]\label{lem:ft}
Let $V(\mathbf{x}(t), t) : \mathcal{D} \times \mathbb{R}_+ \to \mathbb{R}$ 
be a continuously differentiable function and $\mathcal{D} \subset \mathbb{R}^n$ be a domain containing the origin. 
If there exists a real constant $b > 0$ such that $V(0, t) = 0$ and $V(\mathbf{x}(t), t) > 0 ~\in \mathcal{D} \setminus \{0\}$, with $\dot{V} \leq -bV - 2 \dfrac{\dot{\Theta}}{\Theta} ~V \in\ \mathcal{D}$ on $[t_0, \infty)$, then the origin of 
\eqref{system}  is prescribed-time stable with the prescribed time $T$ given by user through function $\Theta(t)$ satisfying $\dfrac{\dot \Theta (t)}{\Theta(t)}>0$. If $\mathcal{D} = \mathbb{R}^n$, then the origin of the system is globally prescribed-time stable with the prescribed time given by the function $\Theta(t)$. In addition, for 
$t \in [t_0, t_1)$, it holds that
\begin{equation}
    V(t) \leq \Theta^{-2} \exp^{-b(t - t_0)} \left(\Theta(t_0)\right)^2V(t_0),
\end{equation}
and, for $t \in [t_1, \infty)$, it holds that $V(t) \equiv 0$. 
\end{lemma}
We are now prepared to formally state the problems addressed in this paper.
\begin{problem}\label{pr:main}
Subject to \Cref{asm:seeker}, design a one-shot nonlinear cooperative predefined-time convergent joint estimation-guidance-control architecture to achieve simultaneous interception of a stationary target.
\end{problem}
In essence, \Cref{pr:main} decomposes into cohesive subproblems to guarantee a cooperative simultaneous interception. The first subproblem involves the design of a predefined-time distributed estimator that reconstructs the target states under heterogeneous sensing constraints. The estimated states are then employed within the guidance subsystem, which computes cooperative acceleration commands by enforcing predefined-time consensus on the estimated time-to-go values across the network. Once these commands are issued, the autopilot subsystem must guarantee sufficiently fast inner-loop dynamics to track these cooperative guidance commands almost instantaneously, thereby realizing predefined-time consensus in the canard deflections and preserving the overall synchronization established at the guidance level.

\section{Cooperative One-Shot Estimation-Guidance-Control Framework}\label{sec:main}
Designing cooperative salvo guidance laws requires an accurate estimation of time-to-go, although obtaining an exact value is usually not tractable in most scenarios. As a practical approach, time-to-go is generally approximated using ad-hoc solutions, assuming the interceptors are always confined close to the LOS with the target. For cases with large initial heading angles, this small-angle assumption may not be valid. As shown in \cite{sp1}, a refined estimate for the $i$\textsuperscript{th} interceptor's time-to-go can be given by
\begin{align}
   {{t}_\mathrm{go}}_i = \dfrac{r_i}{V_{M_i}} \left[1 + \dfrac{\sin^2 \sigma_i}{k_i}\right],\label{eq:tgo}
\end{align}
where $\sigma_i = \gamma_{M_i} - \theta_i$ is the lead angle, $k_i = 4N_i - 2$, with  $N_i$ being a constant greater than 2.

\begin{lemma}\label{lem:tgodynamics}
The time-to-go \eqref{eq:tgo} dynamics for the \ith~exhibits a relative degree of one with respect to its interceptor’s lateral acceleration.
\end{lemma}
\begin{proof}
On differentiating \eqref{eq:tgo} with respect to time and applying \eqref{eq:enggeo}, the expression 
\begin{equation}
    \dot{{t}}_{\mathrm{go}_i} =-\cos \sigma_i-\frac{\sin^2 \sigma_i\cos\sigma_i}{k_i}+\frac{\sin2\sigma_i\sin \sigma_i}{k_i}+\frac{\ r_ia_{M_i}\sin 2\sigma_i}{k_i{V_M}_i^2}
    \label{tgodynamics}
\end{equation}  
is obtained, showing that the lateral acceleration $a_{M_i}$ appears in the first derivative of $ t_{\mathrm{go}_i}$.
\end{proof}

In cooperative simultaneous interception, the guidance command $a_{M_i}$ for regulating the engagement duration (by manipulating \eqref{eq:tgo}) necessitates access to specific engagement variables, such as relative range and lead angles, as evident from \cref{lem:tgodynamics}. However, such variables are not always available to all interceptors if a subset of interceptors do not have seekers installed on them. Consistent with~\Cref{asm:seeker}, seeker-less interceptors must therefore estimate the required engagement parameters before cooperative guidance design is done.

Let $\mathbf{p} = [x_T,\, y_T]^\top$ denote the target's position vector in the inertial frame, and $\hat{\mathbf{p}}_{i}$ represent the estimate of $\mathbf{p}$ as obtained by the \ith. Suppose $\boldsymbol{\varepsilon}_{i}$ represents the corresponding inter-interceptor relative estimation error. Then, the prescribed-time distributed observer (PTDO) for the \ith~to estimate $\mathbf{p}$ is given by

\begin{align}
\dot{\hat{\mathbf{p}}}_i(t) =
-\left(\alpha - \beta\frac{\dot f(t,t_p)}{f(t,t_P)}\right)\boldsymbol{\varepsilon_i},
\label{eq:Stateobserver}
\end{align}
where  $\alpha\in\mathbb{R}_+$ and $\beta\geq \dfrac{2\lambda_{\max}(\mathbf{M})}{\lambda_1(\mathbf{C})}$ are  design parameters with
\begin{align}
    \boldsymbol{\varepsilon_i} =& a_{i0} (\hat{\mathbf{p}}_i -{\mathbf{p}})+ \sum_{j=1}^N a_{ij} (\hat{\mathbf{p}}_i -\hat{\mathbf{p}}_j).
    \label{relative estimation error}
\end{align}
\begin{remark}
Seeker-equipped agents obtain relative range and bearing measurements of the target, and by integrating these with their self-positions derived from the inertial navigation system, they can accurately determine the target’s position $(x_T, y_T)$ in the inertial reference frame for subsequent tasks.
\end{remark}
\begin{theorem}[Distributed Estimation of Target's States]\label{thm:obs}
Consider the cooperative multi-interceptor target engagement scenario given by \eqref{eq:enggeo}, consisting of seeker-equipped and seeker-less agents. If the \ith~employs a distributed observer \eqref{eq:Stateobserver} over a directed sensing graph $\mathscr{S}$, then its own local target's state estimate converges to the 
true target's position within a predefined time $t_p$ regardless of the initial estimates, that is, 
\begin{align}
    \lim_{t \to t_p^-} \left(\hat{\mathbf{p}}_i(t)-\mathbf{p}(t)\right) &= 0.\label{observerstab}
\end{align}
\end{theorem}
\begin{proof}
Denote $\hat{\mathbf{p}} = [\hat{\mathbf{p}}_{1}^\top, \ldots, \hat{\mathbf{p}}_{N}^\top]^\top
$, $\boldsymbol{\varepsilon} = [\boldsymbol{\varepsilon}_{1}^\top, \ldots, \boldsymbol{\varepsilon}_{N}^\top]^\top
$ and let the estimation error for the  $i$\textsuperscript{th} interceptor be \(\tilde{\mathbf{p}}_{i} = \hat{\mathbf{p}}_{i} - \mathbf{p}\) such that $\tilde{\mathbf{p}} = [\tilde{\mathbf{p}}_{1}^\top, \ldots, \tilde{\mathbf{p}}_{N}^\top]^\top
$. Then,
\begin{equation}
\dot{\tilde{\mathbf{p}}} = - (\mathbf{I}_N \otimes \mathbf{I}_{2})\left(\alpha-\beta\frac{\dot f(t,t_p)}{f(t,t_p)}\right)\boldsymbol{\varepsilon}.
\end{equation}
From \eqref{relative estimation error}, it follows that \(\boldsymbol{\varepsilon} = (\mathcal{L}_{II} \otimes \mathbf{I}_{2}) \tilde{\mathbf{p}}\), whose dynamics becomes
\begin{equation}
\dot{\boldsymbol{\varepsilon}} = - (\mathcal{L}_{II} \otimes \mathbf{I}_{2})\left(\alpha-\beta\frac{\dot f(t,t_p)}{f(t,t_p)}\right)\boldsymbol{\varepsilon}.
\label{errorequation}
\end{equation}
Let us now define an auxiliary function $\Theta_p(t)$ such that 
\begin{equation}
    \Theta_p(t)=\frac{1}{f(t,t_p)}\label{eq:auxfn}.
\end{equation}
On differentiating \eqref{eq:auxfn}, we get,
\begin{equation}
    \frac{\dot \Theta_p(t)}{\Theta_p(t)}=-\frac{\dot f(t,t_p)}{f(t,t_p)},
\end{equation}
where the right-hand side in the above equation is always positive since the ratio $\dfrac{\dot f(t,t_p)}{f(t,t_p)}<0~\forall~t\in [0,t_p)$. Consequently, \eqref{errorequation}, in terms of the auxiliary function, becomes

\begin{equation}
\dot{\boldsymbol{\varepsilon}} = - (\mathcal{L}_{II} \otimes \mathbf{I}_{2})\left(\alpha+\beta\frac{\dot \Theta_p(t)}{\Theta_p(t)}\right)\boldsymbol{\varepsilon}.
\label{errorequation1}
\end{equation}
Now, consider a Lyapunov function candidate 
\begin{align}
V_1=\boldsymbol{\varepsilon}^\top(\mathbf{M}\otimes \mathbf{I}_{2})\boldsymbol{\varepsilon},\label{Lyapunov}
\end{align}
whose time derivative is given as
\begin{align}
\dot{V_1}
=2\boldsymbol{\varepsilon}^\top(\mathbf{M}\otimes \mathbf{I}_{2})\dot {\boldsymbol{\varepsilon}}.
\end{align}
It follows from \eqref{errorequation1} that
\begin{align}
\dot{V_1}& 
=2\boldsymbol{\varepsilon}^\top(\mathbf{M}\otimes \mathbf{I}_{2})\big[- (\mathcal{L}_{II} \otimes \mathbf{I}_{2})\left(\alpha+\beta\frac{\dot \Theta_p(t)}{\Theta_p(t)}\right)\boldsymbol{\varepsilon}\big]\notag =-\left(\alpha+\beta\frac{\dot \Theta_p(t)}{\Theta_p(t)}\right)\boldsymbol{\varepsilon}^\top\big[ (\mathbf{M} \mathcal{L}_{II} +\mathcal{L}_{II}^\top\mathbf{M})\otimes \mathbf{I}_{2} \big]\boldsymbol{\varepsilon},
\end{align}
which can be simplified using the results in \Cref{lem:ineq} to
\begin{align}
    \dot{V_1}& 
=-\left(\alpha+\beta\frac{\dot \Theta_p(t)}{\Theta_p(t)}\right)\boldsymbol{\varepsilon}^\top(\mathbf{C}\otimes\mathbf{I}_{2} )\boldsymbol{\varepsilon} .
\end{align}
Upon further simplification based on Rayleigh inequality and \Cref{lem:L}, we obtain
\begin{align}
\dot{V_1}
&\leq -\alpha \lambda_1(\mathbf{C})\boldsymbol{\varepsilon}^\top\boldsymbol{\varepsilon}-\beta\frac{\dot \Theta_p(t)}{\Theta_p(t)}\lambda_1(\mathbf{C})\boldsymbol{\varepsilon}^\top\boldsymbol{\varepsilon}\notag \leq -\alpha \frac{\lambda_1(\mathbf{C})}{\lambda_{\max}(\mathbf{M})}V_1-\beta\frac{\dot \Theta_p(t)}{\Theta_p(t)}\frac{\lambda_1(\mathbf{C})}{\lambda_{\max}(\mathbf{M})}V_1 .\label{vdott}
\end{align}
Since the design parameters are $\alpha\in\mathbb{R}_+$ and $\beta\geq \dfrac{2\lambda_{\max}(\mathbf{M})}{\lambda_1(\mathbf{C})}$, \eqref{vdott} can be expressed as
\begin{align}
    \dot V_1\leq -\hat \alpha V_1-2\frac{\dot \Theta_p(t)}{\Theta_p(t)}V_1 ,
\end{align}
where $\hat\alpha=\alpha \dfrac{\lambda_1(C)}{\lambda_{\max}(M)}>0$. Following \Cref{lem:ft}, we have
\begin{align}\label{vinequality2}
 \Theta_p(t)^2 V_1(t) &\leq \exp\left( -\hat{\alpha}t \right) \Theta_p(0)^2 V_1(0) =\frac{1}{(t_p)^2}\exp\left( -\hat{\alpha}t \right)V_1(0)  ,
\end{align}
which further implies that

\begin{align}
\|\boldsymbol{\varepsilon}(t)\|^2_2 &\leq \frac{1}{\Theta_p(t)^2t_p^2} \exp\left( -\frac{\alpha \lambda_{1}({\mathbf{C}})}{\lambda_{\max}(\mathbf{M})}t \right) \frac{\lambda_{\max}(\mathbf{M})}{\lambda_{\min}(\mathbf{M})} \|\boldsymbol{\varepsilon}(0)\|^2_2 \label{eq:epsilonsquare} 
\end{align}
which  leads to
\begin{align}
\|\boldsymbol{\varepsilon}(t)\|_2 &\leq\dfrac{|f(t,t_p)|}{t_p} \exp\left\{\left( -\frac{\alpha \lambda_{1}({\mathbf{C}})}{2\lambda_{\max}(\mathbf{M})} \right) t\right\}
  \times \sqrt{ \frac{\lambda_{\max}(\mathbf{M})}{\lambda_{\min}(\mathbf{M})}} \left\|\boldsymbol{\varepsilon}(0)\right\|_2  \label{eq:epsilon}
\end{align}
and then that
\begin{align}
\|\tilde{\mathbf{p}}(t)\|_2 &= \left\| (\mathcal{L}_{II} \otimes \mathbf{I}_2)^{-1} \boldsymbol{\varepsilon}(t) \right\|_2 \leq \left\| (\mathcal{L}_{II} \otimes \mathbf{I}_2)^{-1} \right\|_2  \left\|\boldsymbol{\varepsilon}(t)\right\|_2 \\
&\leq \dfrac{|f(t,t_p)|}{t_p} \exp\left\{\left( -\frac{\alpha \lambda_{1}({\mathbf{C}})}{2\lambda_{\max}(\mathbf{M})} \right) t\right\}
  \times \sqrt{ \frac{\lambda_{\max}(\mathbf{M})}{\lambda_{\min}(\mathbf{M})}}\left\| (\mathcal{L}_{II} \otimes \mathbf{I}_2)^{-1} \right\|_2\; \left\|\boldsymbol{\varepsilon}(0)\right\|_2 ,
\end{align}
since $|f(t,t_p)|\to0$ as $t\to t_p^-$, this essentially leads us to arrive at
\begin{equation}
\lim_{t\to t_p^-}\left\|\mathbf{\tilde p}(t)\right\|_2 \to 0 .
\end{equation}
This essentially means that $\lim_{t \to t_p^-} (\hat{\mathbf{p}}_i(t)-\mathbf{p}(t))= 0$ in accordance with the results in \Cref{lem:ft} and consensus in the estimates is preserved over the interval $[t_p,\infty)$, since \eqref{Lyapunov} is identically zero for all $t \ge t_p$. This completes the proof.
\end{proof}
\begin{remark}\label{rmk:correctionboundedness}
    Recall \eqref{errorequation}, where the term $\left(\alpha-\beta\dfrac{\dot f(t,t_p)}{f(t,t_p)}\right)\boldsymbol{\varepsilon}$ in its right-hand side dictates the error dynamics when $\boldsymbol{\varepsilon}\neq 0$, that is, for $t<t_p$. From the triangle inequality, it is apparent that
\begin{align}\label{eq:inequality_observer}
     \left\vert\left\vert\left(\alpha-\beta\dfrac{\dot f(t,t_p)}{f(t,t_p)}\right)\boldsymbol{\varepsilon}\right\vert\right\vert_2&
     \leq |\alpha|\;\|\boldsymbol{\varepsilon}(t)\|_2+\left|\beta\frac{\dot f(t,t_p)}{f(t,t_P)}\right|\|\boldsymbol{\varepsilon}(t)\|_2.
\end{align}
Since $\alpha>0$ and $\beta\geq \dfrac{2\lambda_{\max}(\mathbf{M})}{\lambda_1(\mathbf{C})}>0$, it follows from \eqref{eq:epsilon} and \eqref{eq:inequality_observer} that 
\begin{align}
    \left\vert\left\vert\left(\alpha-\beta\dfrac{\dot f(t,t_p)}{f(t,t_p)}\right)\boldsymbol{\varepsilon}\right\vert\right\vert_2 \leq \dfrac{\left(\alpha|f(t,t_p)|+\beta|\dot f(t,t_p)|\right)}{t_p}\exp\left\{\left( -\frac{\alpha \lambda_{1}({\mathbf{C}})}{2\lambda_{\max}(\mathbf{M})} \right) t\right\}
   \sqrt{ \frac{\lambda_{\max}(\mathbf{M})}{\lambda_{\min}(\mathbf{M})}}\left\|\boldsymbol{\varepsilon}(0)\right\|_2.\label{eq:control_effort_bound}
\end{align}
As both $|f(t,t_{p})|$ and $|\dot f(t,t_{p})|$ are smooth, decaying, and bounded over the interval $[0, t_{p})$, it follows from \eqref{eq:epsilonsquare} and \eqref{eq:control_effort_bound} that $\|\boldsymbol{\varepsilon}(t)\|_2
\in L_{\infty}$\footnote{For any signal $x(t)$, one has $
L_{\infty} \coloneqq \{ x(t) \mid x:\mathbb{R}_{+} \rightarrow \mathbb{R},\; \sup_{t\in\mathbb{R}_{+}} |x(t)| < \infty \}
$.} and $\left\vert\left\vert\left(\alpha-\beta\dfrac{\dot f(t,t_p)}{f(t,t_p)}\right)\boldsymbol{\varepsilon}\right\vert\right\vert_2\in L_{\infty}$ on $[0, t_{p})$, which verifies that this control effort term is uniformly bounded over this interval. Furthermore, from \Cref{lem:ft}, since consensus is preserved for all $t\geq t_{p}$, it implies that this term remains identically zero on $[t_{p},\, \infty)$.
\end{remark}

Note that the variables such as relative range and LOS angle can be computed based on the local estimates of the \ith,~which converge to the true states after $t=t_p$. To this end, we have
\begin{align}
    \hat{r}_i=& \sqrt{(\hat x_{T_i}-x_{I_i})^2+(\hat y_{T_i}-y_{I_i})^2},~~
    \hat{\theta}_i= \tan^{-1}\frac{\hat y_{T_i}-y_{I_i}}{\hat x_{T_i}-x_{I_i}}.
\end{align}
Therefore, the equations of relative motion \eqref{eq:enggeo} are essentially
\begin{subequations}\label{eq:enggeohat}
    \begin{align}
\hat V_{r_i} &=  - V_{M_i} \cos ({\gamma_{M}}_i-\hat\theta_i) = -V_{M_i} \cos\hat \sigma_i , \\
\hat V_{\theta_i} &= - V_{M_i} \sin ({\gamma_{M}}_i-\hat\theta_i) = -V_{M_i} \sin\hat\sigma_i,
\end{align}
\end{subequations}
Let the error in common interception time for the $i$\textsuperscript{th} interceptor be defined as  
\begin{align}
    e_i = t_{\mathrm{go}_i} - t_{\mathrm{go}_d}, \label{eq:ei}
\end{align}
where the desired common interception time $t_{\mathrm{go}_d}$ is not given prior to the engagement. Instead, the interceptors cooperatively converge on it during the engagement. It is also worth noting that achieving consensus in $t_{\mathrm{go}_i}$ values is equivalent to agreement in $e_i$, since 
$t_{\mathrm{go}_i} - t_{\mathrm{go}_j} = e_i - e_j,$ for any pair of interceptors $i,j$. Moreover, each $e_i$ must also converge to zero to ensure a cooperative salvo. The cooperative guidance command for the $i$\textsuperscript{th} interceptor is presented in the following theorem. 

\begin{theorem}[Cooperative Guidance Command for Simultaneous Interception]\label{amtheorem}
   Consider the cooperative multi-interceptor target engagement scenario given by \eqref{eq:enggeo}, consisting of seeker-equipped and seeker-less agents, whose time-to-go values are estimated using \eqref{eq:tgo}. Each interceptor cooperatively exchanges its guidance command,  
    \begin{align}
    a_{M_i} =
    &V_{M_i} \dot{\hat{\theta}}_i+\frac{V_{M_i}^2 k_i \big( \cos\hat{\sigma}_{i} - 1 \big) }{\hat r_i \sin 2\hat{\sigma}_{i} } 
    + \frac{V_{M_i}^2 \sin\hat{\sigma}_{i} }{ 2 \hat r_i } 
     +\frac{V_{M_i}^2 k_i }{\hat  r_i \sin 2\hat{\sigma}_{i} } \left(-\zeta+\eta\frac{\dot f(t,t_e)}{f(t,t_e)}\right)\sum _{j=1}^N\left[\mathcal{L}_I\right]_{ij} e_j
\label{eq:am},
    \end{align}
     for some $\zeta>0,\eta \geq \dfrac{1}{\lambda_2(\widehat{\mathscr{A}})}$ over a directed actuation graph $\mathscr{A}$ to establish group consensus on their time-to-go within the predefined time $t_e>t_p$. This consequently leads to a simultaneous interception of the stationary target at a time determined cooperatively during the engagement.
\end{theorem}
\begin{proof}
Differentiating \eqref{eq:ei} we get, $\dot e_i=\dot t_{{\rm go}_i}+1$. On substituting \eqref{tgodynamics} in \eqref{eq:ei}, for the \ith, one obtains 
\begin{align}
\dot{e}_i &=  1 - \cos{\sigma}_i
    - \frac{\sin^2{\sigma}_i\cos{\sigma}_i}{k_i}
    + \frac{\sin 2{\sigma}_i \sin{\sigma}_i}{k_i}+
    \frac{{r}_i \sin 2{\sigma}_i}{k_i V_{M_i}^2} 
a_{M_i}.\label{eq:eidotactual}
\end{align}
Upon substituting the cooperative guidance command \eqref{eq:am} in \eqref{eq:eidotactual}, one may obtain 

\begin{align}
\dot{e}_i &=  1 - \cos{\sigma}_i
    - \frac{\sin^2{\sigma}_i\cos{\sigma}_i}{k_i}
    + \frac{\sin 2{\sigma}_i \sin{\sigma}_i}{k_i}\nonumber\\
    &+
    \frac{{r}_i \sin 2{\sigma}_i}{k_i V_{M_i}^2}\left[V_{M_i} \dot{\hat{\theta}}_i+\frac{V_{M_i}^2 k_i \big( \cos\hat{\sigma}_{i} - 1 \big) }{\hat r_i \sin 2\hat{\sigma}_{i} } 
    + \frac{V_{M_i}^2 \sin\hat{\sigma}_{i} }{ 2 \hat r_i } 
     + \frac{V_{M_i}^2 k_i }{\hat  r_i \sin 2\hat{\sigma}_{i} } \left(-\zeta+\eta\frac{\dot f(t,t_e)}{f(t,t_e)}\right)\sum _{j=1}^N\left[\mathcal{L}_I\right]_{ij} e_j\right]  \label{eq:eidotinestimated}
\end{align}
On letting 
\begin{align*}
    F_i=& 1 - \cos{\sigma}_i
    - \frac{\sin^2{\sigma}_i\cos{\sigma}_i}{k_i }
    + \frac{\sin 2{\sigma}_i \sin{\sigma}_i}{k_i},~\hat{F}_i= 1 - \cos{\hat\sigma}_i
    - \frac{\sin^2{\hat\sigma}_i\cos{\hat\sigma}_i}{k_i}
    + \frac{\sin 2{\hat\sigma}_i \sin{\hat\sigma}_i}{k_i}\\
    B_i=& \frac{{r}_i \sin 2{\sigma}_i}{k_i V_{M_i}^2},~\hat{B}_i= \frac{{\hat r}_i \sin 2{\hat\sigma}_i}{k_i V_{M_i}^2},~u_{c_i} = \left(-\zeta+\eta\frac{\dot f(t,t_e)}{f(t,t_e)}\right)\sum _{j=1}^N\left[\mathcal{L}_I\right]_{ij} e_j,
\end{align*}
we can write \eqref{eq:eidotinestimated} as
\begin{align}
    \dot{e}_i &= F_i + B_i a_{M_i} = F_i + B_i \left(\dfrac{{u_c}_i-\hat F_i}{\hat B_i}\right).
\end{align}
Let us now define the error variables $\tilde F_i= F_i-\hat F_i$ and $\tilde B_i= B_i-\hat B_i$, which allows us to express the time-to-go error dynamics as
\begin{align}
    \dot e_i &= F_i-\frac{B_i}{B_i- \tilde{B_i}}(F_i-\tilde F_i)+\frac{B_i}{B_i- \tilde{B_i}}{u_c}_i   .
\end{align}
Define $\psi_i=\dfrac{\tilde B_i}{B_i}$ and $\Phi_i=\dfrac{\tilde F_i}{F_i}$, then it follows from \Cref{thm:obs} that $\lim_{t\to t_p} \psi_i\to 0,~\lim_{t\to t_p}\Phi_i\rightarrow 0$, and
\begin{align}
    \dot e_i=F_i\frac{\Phi_i-\psi_i}{1-\psi_i}+\frac{{u_c}_i}{1-\psi_i}.
\end{align}
To streamline the notations further, we treat the variables in vector/matrix form for compactness. To this end, define
\begin{align}
    \mathbf{u}_c = &[u_{c_1}\;\; u_{c_2}\;\; \cdots\;\; u_{c_N}]^\top
= \left(-\zeta+\eta\dfrac{\dot f(t,t_e)}{f(t,t_e)}\right)\mathcal{L}_I(\mathcal{A})\mathbf{e}
\end{align}
with $\mathbf{e}=[e_1\;\; e_2\;\; \cdots\;\; e_N]^\top$. Similarly, $\boldsymbol{\Phi} = [\Phi_1 \; \cdots \; \Phi_N]^\top$,  
$\boldsymbol{\psi} = [\psi_1 \; \cdots \; \psi_N]^\top$, and $\mathbf{F}=\mathrm{diag}(F_1\; \cdots \; F_N)$. Then, one also has
\begin{align}
    \bar{\mathbf{D}}_{\boldsymbol{\psi}} = 
\left[\dfrac{\Phi_1-\psi_1}{1-\psi_1}\;\; \cdots\;\; \frac{\Phi_N-\psi_N}{1-\psi_N}\right]^\top,~~\mathbf{D}_{\boldsymbol{\psi}} = 
\mathrm{diag}\!\left(\frac{1}{1-\psi_1}\;\; \cdots\;\; \frac{1}{1-\psi_N}\right),
\end{align}
which leads us to write
\begin{align}
    \dot{\mathbf{e}} &= \mathbf{F}{\bar{\mathbf{D}}}_{\boldsymbol{\psi}} + \mathbf{D}_{\boldsymbol{\psi}}\mathbf{u}_c.\label{eq:ecl}
\end{align}
Consider a Lyapunov function candidate $V_2 = \dfrac{1}{2}\mathbf{e}^\top\mathbf{e}$,
whose time derivative is given as
\begin{align}
    \dot{V}_2 &= \mathbf{e}^\top\dot{\mathbf{e}}
    = \mathbf{e}^\top\left(\mathbf{F}{\bar{\mathbf{D}}}_{\boldsymbol{\psi}} + \mathbf{D}_{\boldsymbol{\psi}}\mathbf{u}_c\right) = \mathbf{e}^\top\mathbf{F}{\bar{\mathbf{D}}}_{\boldsymbol{\psi}}
    + \mathbf{e}^\top\mathbf{D}_{\boldsymbol{\psi}}
    \left(-\zeta + \eta\frac{\dot{f}(t,t_e)}{f(t,t_e)}\right)\mathcal{L}_I\mathbf{e}.
\end{align}
It is now our goal to ascertain the stability of the closed-loop error dynamics \eqref{eq:ecl}, which is to say that the system, while remaining uniformly ultimately bounded prior to the observer error convergence, still achieves global predefined-time stability.

Since the scaling functions satisfy $\dfrac{\dot{f}(t,t_e)}{f(t,t_e)} < 0$ for $t < t_e$, we can introduce an auxiliary function $\Theta_e(t) = \dfrac{1}{f(t,t_e)}$, which on differentiating with respect to time gives 
$\dfrac{\dot{\Theta_e}(t)}{\Theta_e(t)} = -\dfrac{\dot{f}(t,t_e)}{f(t,t_e)} > 0$. Notice that 
$[\mathbf{D}_{\boldsymbol{\psi}}]_{ii}=\dfrac{1}{1-\psi_i}=\dfrac{B_i}{\hat B_i}
=\dfrac{r_i \sin2\sigma_i}{\hat r_i\sin2\hat \sigma_i}>0$. Define
\begin{align}
    d_{\min} \coloneqq \inf_{0 \le s < t_p} [\mathbf{D}_{\boldsymbol{\psi}}(s)]_{ii} > 0,\quad
d_{\max} \coloneqq \sup_{0 \le s < t_p} [\mathbf{D}_{\boldsymbol{\psi}}(s)]_{ii},
\end{align}
then $d_{\min}\mathbf{I}\le \mathbf{D}_{\boldsymbol{\psi}} \le d_{\max}\mathbf{I}$. 
Moreover, $\lim_{t\to t_p} \mathbf{D}_{\boldsymbol{\psi}}=\mathbf{I}\implies d_{\min}=d_{\max}=1$. Therefore,
\begin{align}
    \dot{V}_2 &\le \mathbf{e}^\top\mathbf{F}{\bar{\mathbf{D}}}_{\boldsymbol{\psi}}
    - d_{\min}\mathbf{e}^\top\big(\zeta+\eta\frac{\dot \Theta_e(t)}{\Theta_e(t)}\big)\mathcal{L}_I\mathbf{e} = \mathbf{e}^\top\mathbf{F}{\bar{\mathbf{D}}}_{\boldsymbol{\psi}}
    - d_{\min}\zeta\,\mathbf{e}^\top\mathbf{e}
    - d_{\min}\eta\frac{\dot \Theta_e(t)}{\Theta_e(t)}\,\mathbf{e}^\top\mathcal{L}_I\mathbf{e} \nonumber\\
    &\le \mathbf{e}^\top\mathbf{F}\mathbf{D}_{\boldsymbol{\psi}}
    \,{\mathbf{D}}_{\boldsymbol{\Phi}}
    - 2d_{\min}\zeta V_2
    - 2d_{\min}\eta\frac{\dot \Theta_e(t)}{\Theta_e(t)}\lambda_2(\widehat{\mathscr{A}})V_2 \nonumber\\
    &\le d_{\min}\left[\frac{d_{\max}}{d_{\min}}\,
    \mathbf{e}^\top\mathbf{F}\mathbf{D}_{\boldsymbol{\Phi}}
    - 2\zeta V_2
    - 2\eta\frac{\dot \Theta_e(t)}{\Theta_e(t)}\lambda_2(\widehat{\mathscr{A}})V_2      \right],\label{eq:V2dot}
\end{align}
where $\mathbf{D}_{\boldsymbol{\Phi}}=[\Phi_1 - \psi_1,\; \ldots,\; \Phi_N - \psi_N]^\top$. Since $\eta \ge \dfrac{1}{\lambda_2(\widehat{\mathscr{A}})}$, letting $\hat{\zeta}=2\zeta$ allows us to simplify \eqref{eq:V2dot} to
\begin{align}
    \dot V_2 &\le d_{\min}\left[\frac{d_{\max}}{d_{\min}}\,
    \mathbf{e}^\top\mathbf{F}\mathbf{D}_{\boldsymbol{\Phi}}
    - \hat{\zeta} V_2
    - 2\frac{\dot \Theta_e(t)}{\Theta_e(t)}V_2\right].\label{pre_v_dot}
\end{align}
Thereafter, note that $\|\mathbf{e}^\top\mathbf{F}\mathbf{D}_{\boldsymbol{\Phi}}\|_2 
    \leq \|\mathbf{e}\|_2 \|\mathbf{F}\|_2 \|\mathbf{D}_{\boldsymbol{\Phi}}\|_2  \leq \sqrt{\mathbf{e}^\top\mathbf{e}}\,\|\mathbf{F}\|_2 \|\mathbf{D}_{\boldsymbol{\Phi}}\|_2 = \sqrt{2V_2}\,\|\mathbf{F}\|_2 \|\mathbf{D}_{\boldsymbol{\Phi}}\|_2$. On using these relations in \eqref{pre_v_dot}, one may obtain
\begin{align}
    \dot{V}_2 &\leq d_{\min}\left[
    \dfrac{d_{\max}}{d_{\min}}\sqrt{2V_2}\,\|\mathbf{F}\|_2\|\mathbf{D}_{\boldsymbol{\Phi}}\|_2
    - \hat{\zeta} V_2
    - 2\dfrac{\dot{\Theta_e}(t)}{\Theta_e(t)}V_2
    \right]\label{v1dot}
\end{align}

Further, define $\sqrt{V_2}= \xi$, $\kappa(t)=\dfrac{d_{\max}\|\mathbf{F}\|_2\|\mathbf{D}_{\boldsymbol{\Phi}}\|_2}{\sqrt{2}}$ and $\upsilon(t)=\left(\hat \zeta+2\dfrac{\dot{\Theta}_e(t)}{\Theta_e(t)}\right)\dfrac{d_{\min}}{2}$. After using these relations to simplify \eqref{v1dot}, one may obtain the differential inequality,
\begin{align}
    \dot \xi(t)+\upsilon(t)\xi(t)\leq\kappa(t) .\label{xiode}
\end{align}
As \eqref{xiode} is a first-order differential inequality in $\xi$, its solution is given as
\begin{align}
\xi(t)&\leq \exp\left\{-\int _0^t \upsilon(s) \,ds\right\} \xi(0)+ \exp\left\{-\int _0^t \upsilon(s) \,ds\right\}\int_0^t \exp\left\{\int _o^\tau \upsilon(s) \,ds\right\}\kappa(\tau)\,d\tau\label{eq:mainxi1}\\ 
&= \exp\left\{-\int _0^t \upsilon(s) \,ds\right\} \xi(0)+ \int_0^t \exp\left\{-\int _\tau^t \upsilon(s) \,ds\right\}\kappa(\tau)\,d\tau\nonumber
\end{align}
which on simplification leads to
\begin{align}
    \xi(t)&\leq\exp\left\{{\dfrac{-\hat \zeta d_{\min}t}{2}}\right\}\left(\left|\dfrac{\Theta_e(0)}{\Theta_e(t)}\right|\right)^{d_{\min}}\xi(0)+\dfrac{1}{\left(|\Theta_e(t)|\right)^{d_{\min}}} \int_0^t \exp\left\{{\dfrac{-\hat \zeta d_{\min}(t-\tau)}{2}}\right\}\left(|{\Theta_e(\tau)}|\right)^{d_{\min}} \kappa(\tau)\,d\tau .\label{eq:xi1}
\end{align}

By design, $\left(|{\Theta_e(t)}|\right)^{d_{\min}} \kappa(t)$ is a continuous function. Thus, if we denote this term by some $\omega(t)$, it follows that $\omega(t)=\left(|{\Theta_e(t)}|\right)^{d_{\min}} \kappa(t)\leq \omega_{\max}$. This fact can be leveraged to simplify \eqref{eq:xi1} further to
\begin{align}
    \xi(t) &\leq \dfrac{1}{\left(|\Theta_e(t)|\right)^{d_{\min}}}\left(\exp\left\{{\dfrac{-\hat \zeta d_{\min}t}{2}}\right\}\left(\left|{\Theta_e(0)}\right|\right)^{d_{\min}}\xi(0)+\dfrac{2\omega_{\max}}{\hat \zeta d_{\min}}\left[1- \exp\left\{{\dfrac{-\hat \zeta d_{\min}t}{2}}\right\}\right]\right)\nonumber\\
    &\leq \dfrac{1}{\left(|\Theta_e(t)|\right)^{d_{\min}}}\left(\exp\left\{{\dfrac{-\hat \zeta d_{\min}t}{2}}\right\}\left(\left|{\Theta_e(0)}\right|\right)^{d_{\min}}\xi(0)+\dfrac{2\omega_{\max}}{\hat \zeta d_{\min}}\left[1- \exp\left\{{\dfrac{-\hat \zeta d_{\min}t_p}{2}}\right\}\right]\right).\label{eq:xiineq}
\end{align}
Recall that $\|\mathbf{e}\|_2=\sqrt{\mathbf{e}^\top\mathbf{e}}=\sqrt{2V_2}=\sqrt{2}\xi$. Then, it follows from \eqref{eq:xiineq} that 
\begin{align}
    \|\mathbf{e}(t)\|_2&\leq \sqrt{2}\left(|f(t,t_e)|\right)^{d_{\min}} \left(\exp\left\{{\dfrac{-\hat \zeta d_{\min}t}{2}}\right\}\dfrac{ \|\mathbf{e}(0)\|_2}{\sqrt{2 }(t_e)^{d_{\min}}}+\dfrac{2\omega_{\max}}{\hat \zeta d_{\min}}\left[1- \exp\left\{{\dfrac{-\hat \zeta d_{\min}t_p}{2}}\right\}\right]\right).\label{eq:errortp}
\end{align}
Owing to the fact that $|f(t,t_e)|$  is monotonically decreasing and that $d_{\min}>0$ and $\hat{\zeta} > 0$ for $t < t_p$, it follows from \eqref{eq:errortp} that $\|e(t)\|_2$ remains ultimately bounded and converges to a compact neighborhood of the origin, the size of which is determined by the user-selected design parameter $\hat{\zeta}$.

For $t \geq t_p$, it is immediate from \Cref{thm:obs} that $\|\mathbf{D}_{\boldsymbol{\Phi}}\|_2 = 0$  which implies that $\kappa(t)=0$ and $d_{\min} = d_{\max} = 1$. Then, \eqref{eq:mainxi1} can be written as 
\begin{align}
    \xi(t) \le \exp\left\{-\left(\int_0^{t_p} \upsilon(s) \,ds + \int_{t_p}^t \upsilon(s) \,ds\right)\right\} \xi(0) + \int_0^{t_p} \exp\left\{-\int_\tau^t \upsilon(s) \,ds\right\}\kappa(\tau)\,d\tau,
\end{align}
which can be simplified to
\begin{align}\label{eq:xi1final}
  \xi(t)\leq &\exp\left\{{\dfrac{-\hat \zeta d_{\min}t_p}{2}}\right\}\left|\dfrac{\Theta_e(0)}{\Theta_e(t_p)}\right|^{d_{\min}}\exp\left\{{\dfrac{-\hat \zeta(t-t_p)}{2}}\right\}\left|\dfrac{\Theta_e(t_p)}{\Theta_e(t)}\right|\xi(0)\nonumber\\
  &+\dfrac{1}{\left(|\Theta_e(t)|\right)^{d_{\min}}} \int_0^{t_p} \exp\left\{{\dfrac{-\hat \zeta d_{\min}(t-\tau)}{2}}\right\}\left(|{\Theta_e(\tau)}|\right)^{d_{\min}} \kappa(\tau)\,d\tau  .
\end{align}
Define an auxiliary variable $\chi=\exp\left\{{\dfrac{-\hat \zeta d_{\min}t_p}{2}}\right\}\dfrac{\left|\Theta_e(0)\right|^{d_{\min}}}{\left|\Theta_e(t_p)\right|^{d_{\min}-1}}>0$. Then the above equation can be expressed as

\begin{align}
    \xi(t)& \leq \exp\left\{{\dfrac{-\hat \zeta(t-t_p)}{2}}\right\}\dfrac{\chi \xi(0)}{\left|\Theta_e(t)\right|}
    +\dfrac{2\omega_{\max}}{\hat \zeta d_{\min}\left|{\Theta_e(t)}\right|^{d_{\min}}}\left[-\exp\left\{{\dfrac{-\hat \zeta d_{\min}t}{2}}\right\}+\exp\left\{{\dfrac{-\hat \zeta d_{\min}(t-t_p)}{2}}\right\}\right],
\end{align}
which can be further simplified to

\begin{align}\label{eq:errorfinal}
   \|\mathbf{e}(t)\|_2 \leq  \sqrt{2}\;\left(\exp\left\{{\dfrac{-\hat \zeta(t-t_p)}{2}}\right\}\dfrac{\chi|f(t,t_e)|\|\mathbf{e}(0)\|_2}{\sqrt{2}}+\dfrac{2 \omega_{\max}\left|{f(t,t_e)}\right|^{d_{\min}}}{\hat \zeta d_{\min}}\left[-\exp\left\{{\dfrac{-\hat \zeta d_{\min}t}{2}}\right\}+\exp\left\{{\dfrac{-\hat \zeta d_{\min}(t-t_p)}{2}}\right\}\right]\right).
\end{align}

From \eqref{eq:errorfinal}, we infer that after $\tilde{\mathbf{p}}$ in \eqref{observerstab} vanishes, the time-to-go error \eqref{eq:ei} keeps on decaying and eventually vanishes because $V_2(t) \equiv 0$ as $t \rightarrow t_e^-$. Together with the results of \Cref{lem:ft}, it is then concluded that the closed-loop (time-to-go) error dynamics is globally predefined-time stable.

This essentially means that the time-to-go error for all the interceptors will vanish within the user-selected predefined-time. Thereafter, the interceptors' time-to-go will be in consensus and decrease. Consequently, the interceptors attain the requisite trajectories leading to simultaneous target interception.
\end{proof}





\begin{remark}
    Similar to \Cref{rmk:correctionboundedness}, it can be readily shown that the error-regulating component $\mathbf{u_c}$ remains uniformly bounded over the interval $[0,t_e)$ and once consensus in time-to-go is achieved at $t=t_e$, $\mathbf {u_c}$ remains identically zero thereafter. 
\end{remark} 
Although the control law includes the estimated quantities $\hat r_i$ and $\hat r_i\sin2\hat\sigma_i$ in the denominator, these estimates do not remain at zero even if they momentarily approach it, unless their corresponding true values are also zero. Since $r_i= 0$
only at the final interception time, this does not lead to unbounded control requirements. A similar observation applies to the term $\hat r_i \sin(2\hat\sigma_i)$, which may suggest a potential loss of control effectiveness at $\hat\sigma_i = 0$ and $\hat\sigma_i = \pi/2$. However, the analysis provided later confirms that the system remains controllable at both points. Prior results in \cite{sp30} further indicate that $\hat\sigma_i = \pi/2$ is not a stable equilibrium and is controllable, whereas $\hat\sigma_i = 0$ corresponds to a stable configuration.

The observer dynamics, together with the error-regulating control components, drive the above estimates away from zero and prevent them from staying arbitrarily small. As a result, the control input remains well posed throughout the engagement and does not exhibit any blow-up behavior. In fact, $a_{M_i}$ decreases smoothly and approaches zero near interception, which will be proved in subsequent analysis.

We now examine the behavior of the lead angle after consensus in time-to-go is achieved at $t = t_e$. Since the observer states converge to their true values at $t = t_p < t_e$, the estimated quantities may be treated as the actual variables for the subsequent analysis. From the results of \cite{sp1,doi:10.2514/1.G005180}, it can be inferred that if $\sigma_i(t_e) \neq 0$ then $r_i=0 \implies\sigma_i=0$ in both 2D and 3D settings. Moreover, similar to the results in \cite{sp1,doi:10.2514/1.G005180}, the proposed cooperative guidance command \eqref{eq:am} can also be modified to 
               \begin{align}
    a_{M_i} =
    &V_{M_i} \dot{\hat{\theta}}_i+\frac{V_{M_i}^2 k_i \left( \cos\hat{\sigma}_{i} - 1 \right) }{\hat r_i |\sin 2\hat{\sigma_i}| } 
    + \frac{V_{M_i}^2 \sin\hat{\sigma_{i}} }{ 2 \hat r_i }
     +\frac{V_{M_i}^2 k_i }{\hat  r_i |\sin 2\hat{\sigma}_{i}| } \left(-\zeta+\eta\frac{\dot f(t,t_e)}{f(t,t_e)}\right)\sum _{j=1}^N\left[\mathcal{L}_I\right]_{ij} e_j,
\label{eq:am3}
    \end{align}  
to cover interception from arbitrary heading angles within the full circular range $(-\pi,\pi]$. We defer the proof as it follows similarly to the proof of \Cref{amtheorem}. 

It is also apparent that under \eqref{eq:am} or \eqref{eq:am3}, $\sigma_i$ decreases to zero in the endgame. Note that after consensus in time-to-go values has been achieved, then the $i$\textsuperscript{th} interceptor's acceleration is given by
 \begin{align}
    a_{M_i} =
    &V_{M_i} \dot{\hat{\theta}}_i+\frac{V_{M_i}^2 k_i \left( \cos{\hat\sigma}_{i} - 1 \right) }{\hat r_i |\sin 2{\hat \sigma_i}| } 
    + \frac{V_{M_i}^2 \sin{\hat \sigma_{i}} }{ 2  \hat r_i },\label{eq:amequivalent}
\end{align}  
which leads to the closed-loop dynamics of the lead angle $\sigma_i$, as
\begin{align}
    \dot{\sigma}_i = \frac{V_{M_i} k_i \left( \cos{\sigma}_{i} - 1 \right) }{ r_i |\sin 2{\sigma_i}| } 
    + \frac{V_{M_i} \sin{\sigma_{i}} }{ 2  r_i } = \frac{V_{M_i}}{2 r_i}
\left[
-\frac{k_i \sin(\sigma_i/2)}{\cos(\sigma_i/2)|\cos\sigma_i|}
+  \sin\sigma_i
\right]
\end{align}
since observer errors have also vanished prior to consensus in time-to-go values. As $\frac{V_{M_i}}{2 r_i}>0$ and $k_i>2$, it can be readily verified that the term 
\begin{align}
    \left[
-\frac{k_i \sin(\sigma_i/2)}{\cos(\sigma_i/2)|\cos\sigma_i|}
+  \sin\sigma_i
\right]<0
\end{align}
for $\sigma_i\in(-\pi,\pi)$. This essentially means that the lead angle will always decrease after consensus in time-to-go has been established. Moreover, the lead angle becomes small near interception (note $r\neq 0$ yet) due to interceptors all aligned on the collision triangle, and thus, in the endgame, one can write using small-angle assumptions
\begin{align}
    \dot{\sigma}_i \approxeq  \frac{V_{M_i} {\sigma_{i}} }{ 2  r_i } ,~~\dot{r}_i\approxeq -V_{M_i}.\label{eq:sma}
\end{align}
Then, it follows from \eqref{eq:sma} that $\dfrac{d\sigma_i}{dr_i} = -\dfrac{\sigma_i}{2r_i}$ results in $r_i = r_i(t_e)\left(\dfrac{\sigma_i(t_e)}{\sigma_i}\right)^2$. Equivalently,
$r_i\,\sigma_i^2 = r_i(t_e)\,\sigma_i^2(t_e) > 0$ for any initial condition with $r_i(t_e)>0$ and $\sigma_i^2(t_e)\neq 0$. Thus, within the validity of the approximation \eqref{eq:sma}, the state $\sigma_i= 0$ is not reachable from any nonzero initial lead angle at a finite positive range because $r_i\sigma_i^2$ is constrained to remain strictly positive. In other words, the reduced endgame dynamics \eqref{eq:sma} force the trajectory to evolve along a one-dimensional invariant manifold in the $(r_i,\sigma_i)$-plane, on which the lead angle decreases monotonically but does not reach zero for any $r_i>0$. Physically, this is consistent with the fact that, after time-to-go consensus, the interceptors remain on (or close to) the collision triangle, so that $\sigma_i$ becomes small as $r_i\to 0$, and $\sigma_i=0$ precisely at interception (when $r_i=0$). The impact state $(0,0)$ in the $(r_i,\sigma_i)$-plane lies outside the precise regime of the small-angle approximation \eqref{eq:sma}. Hence, the invariant $r_i\sigma_i^2 =$ (constant) is interpreted as describing the asymptotic approach to interception rather than the impact state itself.

These inferences also confirm the non-singularity of the proposed command \eqref{eq:am3} since $\sigma_i\neq 0$ unless $r_i=0$ provided the initial lead angles are non-zero. For the case of full target information if in case $\sigma_i(0)=0$, then the $i$\textsuperscript{th} interceptor can only take a straight line trajectory if the time-to-go agreement is such that all other interceptors also align on a similar trajectory. In such a scenario, no maneuver is needed, and $a_{M_i}$ is identically zero. Excluding this pathological case, non-singularity of \eqref{eq:am3}, and hence \eqref{eq:am} in \Cref{amtheorem}, is guaranteed.

Up to this point, the closed-loop, predefined-time, stable estimation–guidance strategy has been rigorously developed. Subsequently, we will design an autopilot with dynamics given in \eqref{eq:inter_dynamics}  to ensure that the interceptor’s actual response accurately tracks the commanded lateral acceleration through appropriate canard deflection commands generated by the control system.

It is desirable that the interceptor’s autopilot introduce minimal lag, allowing the actual lateral acceleration to closely follow the commanded acceleration from the guidance system almost instantaneously. In other words, the autopilot should enable the interceptor to achieve the desired acceleration response without noticeable delay. Therefore, employing a predefined time is advantageous, as it enables the convergence time to be explicitly prescribed during design while remaining independent of the initial conditions. Such a design is also consistent with our previous development of estimation and guidance frameworks.
\begin{lemma}
    The dynamics of $i$\textsuperscript{th} interceptor's achieved lateral acceleration $ a_{M_i}^{ac}$ has a relative degree of one with respect to its canard deflection command.
\end{lemma}
\begin{proof}
    From \eqref{eq:alphai_dot}, the lateral acceleration for the $i$\textsuperscript{th} interceptor is given by
    \begin{align}
       a_{M_i}^{ac}= L_{{\alpha_i}}^\beta k_{1_i}(\alpha_i)+L_{\delta_{c_i}}k_{2_i}(\alpha_i+\delta_{c_i}),
    \end{align}
  whose time derivative is
     \begin{align}
       \dot a_{M_i}^{ac}= 
L_{\alpha_i}^{\beta} k'_{1_i}(\alpha_i)\dot{\alpha}_i
+
L_{\delta_{c_i}}\, k'_{2_i}(\alpha_i + \delta_{c_i})
(\dot{\alpha}_i + \dot{\delta}_{c_i}).
    \end{align}
On substituting \eqref{eq:alphai_dot} and \eqref{eq:deltaci_dot} in the above expression, we obtain
     \begin{align}\label{eq:ami_dot}
       \dot a_{M_i}^{ac}= 
\left[L_{\alpha_i}^{\beta} k'_{1_i}(\alpha_i)+L_{\delta_{c_i}}\, k'_{2_i}(\alpha_i + \delta_{c_i})\right]\left(q_i-\dfrac{a_{M_i}^{ac}}{V_i}\right)
-\dfrac{L_{\delta_{c_i}}}{\tau_{c_i}} k'_{2_i}(\alpha_i + \delta_{c_i})\delta_{c_i}
+
\dfrac{L_{\delta_{c_i}}}{\tau_{c_i}} k'_{2_i}(\alpha_i + \delta_{c_i})\delta_{c_i}^c,
    \end{align}
indicating that the commanded canard deflection appears in the first derivative of $a_{M_i}^{ac}.$
\end{proof}
For the $i$\textsuperscript{th} interceptor, let us define the tracking error in the lateral acceleration as

\begin{align}\label{eq:sliding_manifold}
    s_i=a_{M_i}^{ac}-a_{M_{i}}.
\end{align}
Under the assumption that the commanded acceleration (which is desired acceleration $a_{M_{i}}$) does not show abrupt changes, it follows that $a_{M_{i}}$ can be regarded as a bounded uncertainty. Thus, $|\dot{a}_{M_i}| \leq \Upsilon < \infty$ for some finite value of $\Upsilon$.
 \begin{theorem}\label{thm:autopilot}
     Consider the $i$\textsuperscript{th} interceptor's dynamics described by \eqref{eq:inter_dynamics}. The commanded acceleration is tracked by the $i$\textsuperscript{th} interceptor's autopilot within a predefined time $t_a$, if the canard deflection command
     is designed as     
     
     \begin{align}\label{eq:comm_deflec}
    \delta_{c_i}^c=-\dfrac{\left[ L_{\alpha_i}^{\beta} k'_{1_i}(\alpha_i) + L_{\delta_{c_i}} k'_{2_i}\left(\alpha_i + \delta_{c_i}\right) \right]
\left( q_i - \dfrac{a_{M_i}^{ac}}{V_i} \right)
- \dfrac{L_{\delta_{c_i}}}{\tau_{c_i}}  k'_{2_i}\!\left(\alpha_i + \delta_{c_i}\right)  \delta_{c_i}+\left(\rho_1-\rho_2\dfrac{\dot f(t,t_a)}{f(t,t_a)}\right)s_i+\triangle\;\sign(s_i)}{\dfrac{L_{\delta_{c_i}}}{\tau_{c_i}}  k'_{2_i}\left(\alpha_i + \delta_{c_i}\right)}, 
\end{align}
for some $\triangle\geq\Upsilon$, $\rho_1>0$ and $\rho_2\geq2$.
\end{theorem}
\begin{proof}
 On differentiating \eqref{eq:sliding_manifold} with respect to time and substituting \eqref{eq:ami_dot} in it, one can obtain
\begin{align}
 \dot s_i&=\left[ L_{\alpha_i}^{\beta} k'_{1_i}(\alpha_i) + L_{\delta_{c_i}} k'_{2_i}\left(\alpha_i + \delta_{c_i}\right) \right]
\left( q_i - \frac{a_{M_i}^{ac}}{V_i} \right)
- \frac{L_{\delta_{c_i}}}{\tau_{c_i}}  k'_{2_i}\!\left(\alpha_i + \delta_{c_i}\right)  \delta_{c_i}+\frac{L_{\delta_{c_i}}}{\tau_{c_i}}  k'_{2_i}\left(\alpha_i + \delta_{c_i}\right)\delta_{c_i}^c-\dot a_{M_{i}}. \label{eq:si_dot}
\end{align}
    On substituting \eqref{eq:comm_deflec} into \eqref{eq:si_dot}, the closed-loop tracking error dynamics becomes
    \begin{align}
        \dot s_i=-\left(\rho_1-\rho_2\dfrac{\dot f(t,t_a)}{f(t,t_a)}\right)s_i-\triangle\;\text{sign}(s_i)-\dot a_{M_{i}}.
    \end{align}
    Consider the Lyapunov function candidate $V_3=|s_i|$, and on differentiating it with respect to time, one can obtain
    \begin{align}
            \dot V_3 =\dot s_i\; \sign(s_i) =&-\left(\rho_1-\rho_2\dfrac{\dot f(t,t_a)}{f(t,t_a)}\right)|s_i|-\triangle-\dot a_{M_{i}} \text{sign}(s_i) \notag\\
 &\leq -\left(\rho_1-\rho_2\dfrac{\dot f(t,t_a)}{f(t,t_a)}\right)|s_i|-\triangle+|\dot a_{M_{i}}| \notag\\
          &\leq -\left(\rho_1-\rho_2\dfrac{\dot fV(t,t_a)}{f(t,t_a)}\right)|s_i|-(\triangle- \Upsilon).\label{eq:V3dota}
\end{align}            
From \Cref{remark:scaling_function} we know that the scaling functions satisfy $\dfrac{\dot{f}(t,t_a)}{f(t,t_a)} < 0$ for $t < t_a$, thus we can introduce an auxiliary function $\Theta_a(t) = \dfrac{1}{f(t,t_a)}$, which on differentiating gives 
$\dfrac{\dot{\Theta}_a(t)}{\Theta_a(t)} = -\dfrac{\dot{f}(t,t_a)}{f(t,t_a)} > 0$. 
This allows us to simplify \eqref{eq:V3dota} to
\begin{align}\label{eq:va_pre_final}
            \dot V_3 &\leq -\rho_1|s_i|-\rho_2\dfrac{\dot \Theta_a(t,t_a)}{\Theta_a(t,t_a)}|s_i|.
\end{align}
Upon letting the design parameters as $\rho_1>0$ and $\rho_2\geq2$, \eqref{eq:va_pre_final} can be expressed as
\begin{align}
    \dot V_3 &\leq -\rho_1 V_3-2\dfrac{\dot \Theta_a(t,t_a)}{\Theta_a(t,t_a)}V_3,\label{eq:va_dot_final}
\end{align}
wherefrom results from \Cref{lem:ft} can be applied to solve for \eqref{eq:va_dot_final} as
\begin{align}
\Theta_a^2(t)V_3(t)&\leq\exp(-\rho_1t)\Theta_a^2(0)V_3(0),
\end{align}
which leads to
\begin{align}
   |s_i(t)|&\leq\dfrac{f(t,t_a)^2}{t_p^2}\exp(-\rho_1t)|s_i(0)|.
\end{align}
This implies that $s_i$ will converge to zero at $t\rightarrow t_a^-$, which further implies that the interceptor's autopilot tracks the commanded acceleration within the predefined time $t_a$. 
\end{proof}
\begin{remark}
    The denominator in \eqref{eq:comm_deflec} remains non-zero since $L_{\delta_{c_i}}$ and $\tau_{c_i}$ are strictly non-zero constants, and $k_{2_i}$ is defined through a bounded nonlinear function that does not vanish over the operating range, as commonly employed in \cite{sp27,sp28}. Consequently, the proposed canard deflection command remains well-defined.
\end{remark}
\begin{figure}[h!]
   \centering
   \includegraphics[width=\linewidth]{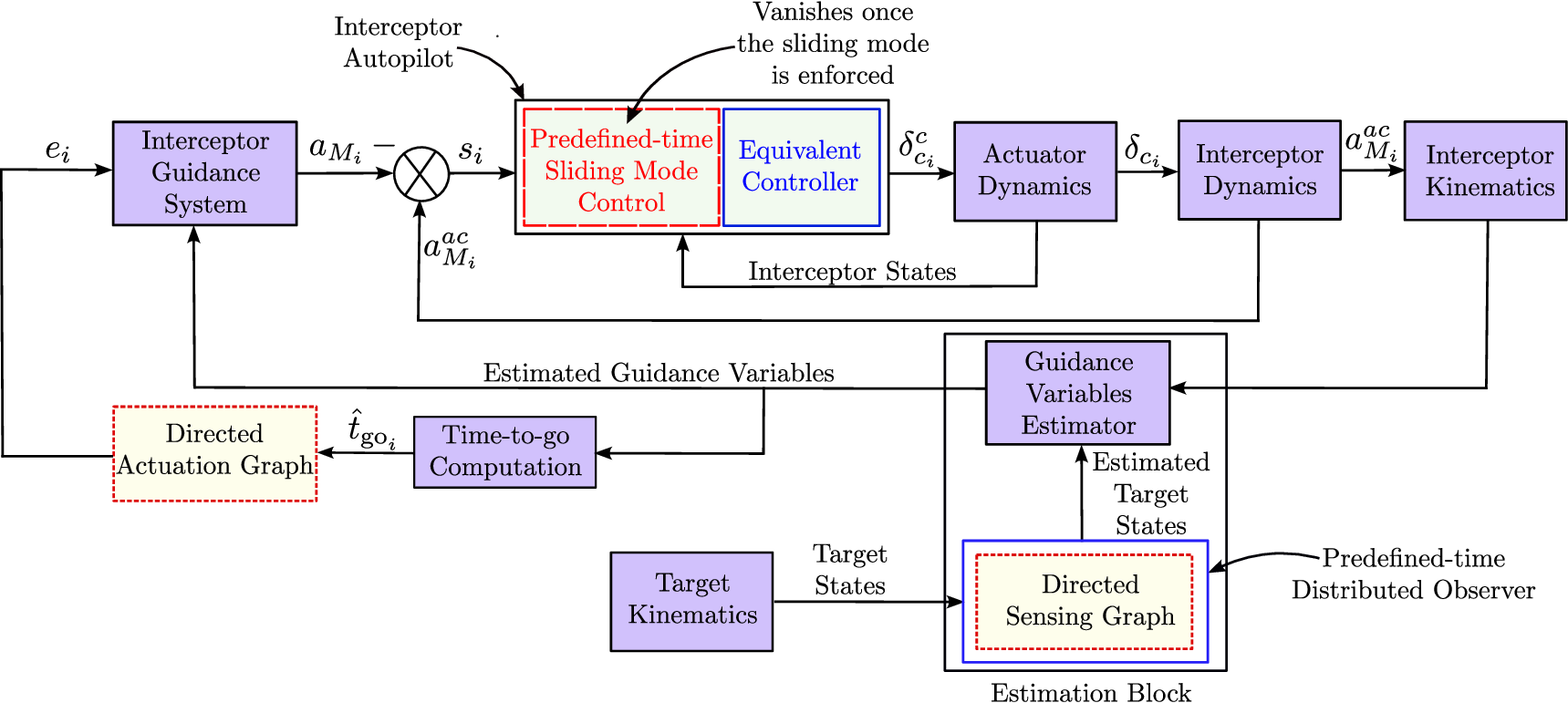}
   \caption{Schematic of the proposed one-shot nonlinear cooperative estimation-guidance-control framework for the $i$\textsuperscript{th} interceptor.}
  \label{bd1}
  \end{figure}

The overall design schematic for $i$\textsuperscript{th} interceptor is illustrated in \Cref{bd1}. The architecture comprises two interconnected parts. The first part discusses the design of an estimation-guidance scheme, where target states are initially estimated using a predefined-time distributed observer operating over a directed sensing graph. These estimated target states, together with the interceptor’s kinematics, are then employed to estimate the required guidance variables and determine the corresponding time-to-go. The desired interception time is determined cooperatively among the interceptors over a directed actuation graph, and the corresponding error in time-to-go, along with the relevant estimated guidance variables, is used to formulate a corrective control that drives the time-to-go error to zero within the user-specified time. This corrective component, combined with the nominal control input, produces the desired lateral acceleration $a_{M_i}$ to achieve synchronized interception. 

The second part addresses the inner control loop. The commanded lateral acceleration $a_{M_i}$ is not applied instantaneously due to the interceptor's dynamics and actuator limitations. The autopilot regulates the canard control surfaces to track the desired lateral acceleration $a_{M_i}$. The tracking error $s_i$ is used by the autopilot at every instant to generate the canard deflection command $\delta_{c_i}^c$, which is passed to the actuator dynamics and then applied to the interceptor, producing the aerodynamic forces and moments required to drive $s_i$ to zero and steer the interceptor along the desired collision trajectory.

\section{Performance Evaluations}\label{sec:simulation}
The performance of the proposed strategy is evaluated in this section. First, the efficacy of the estimation–guidance architecture is examined under various engagement scenarios. Subsequently, the one-shot nonlinear cooperative estimation-guidance-control framework is assessed to demonstrate the overall performance.

\subsection{Cooperative Salvo using Spatially Offset Interceptors}
\begin{figure}[h!]
    \centering
    \begin{subfigure}[b]{0.47\linewidth}
        \centering
         \resizebox{.5\linewidth}{!}{%
        \begin{tikzpicture}[
  ->, >=Stealth, thick,
  font=\bfseries\large  
]
  \node[circle, draw=red, text=red, minimum size=1cm, inner sep=0pt] (0) at (2.5,4) {0};
  \node[circle, draw=Tan, text=blue, minimum size=1cm, inner sep=0pt, fill = Tan] (1) at (1,2) {1};
  \node[circle, draw=blue, text=blue, minimum size=1cm, inner sep=0pt] (2) at (4,2) {2};
  \node[circle, draw=blue, text=blue, minimum size=1cm, inner sep=0pt] (3) at (1,0) {3};
  \node[circle, draw=blue, text=blue, minimum size=1cm, inner sep=0pt] (4) at (4,0) {4};

  \draw[->] (0) -- (1);              
  \draw[->] (1) -- (2);              
  \draw[->] (1) -- (3);              
  \draw[->] (2) -- (4);              
  \draw[->] (3) -- (2);              
  \draw[->] (4) -- (1);              

  \node[text=red] at (2.5,4.8) {Target};
\node[text=blue, anchor=east] at (1.west) {$I_1$};
\node[text=blue, anchor=west] at (2.east) {$I_2$};
\node[text=blue] at (1,-0.9) {$I_3$};
\node[text=blue] at (4,-0.9) {$I_4$};

\end{tikzpicture}%
}
        \caption{Sensing Graph $\mathscr{S}$}
        \label{sensing_graph}
    \end{subfigure}
    \hfill
    \begin{subfigure}[b]{0.47\linewidth}
        \centering
        \resizebox{.5\linewidth}{!}{%
        \begin{tikzpicture}[
  ->, >=Stealth, thick,
  font=\bfseries\large  
]
  \node[circle, draw=blue, text=blue, minimum size=1cm, inner sep=0pt] (1) at (1,2) {1};
  \node[circle, draw=blue, text=blue, minimum size=1cm, inner sep=0pt] (2) at (4,2) {2};
  \node[circle, draw=blue, text=blue, minimum size=1cm, inner sep=0pt] (3) at (1,0) {3};
  \node[circle, draw=blue, text=blue, minimum size=1cm, inner sep=0pt] (4) at (4,0) {4};
  \draw[->] (1) -- (2);              
    \draw[->] (1) -- (3);              

  \draw[->] (2) -- (4);              
  \draw[->] (4) -- (1);              
  \draw[->] (3) -- (2);              
\node[text=blue, anchor=east] at (1.west) {$I_1$};
\node[text=blue, anchor=west] at (2.east) {$I_2$};
\node[text=blue] at (1,-0.9) {$I_3$};
\node[text=blue] at (4,-0.9) {$I_4$};
\end{tikzpicture}%
}
        \caption{Actuation Graph $\mathscr{A}$}
        \label{actuation_graph}
    \end{subfigure}
    \caption{Interceptors' Communication Topologies}
    \label{fig:graphs}
\end{figure}
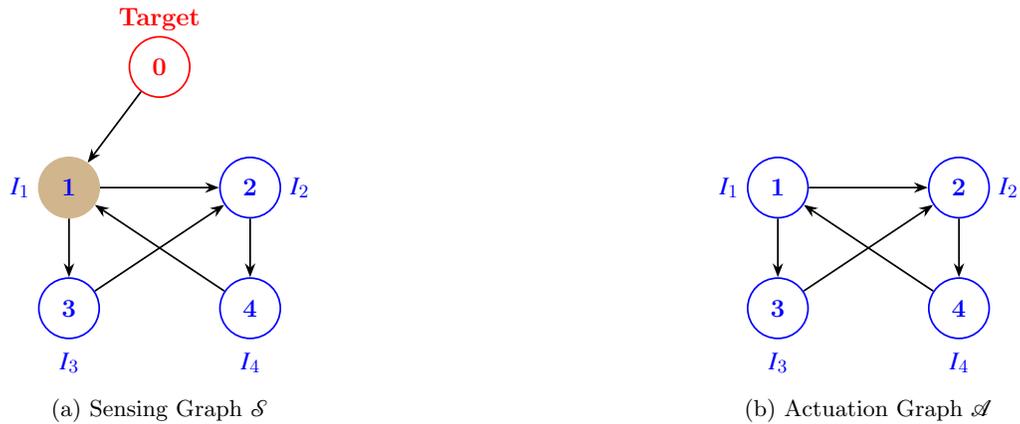

\begin{figure}[h!]
    \centering
    
    \begin{subfigure}[t]{0.45\textwidth}
        \centering
        \includegraphics[width=\linewidth]{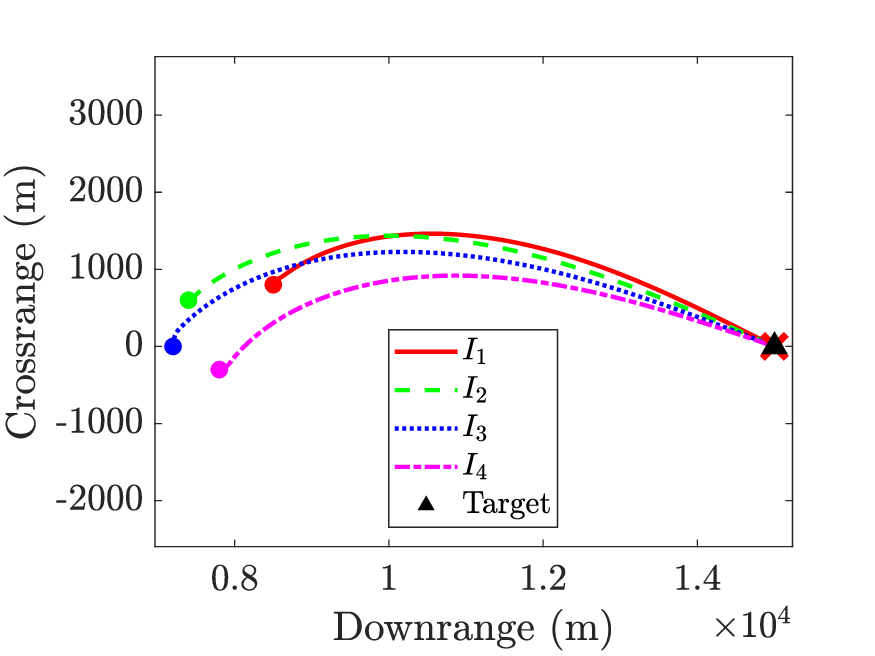}
        \caption{Interceptors' trajectories.}
        \label{fig:Trajectory}
    \end{subfigure}
    \begin{subfigure}[t]{0.45\textwidth}
        \centering
        \includegraphics[width=\linewidth]{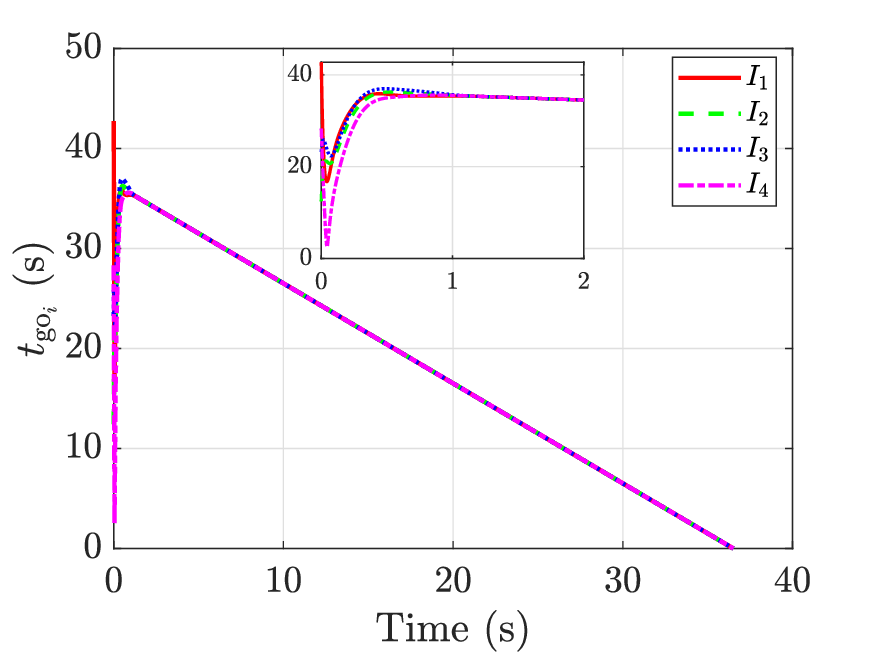}
        \caption{Interceptors' time-to-go values.}
        \label{fig:Time-to-go}
    \end{subfigure}
    
    \begin{subfigure}[t]{0.9\textwidth}
        \centering
        \includegraphics[width=0.49\linewidth]{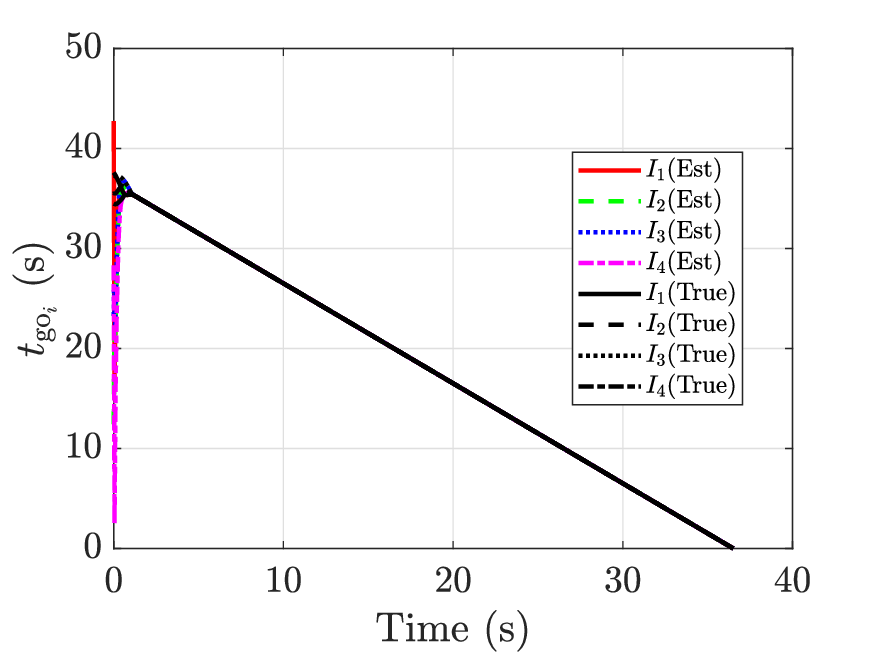}
        \includegraphics[width=0.49\linewidth]{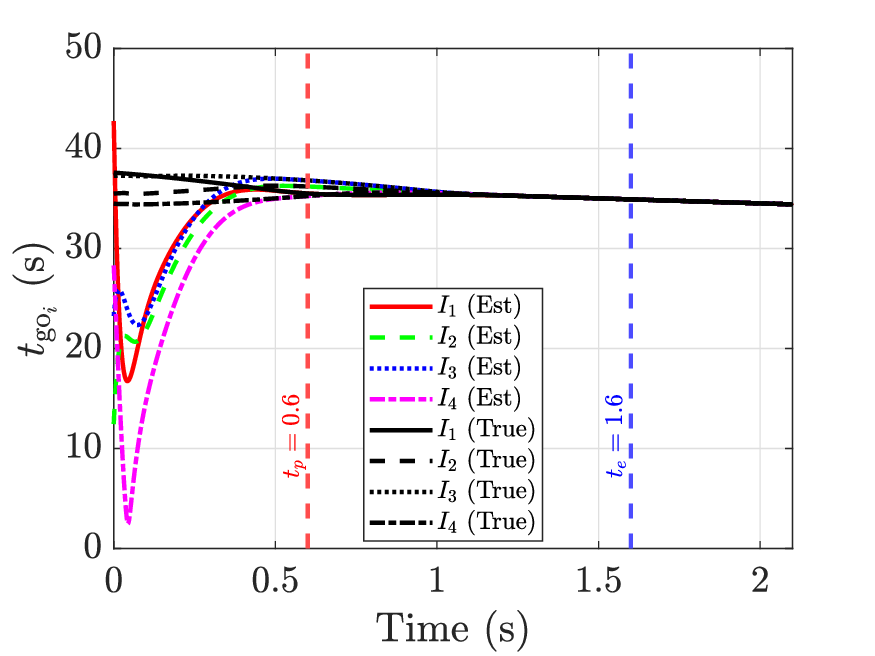}
        \caption{True vs estimated time-to-go (left) and zoomed view (right).}
        \label{fig:truevsestim_combined}
    \end{subfigure}
    
    \begin{subfigure}[t]{0.45\textwidth}
        \centering
        \includegraphics[width=\linewidth]{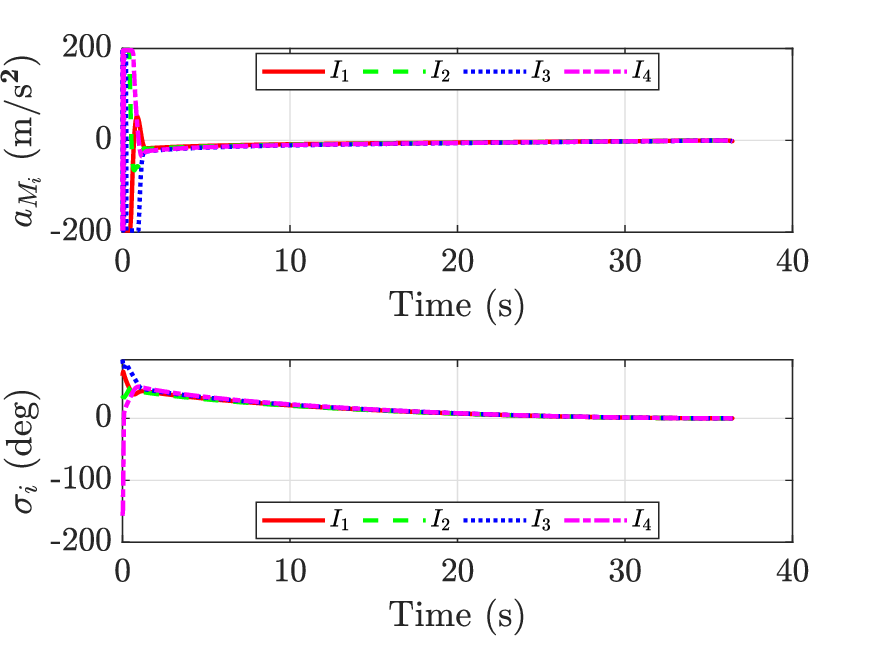}
        \caption{Interceptors' lateral acceleration and lead angle profiles.}
        \label{fig:acceleration}
    \end{subfigure}
    \begin{subfigure}[t]{0.45\textwidth}
        \centering
        \includegraphics[width=\linewidth]{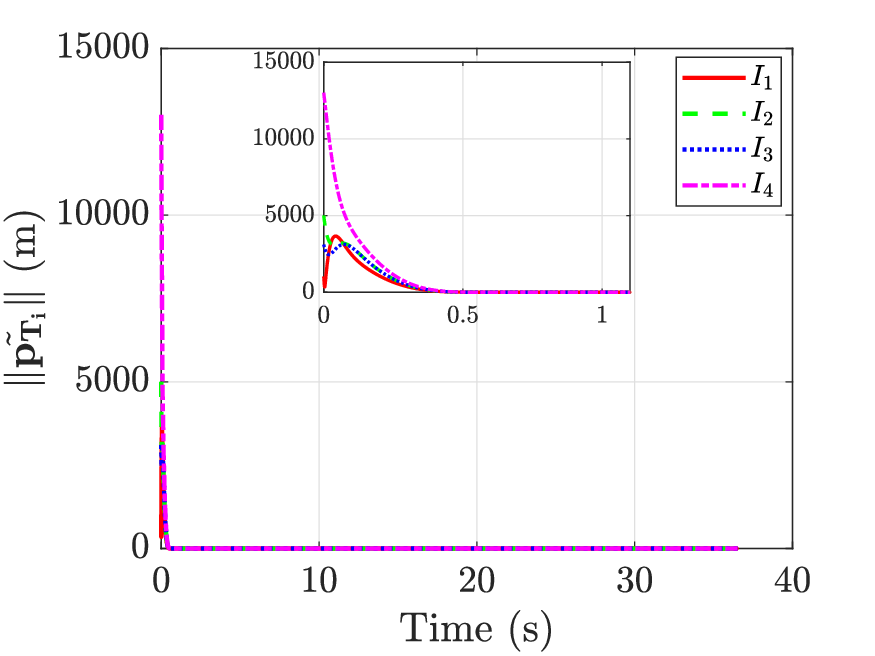}
        \caption{Target's states estimation error.}
        \label{fig:Observer_error}
    \end{subfigure}
    
    \caption{Performance evaluation of the proposed method in a typical case.}
    \label{fig:stationary}
\end{figure}

\begin{figure}[h!]
    \centering

    \begin{subfigure}[t]{0.32\textwidth}
        \centering
        \includegraphics[width=\linewidth]{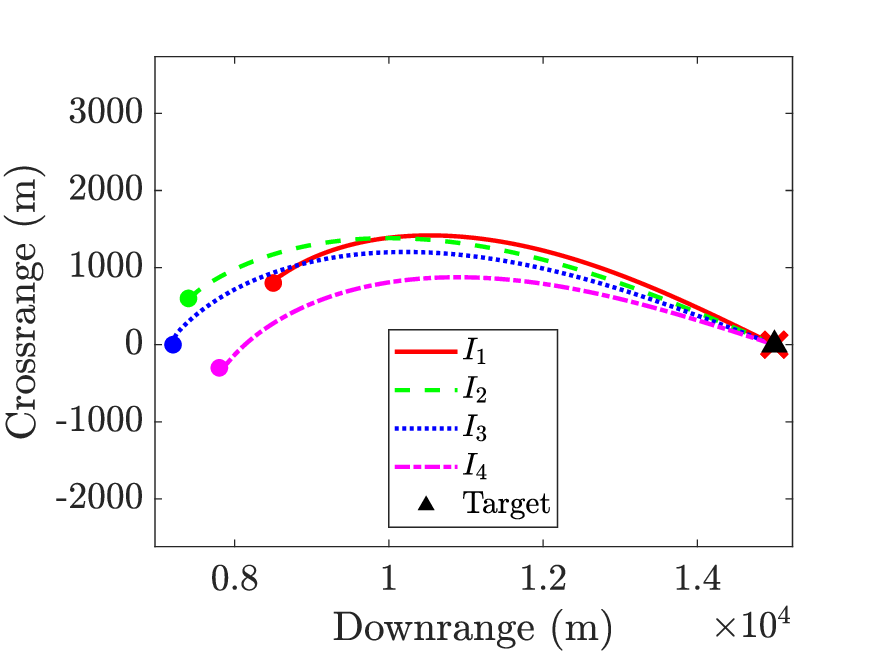}
        \caption{Interceptors' trajectories.}
        \label{fig:Trajectory_true}
    \end{subfigure}
    \begin{subfigure}[t]{0.32\textwidth}
        \centering
        \includegraphics[width=\linewidth]{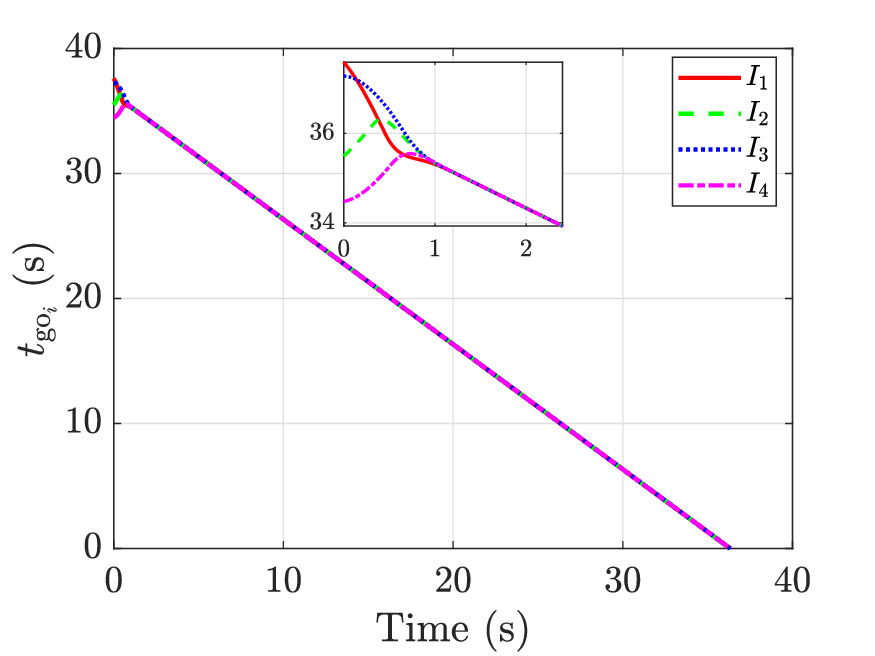}
        \caption{Interceptors' time-to-go values.}
        \label{fig:Time-to-go_true}
    \end{subfigure}
    \begin{subfigure}[t]{0.32\textwidth}
        \centering
        \includegraphics[width=\linewidth]{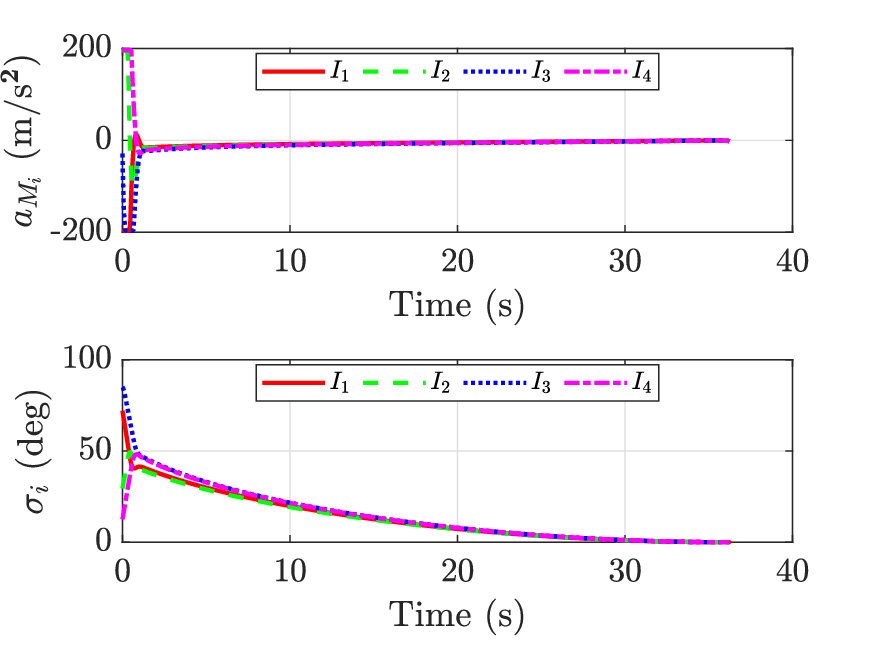}
        \caption{Interceptors' lateral acceleration and lead angle profiles.}
        \label{fig:acceleration_true}
    \end{subfigure}

    \caption{Comparison of the proposed strategy when target information is available to all interceptors}
    \label{fig:stationary_true}
\end{figure}
\begin{figure}[h!]  
    \centering
    \begin{subfigure}[t]{0.45\textwidth}
        \centering
        \includegraphics[width=\linewidth]{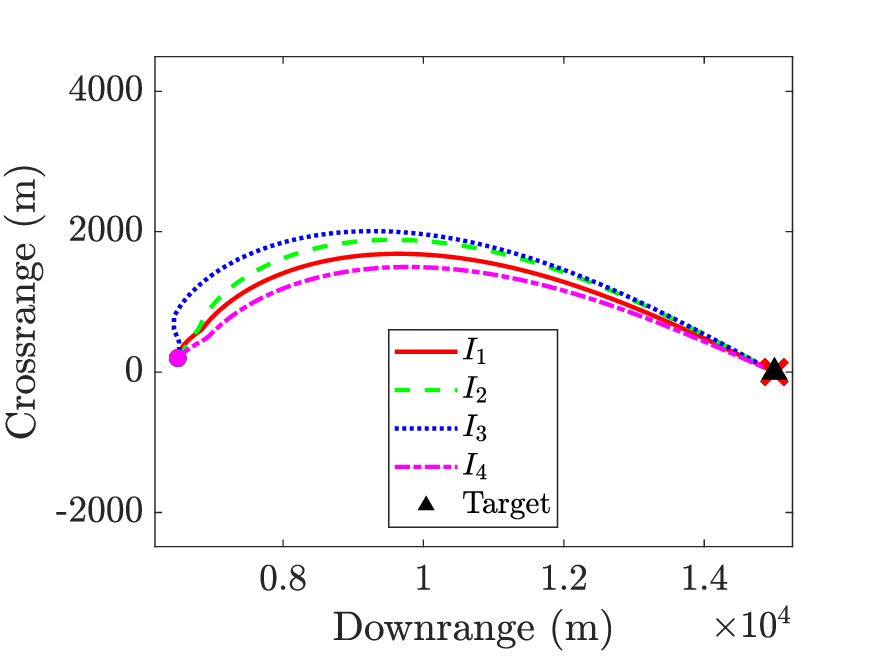}
        \caption{Interceptors' trajectories.}
        \label{fig:Trajectory2}
    \end{subfigure}
    \begin{subfigure}[t]{0.45\textwidth}
        \centering
        \includegraphics[width=\linewidth]{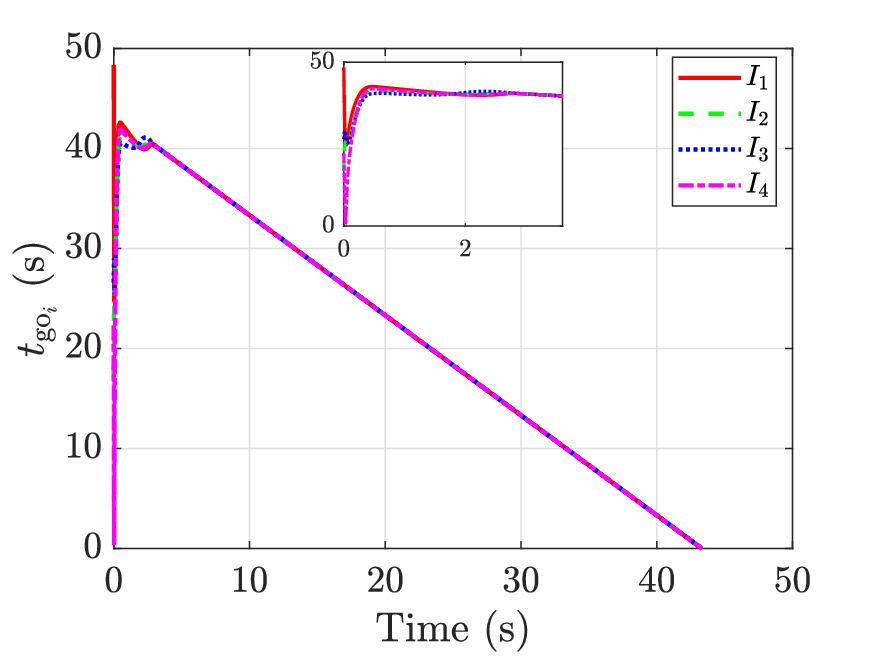}
        \caption{Interceptors' time-to-go values.}
        \label{fig:Time-to-go2}
    \end{subfigure}
    
    \begin{subfigure}[t]{0.9\textwidth}
        \centering
        \includegraphics[width=0.49\linewidth]{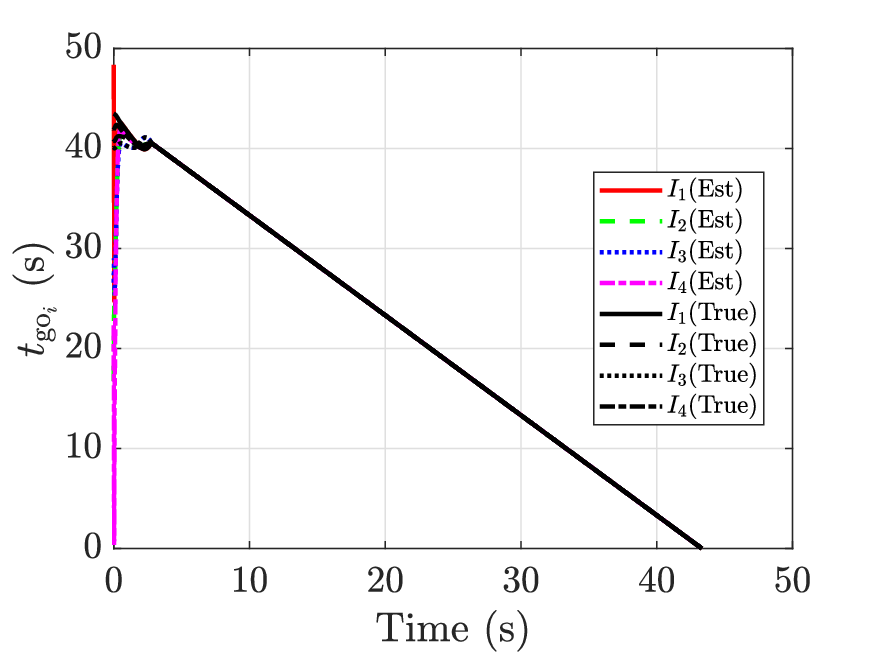}
        \includegraphics[width=0.49\linewidth]{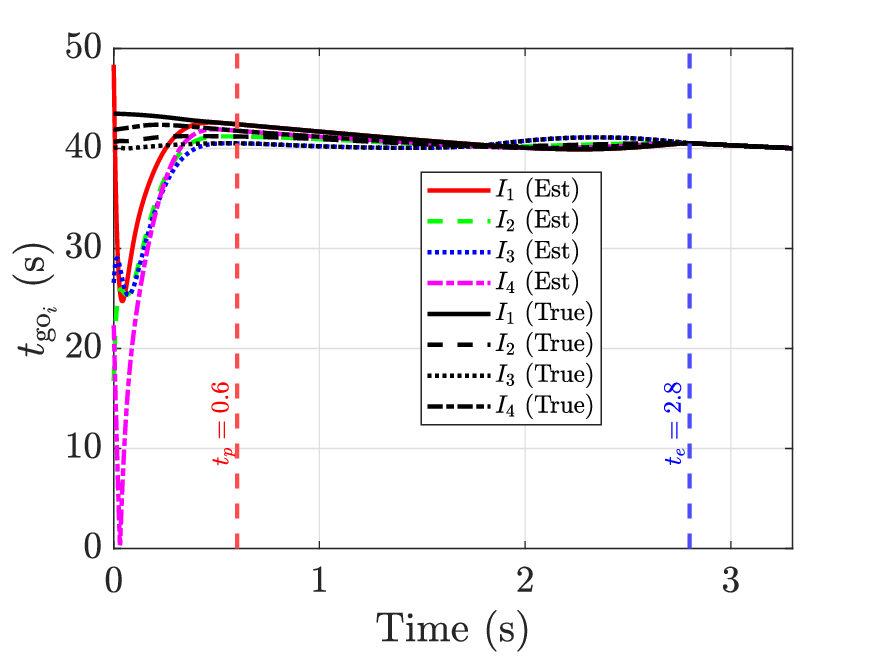}
        \caption{True vs estimated time-to-go (left) and zoomed view (right)}
        \label{fig:truevsestim_combined2}
    \end{subfigure}
    
    \begin{subfigure}[t]{0.45\textwidth}
        \centering
        \includegraphics[width=\linewidth]{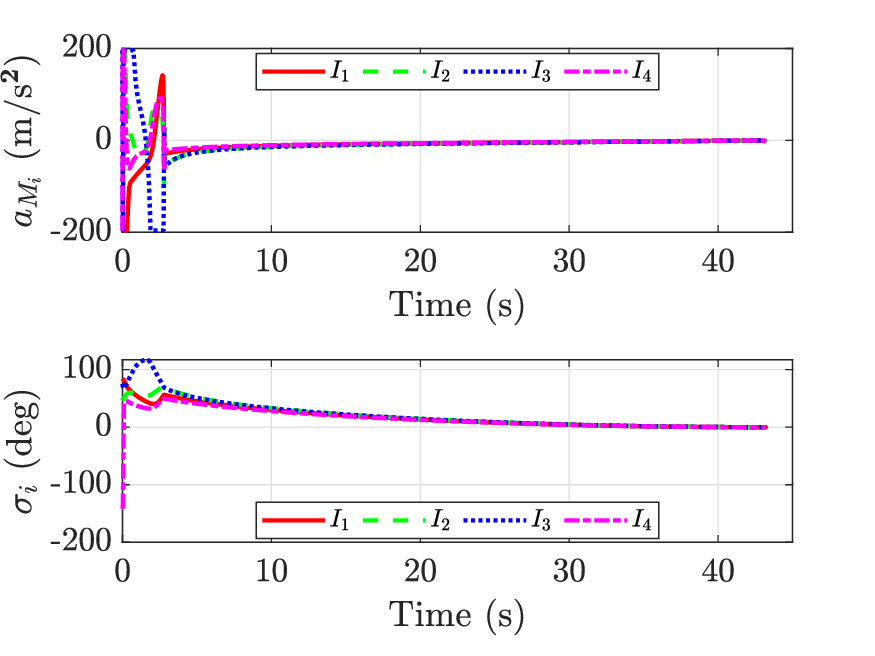}
        \caption{Interceptors' lateral acceleration and lead angle profiles.}
        \label{fig:acceleration2}
    \end{subfigure}
    \begin{subfigure}[t]{0.45\textwidth}
        \centering
        \includegraphics[width=\linewidth]{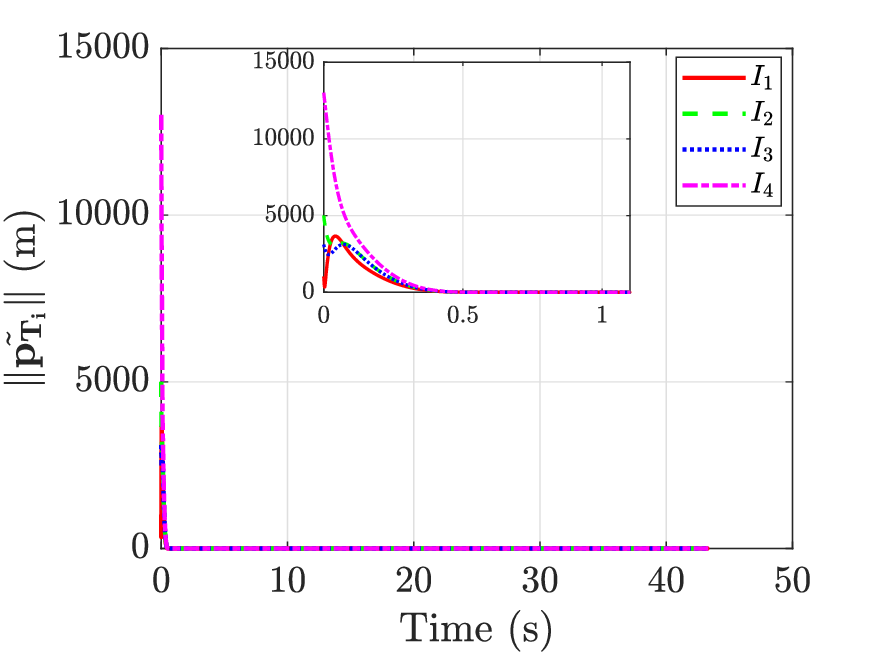}
        \caption{Target's state estimation error.}
        \label{fig:Observer_error2}
    \end{subfigure}
    
    \caption{Performance evaluation of the proposed method with different engagement geometries.}
    \label{fig:stationary2}
\end{figure}

The efficacy of the proposed method is demonstrated through a simulation scenario involving one stationary target and four interceptors ($I_1$–$I_4$). Among them, only $I_1$ is seeker-equipped and can directly measure the target’s position, while the remaining interceptors are seeker-less and rely on information sharing according to the sensing topology $\mathscr{S}$ in \Cref{sensing_graph}, and the actuation graph $\mathscr{A}$ is shown in \Cref{actuation_graph} over which time-to-go information is exchanged. It is worth noting that, while the present case considers identical communication links among interceptors in both the actuation and sensing topologies, the proposed strategy is inherently applicable to dissimilar communication link configurations, which will be demonstrated in subsequent analysis. In all the plots to follow, interceptor launch locations are indicated by circular markers, interception points are denoted by cross markers, and the target position is represented by a black triangular marker.

The stationary target is located at $(15000,0)~\mathrm{m}$. The interceptor speeds are given as 
$ \begin{bmatrix} 190 & 220 & 230 & 210 \end{bmatrix}~\mathrm{m/s} $, 
with respective initial heading angles 
$ \begin{bmatrix} 65^\circ & 25^\circ & 85^\circ & 15^\circ \end{bmatrix} $. 
The interceptors are assumed to be spatially located at
$ \begin{bmatrix} (8.5,0.8) & (7.4,0.6) & (7.2,0) & (7.8,-0.3) \end{bmatrix}$ km. 
Each interceptor operates under actuator constraints limiting the maximum achievable lateral acceleration to $20$ g, where g is the gravitational acceleration. 


Based on the target's true position, the initial time-to-go estimates are 
$ \begin{bmatrix} 37.59 & 35.49 & 37.28 & 34.48 \end{bmatrix}~\mathrm{s} $. 
The initial target's position estimates maintained by the interceptors are 
$ \begin{bmatrix} (15.96,0.34) & (10,0.1) & (12,-0.8) & (2,0.5) \end{bmatrix} \times 10^3~\mathrm{m} $, 
which yield corresponding initial time-to-go estimates of 
$ \begin{bmatrix} 42.47 & 12.45 & 23.26 & 28.30 \end{bmatrix}~\mathrm{s} $. The observer and controller gains are selected in accordance with the design criteria established in the preceding analysis. In addition, the predefined observer convergence time is taken as $t_p = 0.6~\mathrm{s}$, and the consensus time is set to $t_e = 1.6~\mathrm{s}$. Actuator constraints limit the growth of the time-scaling gain, and in combination with the chosen controller gains, they determine the minimum achievable convergence time. We note that this aspect warrants further study. The parameter $N$ is fixed at $3$.

\Cref{fig:stationary} presents the stationary target engagement scenario, where the interceptors accomplish cooperative simultaneous interception at $36.51~\mathrm{s}$ despite having incomplete information. The interceptors are initialized from different starting positions (\Cref{fig:Trajectory}), with only $I_1$ possessing direct access to the target's position. The other interceptors, being seeker-less, estimate the target position in a distributed manner through their interaction with $I_1$ via the sensing topology $\mathscr{S}$. As shown in \Cref{fig:Time-to-go} and \Cref{fig:truevsestim_combined}, this information asymmetry does not hinder coordination, since the interceptors achieve consensus on a common time-to-go within the prescribed convergence time $t_e = 1.6~\mathrm{s}$, thereby attaining the geometry required for simultaneous interception. The corresponding lateral accelerations and lead angles are illustrated in \Cref{fig:acceleration}. It is evident that, before consensus on time-to-go is reached, the lead angles exhibit fluctuations, and the lateral acceleration demands are relatively large. However, once the observer error converges to zero within the predefined time $t_p = 0.6~\mathrm{s}$ (\Cref{fig:Observer_error}), and consensus on time-to-go is attained at $t_e = 1.6~\mathrm{s}$, both the lead angles and lateral accelerations settle to stable values and approach zero near interception.

\Cref{fig:stationary_true} illustrates the simultaneous interception of the target by the interceptors at 36.33 s when all the interceptors are equipped with seekers, i.e., full target information is available to all the interceptors. From \Cref{fig:Time-to-go_true}, it is evident that the agents achieve consensus in the time-to-go well before $t_e$. Furthermore, \Cref{fig:acceleration_true} shows that no noticeable fluctuations are present in the full-information case, whereas the incomplete-information case exhibits initial fluctuations due to the transient behavior of the observer dynamics, and once the observer converges, these fluctuations subside, and the system's behavior stabilizes.

It is worth noting that, even under incomplete information, the achieved interception time is nearly identical to that of the full-information case, thereby demonstrating the effectiveness of the proposed strategy. Moreover, since sensing responsibilities are assigned only to a subset of interceptors, i.e., only a few agents are equipped with seekers, the proposed framework offers both resource efficiency and cost effectiveness while maintaining similar performance.

\subsection{Cooperative Salvo using Co-located Interceptors}

To further assess the effectiveness of the proposed method, the result in \Cref{fig:stationary2} shows a scenario when the engagement geometry is such that the interceptors are co-located at $(7.5,0.2)$ km. Their  speeds are now taken as $ \begin{bmatrix} 215 & 220 & 230 & 210 \end{bmatrix}$ m/s, 
with respective initial heading angles 
$ \begin{bmatrix} 85^\circ & 45^\circ & 65^\circ & 35^\circ \end{bmatrix} $. The initial target's position estimates, maximum achievable lateral acceleration, observer gain, controller gain, and observer convergence time are unchanged. The consensus time $t_e$ is set to 2.8 $~\mathrm{s}$.

The true initial time-to-go values are 
$[\,43.484 \;\; 40.670 \;\; 40.068 \;\; 41.910\,]~\mathrm{s}$. Based on the position  estimates, the initial time-to-go values are $[\,48.358 \;\; 16.757 \;\; 26.579 \;\; 22.320\,]~\mathrm{s}$. 
\Cref{fig:stationary2} depicts the scenario where the interceptors are launched from the same position with different heading angles, achieving cooperative simultaneous interception at $43.3~\mathrm{s}$. The corresponding interceptor trajectories are shown in \Cref{fig:Trajectory2}. The evolution of time-to-go estimates reaching consensus within $t_e = 2.8~\mathrm{s}$ is illustrated in \Cref{fig:Time-to-go2}. The lateral accelerations and lead angles are depicted in \Cref{fig:acceleration2}, while the observer error convergence to zero within $t_p = 0.6~\mathrm{s}$ is shown in \Cref{fig:Observer_error2}. Similar to the previous case, the lead angles undergo fluctuations, and the lateral acceleration demands remain relatively large before consensus on time-to-go is established. Once consensus is attained at $t_e = 2.8~\mathrm{s}$, both the lead angle and the lateral acceleration demands reduce toward zero in the endgame. 
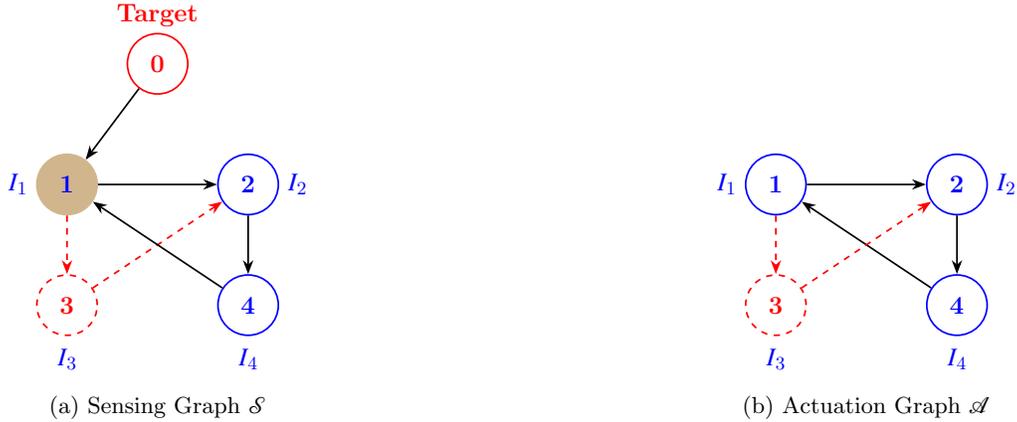
\begin{figure}[h!]
    \centering
    \begin{subfigure}[b]{0.47\linewidth}
        \centering
         \resizebox{.5\linewidth}{!}{%
        \begin{tikzpicture}[
  ->, >=Stealth, thick,
  font=\bfseries\large  
]
  \node[circle, draw=red, text=red, minimum size=1cm, inner sep=0pt] (0) at (2.5,4) {0};
  \node[circle, draw=Tan, text=blue, minimum size=1cm, inner sep=0pt, fill = Tan] (1) at (1,2) {1};
  \node[circle, draw=blue, text=blue, minimum size=1cm, inner sep=0pt] (2) at (4,2) {2};
  \node[circle, draw=red,dashed, text=red, minimum size=1cm, inner sep=0pt] (3) at (1,0) {3};
  \node[circle, draw=blue, text=blue, minimum size=1cm, inner sep=0pt] (4) at (4,0) {4};

  \draw[->] (0) -- (1);              
  \draw[->] (1) -- (2);              
    \draw[dashed,red,->] (1) -- (3);              
  \draw[->] (2) -- (4);              
  \draw[dashed,red,->] (3) -- (2);              
  \draw[->] (4) -- (1);              

  \node[text=red] at (2.5,4.8) {Target};
\node[text=blue, anchor=east] at (1.west) {$I_1$};
\node[text=blue, anchor=west] at (2.east) {$I_2$};
\node[text=blue] at (1,-0.9) {$I_3$};
\node[text=blue] at (4,-0.9) {$I_4$};

\end{tikzpicture}%
}
        \caption{Sensing Graph $\mathscr{S}$}
        \label{sensing_graph_fail1}
    \end{subfigure}
    \hfill
    \begin{subfigure}[b]{0.47\linewidth}
        \centering
        \resizebox{.5\linewidth}{!}{%
        \begin{tikzpicture}[
  ->, >=Stealth, thick,
  font=\bfseries\large  
]
  \node[circle, draw=blue, text=blue, minimum size=1cm, inner sep=0pt] (1) at (1,2) {1};
  \node[circle, draw=blue, text=blue, minimum size=1cm, inner sep=0pt] (2) at (4,2) {2};
  \node[circle, draw=red,dashed, text=red, minimum size=1cm, inner sep=0pt] (3) at (1,0) {3};
  \node[circle, draw=blue, text=blue, minimum size=1cm, inner sep=0pt] (4) at (4,0) {4};
  \draw[->] (1) -- (2);              
    \draw[dashed,red,->] (1) -- (3);              

  \draw[->] (2) -- (4);              
  \draw[->] (4) -- (1);              
  \draw[dashed,red,->] (3) -- (2);              
\node[text=blue, anchor=east] at (1.west) {$I_1$};
\node[text=blue, anchor=west] at (2.east) {$I_2$};
\node[text=blue] at (1,-0.9) {$I_3$};
\node[text=blue] at (4,-0.9) {$I_4$};
\end{tikzpicture}%
}
        \caption{Actuation Graph $\mathscr{A}$}
        \label{actuation_graph_fail1}
    \end{subfigure}
    \caption{Interceptors' Communication Topologies when Interceptor 3 Fails at t=1.5 $~\mathrm{s}$.}
    \label{fig:graphs_fail1}
\end{figure}

\begin{figure}[h!]  
    \centering
    \begin{subfigure}[t]{0.45\textwidth}
        \centering
        \includegraphics[width=\linewidth]{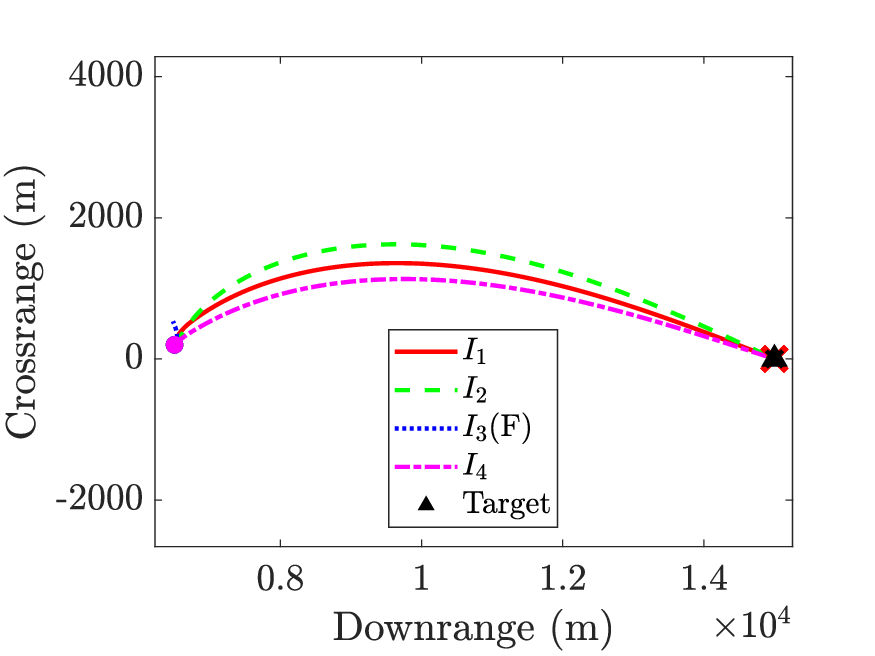}
        \caption{Interceptors' trajectories.}
        \label{fig:Trajectory3}
    \end{subfigure}
    \begin{subfigure}[t]{0.45\textwidth}
        \centering
        \includegraphics[width=\linewidth]{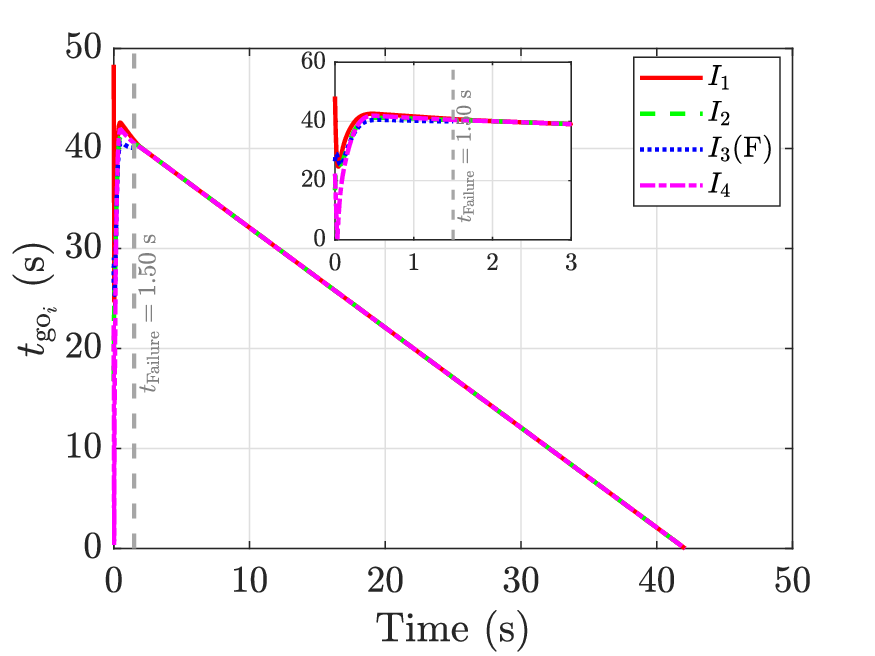}
        \caption{Interceptors' time-to-go values.}
        \label{fig:Time-to-go3}
    \end{subfigure}
    
    \begin{subfigure}[t]{0.9\textwidth}
        \centering
        \includegraphics[width=0.49\linewidth]{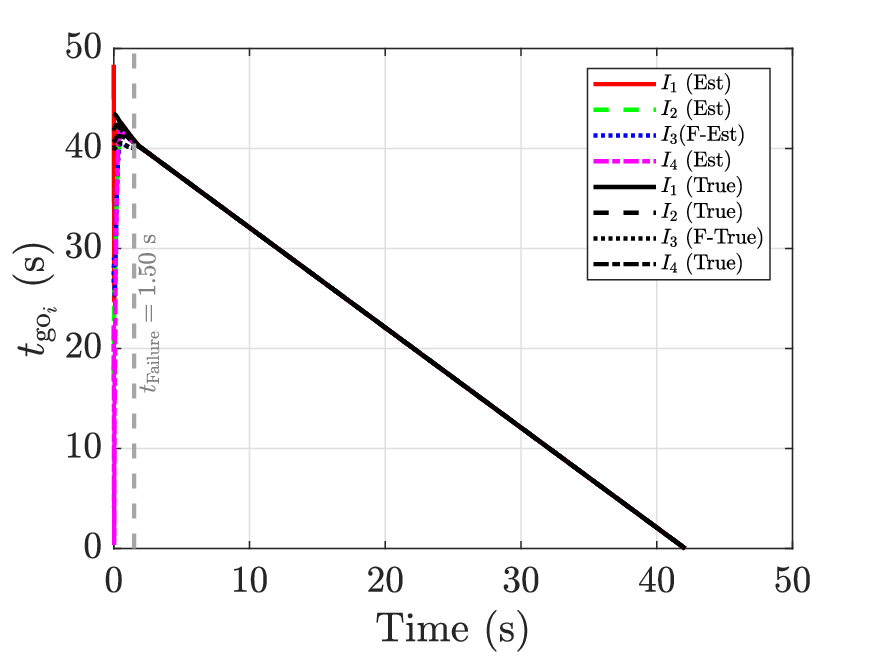}
        \includegraphics[width=0.49\linewidth]{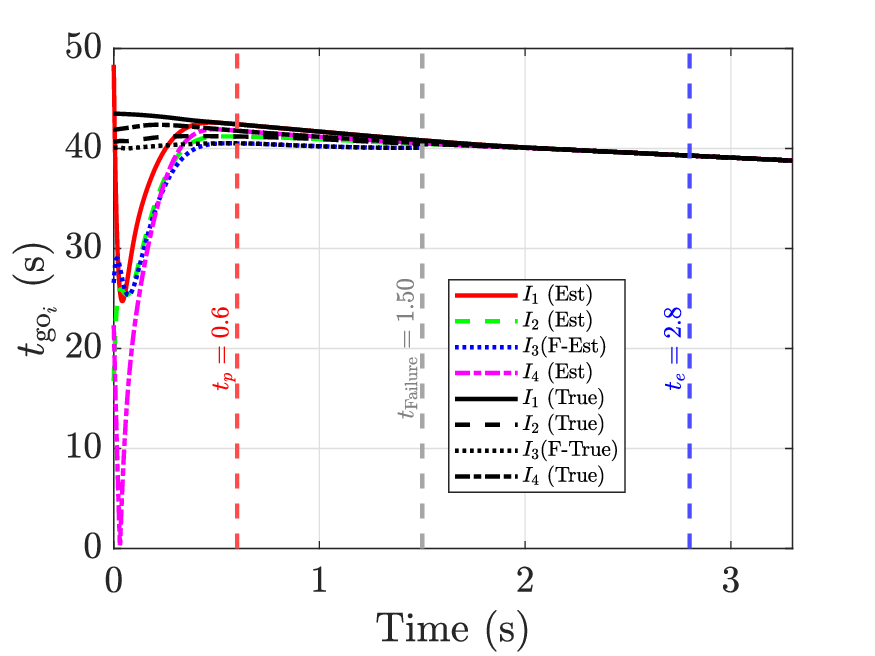}
        \caption{True vs Estimated Time-to-go (left) and Zoomed view (right).}
        \label{fig:truevsestim_combined3}
    \end{subfigure}
    
    \begin{subfigure}[t]{0.45\textwidth}
        \centering
        \includegraphics[width=\linewidth]{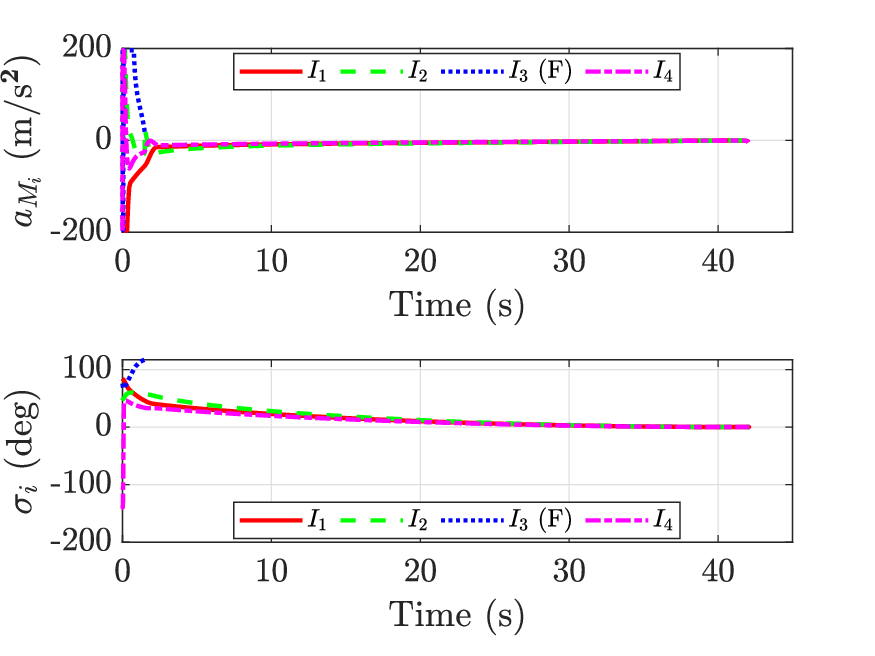}
        \caption{Interceptors' lateral acceleration and lead angle profiles.}
        \label{fig:acceleration3}
    \end{subfigure}
    \begin{subfigure}[t]{0.45\textwidth}
        \centering
        \includegraphics[width=\linewidth]{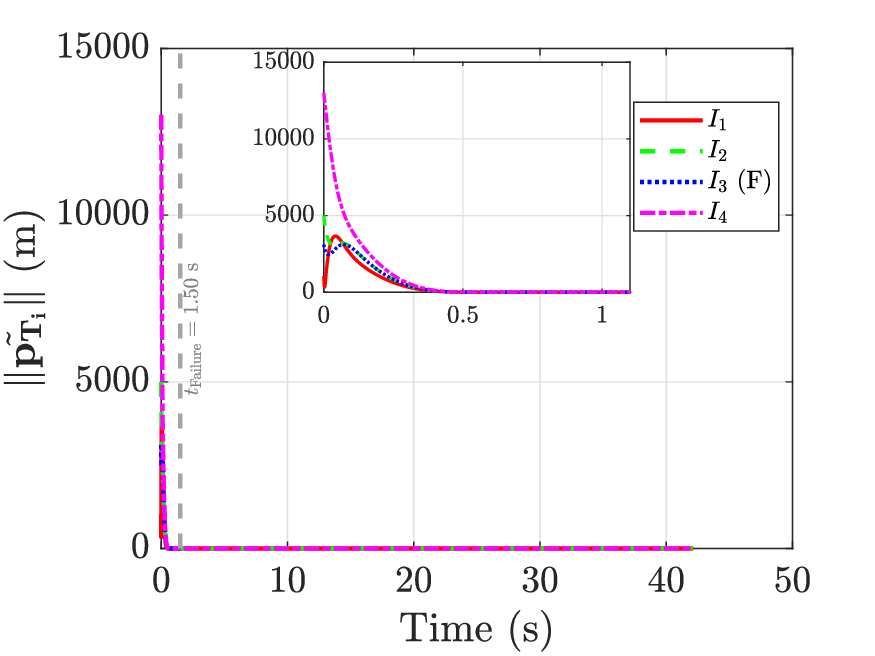}
        \caption{Target's state estimation error.}
        \label{fig:Observer_error3}
    \end{subfigure}
    
    \caption{Performance evaluation of the proposed strategy when interceptor 3 fails.}
    \label{fig:stationary3}
\end{figure}

Consider the scenario in which {interceptor~3} fails at \( t = 1.5 \mathrm{~s}\) under the same engagement conditions, due to reasons such as an unexpected onboard hardware malfunction or actuator outage. The corresponding sensing and actuation topologies are shown in \Cref{fig:graphs_fail1}, where the dotted line denotes the communication link that becomes inactive as a result of the agent failure. From \Cref{fig:stationary3}, it can be observed that simultaneous interception by the remaining agents is still achieved at approximately \( 42.1\,\text{s} \), provided the target maintains a spanning tree in the sensing graph \( \mathscr{S} \) and the actuation graph \( \mathscr{A} \) remains strongly connected. This validates the efficacy and robustness of the proposed strategy against agent failure.

\subsection{Cooperative Salvo with Large Number of Interceptors}

\begin{figure}[h!]
    \centering
    \begin{subfigure}[b]{0.47\linewidth}
        \centering
         \resizebox{.5\linewidth}{!}{%
\begin{tikzpicture}[
    ->, >=Stealth, thick,
    font=\bfseries\large,
    circnode/.style={circle, draw=red, text=red, minimum size=1cm, inner sep=0pt} 
]

  \node[circnode] (0) at (3,6) {0};
   \node[circle, draw=Tan, text=blue, minimum size=1cm, inner sep=0pt, fill = Tan]  (1) at (1,4) {1};
   \node[circle, draw=Tan, text=blue, minimum size=1cm, inner sep=0pt, fill = Tan]  (4) at (5,4) {4};
   \node[circle, draw=blue, text=blue, minimum size=1cm, inner sep=0pt]  (2) at (1,2) {2};
  \node[circle, draw=blue, text=blue, minimum size=1cm, inner sep=0pt]  (5) at (5,2) {5};
  \node[circle, draw=blue, text=blue, minimum size=1cm, inner sep=0pt]  (3) at (1,0) {3};
  \node[circle, draw=blue, text=blue, minimum size=1cm, inner sep=0pt]  (6) at (5,0) {6};

  \draw (0) -- (1);
  \draw (0) -- (4);

  \draw (1) -- (2);
  \draw (1) -- (4);

  \draw (2) -- (3);
  \draw (2) -- (6);
  \draw (2) -- (4);

  \draw (3) -- (5);

  \draw (4) -- (5);

  \draw (5) -- (2);
    \draw (5) -- (1);
  \draw (5) -- (6);

  \draw (6) -- (3);

  \node[text=red] at (3,6.7) {Target};
\node[text=blue, anchor=east] at ([xshift=-0.4cm]1.west) {$I_1$};
\node[text=blue, anchor=west] at ([xshift=0.4cm]4.east) {$I_4$};
\node[text=blue, anchor=east] at ([xshift=-0.4cm]2.west) {$I_2$};
\node[text=blue, anchor=west] at ([xshift=0.4cm]5.east) {$I_5$};
\node[text=blue, anchor=east] at ([xshift=-0.4cm]3.west) {$I_3$};
\node[text=blue, anchor=west] at ([xshift=0.4cm]6.east) {$I_6$};

\end{tikzpicture}%
}
        \caption{Sensing Graph $\mathscr{S}$}
        \label{sensing_graph_2}
    \end{subfigure}
    \hfill
    \begin{subfigure}[b]{0.47\linewidth}
        \centering
        \resizebox{.5\linewidth}{!}{%
\begin{tikzpicture}[
    ->, >=Stealth, thick,
    font=\bfseries\large,
    circnode/.style={circle, draw=blue, text=blue, minimum size=1cm, inner sep=0pt}
]

   \node[circle, draw=blue, text=blue, minimum size=1cm, inner sep=0pt]  (1) at (1,4) {1};
   \node[circle, draw=blue, text=blue, minimum size=1cm, inner sep=0pt]  (4) at (5,4) {4};
   \node[circle, draw=blue, text=blue, minimum size=1cm, inner sep=0pt]  (2) at (1,2) {2};
  \node[circle, draw=blue, text=blue, minimum size=1cm, inner sep=0pt]  (5) at (5,2) {5};
  \node[circle, draw=blue, text=blue, minimum size=1cm, inner sep=0pt]  (3) at (1,0) {3};
  \node[circle, draw=blue, text=blue, minimum size=1cm, inner sep=0pt]  (6) at (5,0) {6};


  \draw (1) -- (4);
  \draw (1) -- (5);
  
  \draw (2) -- (5);
  \draw (2) -- (4);
  \draw (2) -- (1);

  \draw (3) -- (2);
    \draw (3) -- (6);

  \draw (4) -- (5);

  \draw (5) -- (3);
  \draw (5) -- (6);

  \draw (6) -- (2);

\node[text=blue, anchor=east] at ([xshift=-0.4cm]1.west) {$I_1$};
\node[text=blue, anchor=west] at ([xshift=0.4cm]4.east) {$I_4$};
\node[text=blue, anchor=east] at ([xshift=-0.4cm]2.west) {$I_2$};
\node[text=blue, anchor=west] at ([xshift=0.4cm]5.east) {$I_5$};
\node[text=blue, anchor=east] at ([xshift=-0.4cm]3.west) {$I_3$};
\node[text=blue, anchor=west] at ([xshift=0.4cm]6.east) {$I_6$};

\end{tikzpicture}
}
        \caption{Actuation Graph $\mathscr{A}$}
        \label{actuation_graph_2}
    \end{subfigure}
    \caption{Interceptors' Communication Topologies.}
    \label{fig:graphs2}
\end{figure}
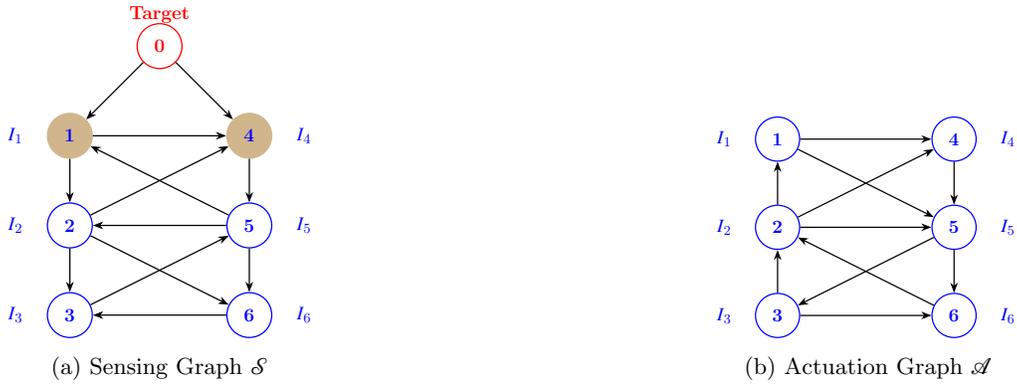

To evaluate the effectiveness of the proposed framework in scenarios involving multiple seeker-equipped interceptors and an expanded network of cooperating agents, consider an engagement scenario comprising one target and six interceptors (denoted as $I_1$ to $I_6$). Among these, interceptors $I_1$ and $I_4$ are equipped with onboard seekers that enable direct measurement of the target state, whereas the remaining interceptors operate without seekers and rely solely on networked information exchange. 

The sensing topology for this engagement is depicted in \Cref{sensing_graph_2}, while the corresponding actuation graph is illustrated in \Cref{actuation_graph_2}.
In contrast to the previous cases presented in \Cref{fig:graphs2}, the current scenario features dissimilar communication links in both the sensing and actuation topologies. $\begin{bmatrix} 190 & 220 & 230 & 210 & 200 & 215 \end{bmatrix}\,\text{m/s}$ and $\begin{bmatrix} 45^\circ & 20^\circ & 35^\circ &15^\circ& 20^\circ & 25^\circ \end{bmatrix}$ are the initial launch speeds and heading angles of the interceptors, respectively. Their launch coordinates are given by $[(8.5,8) \; (7.4,6) \; (7.2,0) \; (7.8,-0.3) \; (8,0.2) \; (7.6,-0.4)]$ km.  The initial target's position, maximum achievable lateral acceleration, controller, and observer convergence times are assumed to be the same as in the previous case. The observer and controller gains are chosen according to the design requirements.  

\begin{figure}[h!]  
    \centering
    \begin{subfigure}[t]{0.45\textwidth}
        \centering
        \includegraphics[width=\linewidth]{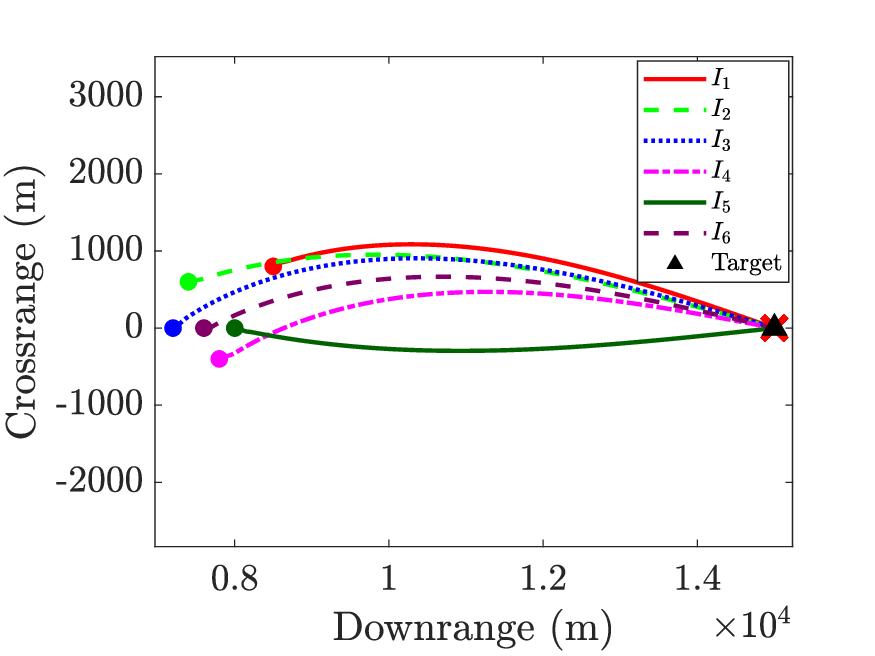}
        \caption{Interceptor's trajectories.}
        \label{fig:Trajectory4}
    \end{subfigure}
    \begin{subfigure}[t]{0.45\textwidth}
        \centering
        \includegraphics[width=\linewidth]{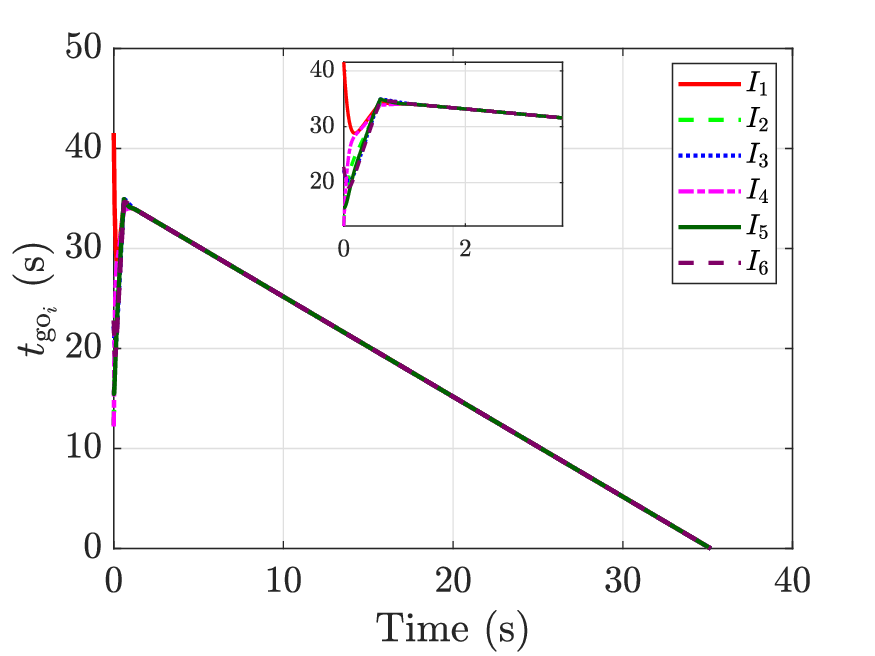}
        \caption{Interceptors' time-to-go values.}
        \label{fig:Time-to-go4}
    \end{subfigure}
    
    \begin{subfigure}[t]{0.9\textwidth}
        \centering
        \includegraphics[width=0.49\linewidth]{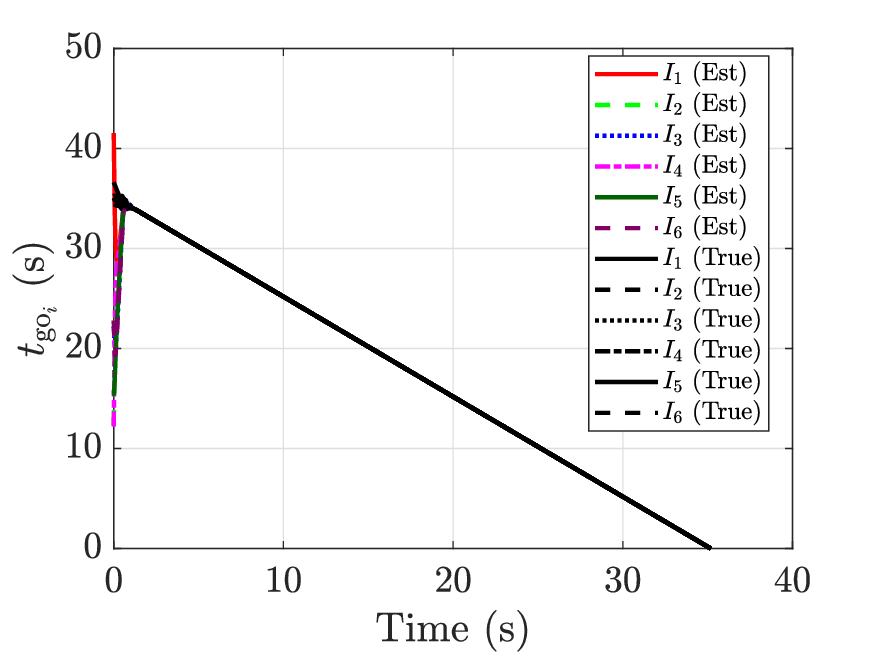}
        \includegraphics[width=0.49\linewidth]{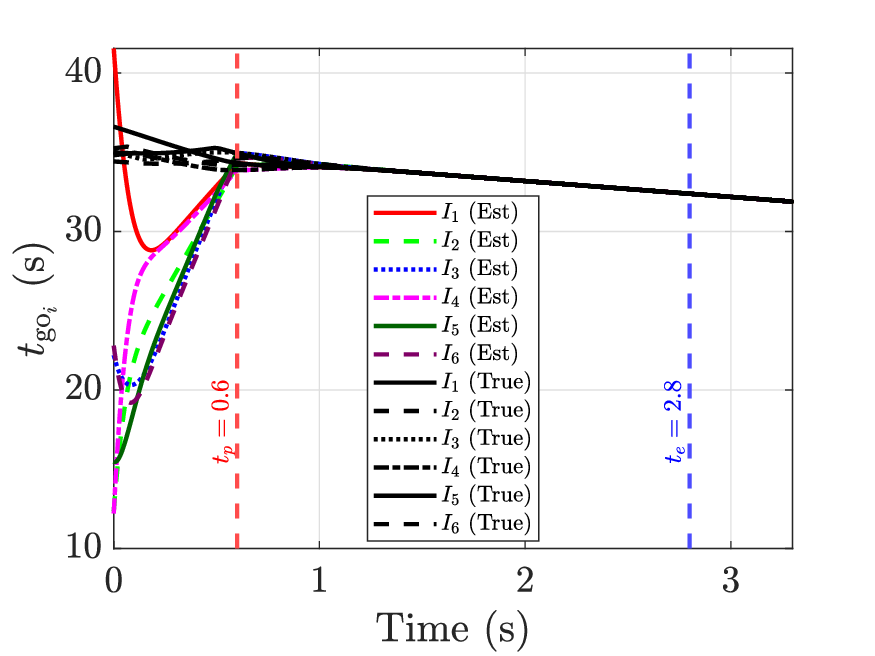}
        \caption{True vs Estimated Time-to-go (left) and Zoomed view (right).}
        \label{fig:truevsestim_combined4}
    \end{subfigure}
    
    \begin{subfigure}[t]{0.45\textwidth}
        \centering
        \includegraphics[width=\linewidth]{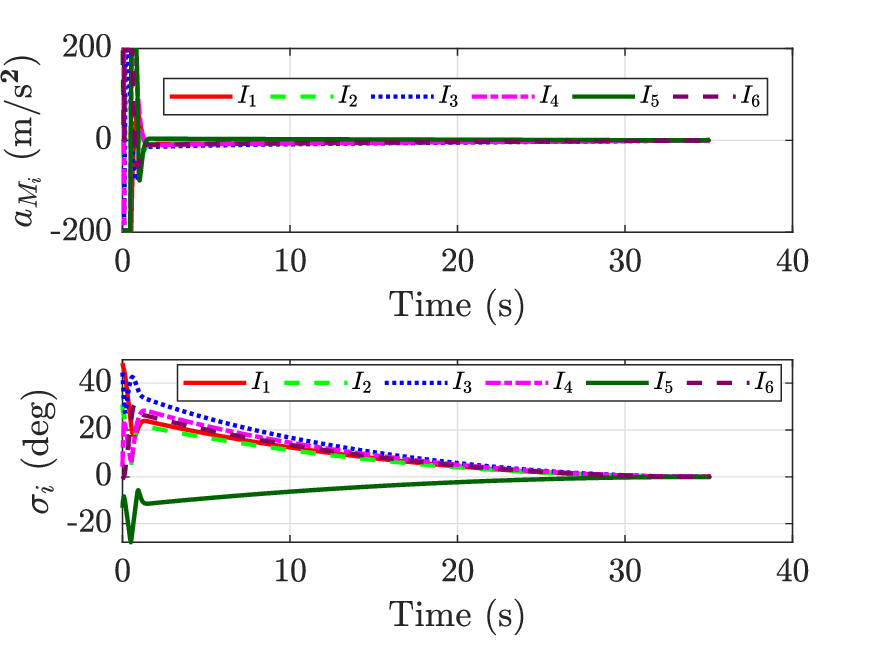}
        \caption{Interceptors' lateral acceleration and lead angle profiles.}
        \label{fig:acceleration4}
    \end{subfigure}
    \begin{subfigure}[t]{0.45\textwidth}
        \centering
        \includegraphics[width=\linewidth]{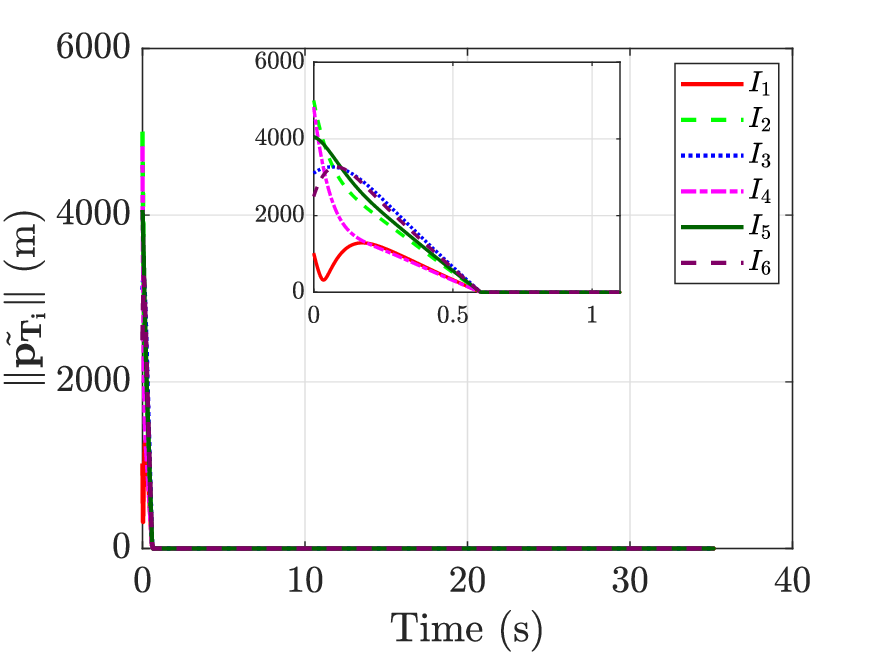}
        \caption{Target's state Estimation Error.}
        \label{fig:Observer_error4}
    \end{subfigure}
    
    \caption{Performance evaluation of the proposed method with an increased number of interceptors.}
    \label{fig:stationary4}
\end{figure}

Similar to previous cases. the initial target's position estimates for observer are chosen arbitrarily, which yields time-to-go estimates of  $[36.610,\; 35.250,\; 35.029,\; 34.813,\; 35.000,\; 34.419]~\text{s}$, while the corresponding true initial time-to-go values are $[41.545,\; 12.352,\; 22.915,\; 12.213,\; 15.482,\; 22.794]~\text{s}.$ \Cref{fig:stationary4} illustrates the engagement scenario with a large number of interceptors achieving cooperative simultaneous interception at $35.18~\mathrm{s}$. The corresponding interceptor trajectories are presented in \Cref{fig:Trajectory4}. The convergence of the estimated time-to-go values reaching consensus by $t_e = 2.8~\mathrm{s}$ is shown in \Cref{fig:Time-to-go4}. The associated lateral acceleration commands and lead-angle profiles are depicted in \Cref{fig:acceleration4}, while \Cref{fig:Observer_error4} shows the observer error converging to zero within $t_p = 0.6~\mathrm{s}$. Consistent with the earlier scenario, the lead angles exhibit noticeable fluctuations for trajectory corrections, and the lateral acceleration demands remain relatively high until a consensus in time-to-go is established. Once consensus is achieved at $t_e = 2.8~\mathrm{s}$, both the lead angles and the lateral acceleration commands show trends similar to what is observed previously.

\begin{figure}[h!]
    \centering
    \begin{subfigure}[b]{0.47\linewidth}
        \centering
         \resizebox{.5\linewidth}{!}{%
\begin{tikzpicture}[
    ->, >=Stealth, thick,
    font=\bfseries\large,
    circnode/.style={circle, draw=red, text=red, minimum size=1cm, inner sep=0pt} 
]

  \node[circnode] (0) at (3,6) {0};
   \node[circle, draw=Tan, text=blue, minimum size=1cm, inner sep=0pt, fill = Tan]  (1) at (1,4) {1};
   \node[circle, draw=Tan, text=blue, minimum size=1cm, inner sep=0pt, fill = Tan]  (4) at (5,4) {4};
   \node[circle, draw=blue, text=blue, minimum size=1cm, inner sep=0pt]  (2) at (1,2) {2};
  \node[circle, draw=blue, text=blue, minimum size=1cm, inner sep=0pt]  (5) at (5,2) {5};
  \node[circle, draw=blue, text=blue, minimum size=1cm, inner sep=0pt]  (3) at (1,0) {3};
  \node[circle, draw=blue, text=blue, minimum size=1cm, inner sep=0pt]  (6) at (5,0) {6};

  \draw (0) -- (1);
  \draw[dashed,red,->] (0) -- (4);

  \draw (1) -- (2);
  \draw[dashed,red,->] (1) -- (4);

  \draw[dashed,red,->] (2) -- (3);
  \draw (2) -- (6);
  \draw (2) -- (4);

  \draw (3) -- (5);

  \draw (4) -- (5);

  \draw[dashed,red,->] (5) -- (2);
    \draw (5) -- (1);
  \draw (5) -- (6);

  \draw (6) -- (3);

  \node[text=red] at (3,6.7) {Target};
\node[text=blue, anchor=east] at ([xshift=-0.4cm]1.west) {$I_1$};
\node[text=blue, anchor=west] at ([xshift=0.4cm]4.east) {$I_4$};
\node[text=blue, anchor=east] at ([xshift=-0.4cm]2.west) {$I_2$};
\node[text=blue, anchor=west] at ([xshift=0.4cm]5.east) {$I_5$};
\node[text=blue, anchor=east] at ([xshift=-0.4cm]3.west) {$I_3$};
\node[text=blue, anchor=west] at ([xshift=0.4cm]6.east) {$I_6$};

\end{tikzpicture}%
}
        \caption{Sensing Graph $\mathscr{S}$}
        \label{sensing_graph_fail2}
    \end{subfigure}
    \hfill
    \begin{subfigure}[b]{0.47\linewidth}
        \centering
        \resizebox{.5\linewidth}{!}{%
\begin{tikzpicture}[
    ->, >=Stealth, thick,
    font=\bfseries\large,
    circnode/.style={circle, draw=blue, text=blue, minimum size=1cm, inner sep=0pt}
]

   \node[circle, draw=blue, text=blue, minimum size=1cm, inner sep=0pt]  (1) at (1,4) {1};
   \node[circle, draw=blue, text=blue, minimum size=1cm, inner sep=0pt]  (4) at (5,4) {4};
   \node[circle, draw=blue, text=blue, minimum size=1cm, inner sep=0pt]  (2) at (1,2) {2};
  \node[circle, draw=blue, text=blue, minimum size=1cm, inner sep=0pt]  (5) at (5,2) {5};
  \node[circle, draw=blue, text=blue, minimum size=1cm, inner sep=0pt]  (3) at (1,0) {3};
  \node[circle, draw=blue, text=blue, minimum size=1cm, inner sep=0pt]  (6) at (5,0) {6};

  \draw[dashed,red,->] (1) -- (4);
  \draw (1) -- (5);
  
 \draw[dashed,red,->] (2) -- (5);
  \draw (2) -- (4);
  \draw (2) -- (1);

 \draw[dashed,red,->] (3) -- (2);
    \draw (3) -- (6);

  \draw (4) -- (5);

  \draw (5) -- (3);
  \draw (5) -- (6);

  \draw (6) -- (2);

\node[text=blue, anchor=east] at ([xshift=-0.4cm]1.west) {$I_1$};
\node[text=blue, anchor=west] at ([xshift=0.4cm]4.east) {$I_4$};
\node[text=blue, anchor=east] at ([xshift=-0.4cm]2.west) {$I_2$};
\node[text=blue, anchor=west] at ([xshift=0.4cm]5.east) {$I_5$};
\node[text=blue, anchor=east] at ([xshift=-0.4cm]3.west) {$I_3$};
\node[text=blue, anchor=west] at ([xshift=0.4cm]6.east) {$I_6$};

\end{tikzpicture}
}
        \caption{Actuation Graph $\mathscr{A}$}
        \label{actuation_graph_fail2}
    \end{subfigure}
    \caption{Interceptors' Communication Topologies after Seeker Failure and Multiple Link Failures at $t=0.5~\mathrm{s}$.}
    \label{fig:graphs_fail2}
\end{figure}
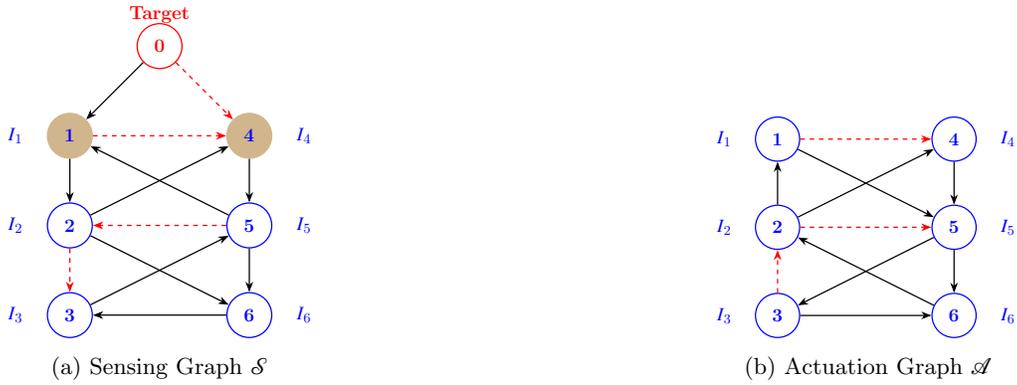
\begin{figure}[h!]  
    \centering
    \begin{subfigure}[t]{0.45\textwidth}
        \centering
        \includegraphics[width=\linewidth]{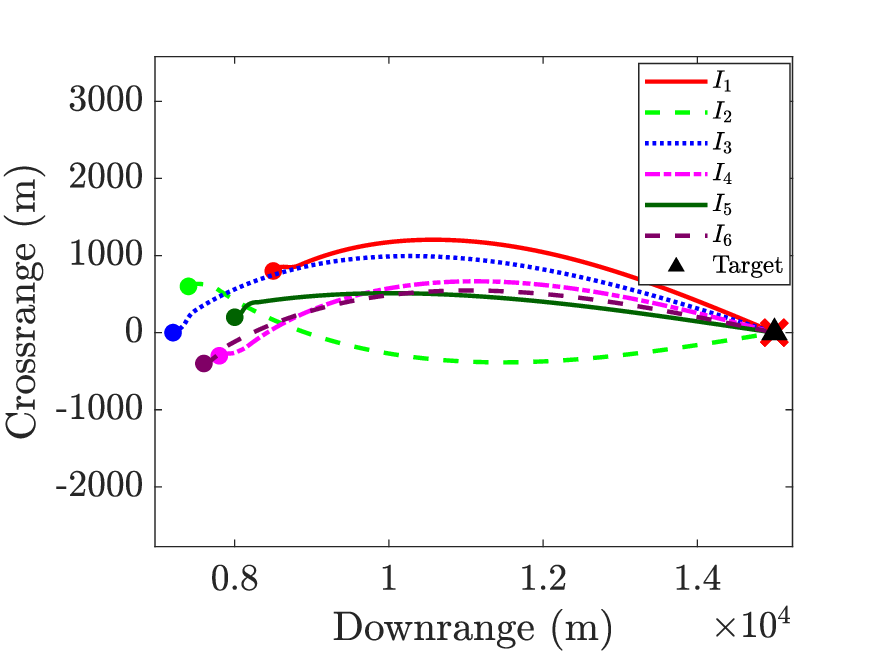}
        \caption{Interceptors' Trajectories.}
        \label{fig:Trajectory5}
    \end{subfigure}
    \begin{subfigure}[t]{0.45\textwidth}
        \centering
        \includegraphics[width=\linewidth]{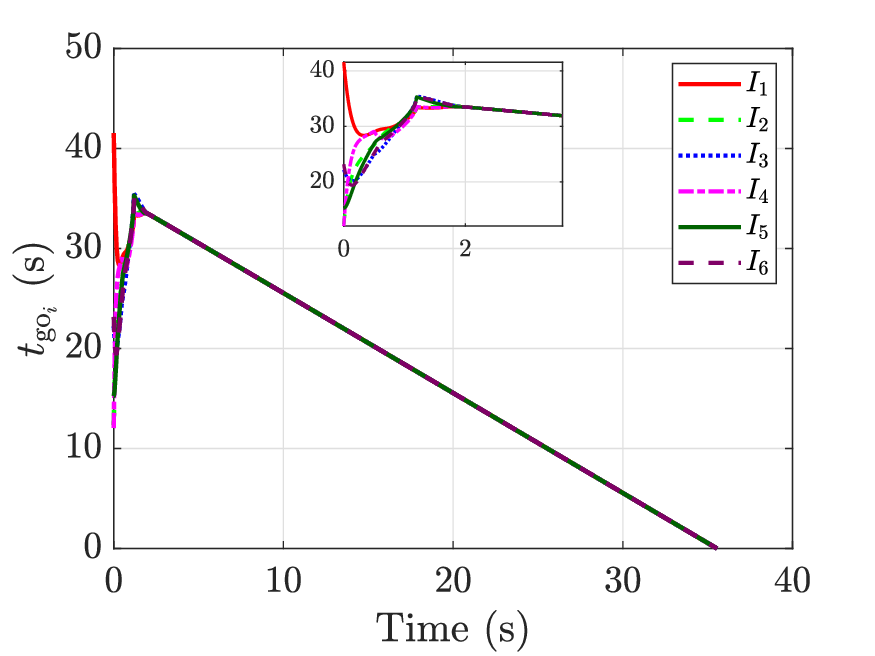}
        \caption{Interceptors' time-to-go values.}
        \label{fig:Time-to-go5}
    \end{subfigure}
    
    \begin{subfigure}[t]{0.9\textwidth}
        \centering
        \includegraphics[width=0.49\linewidth]{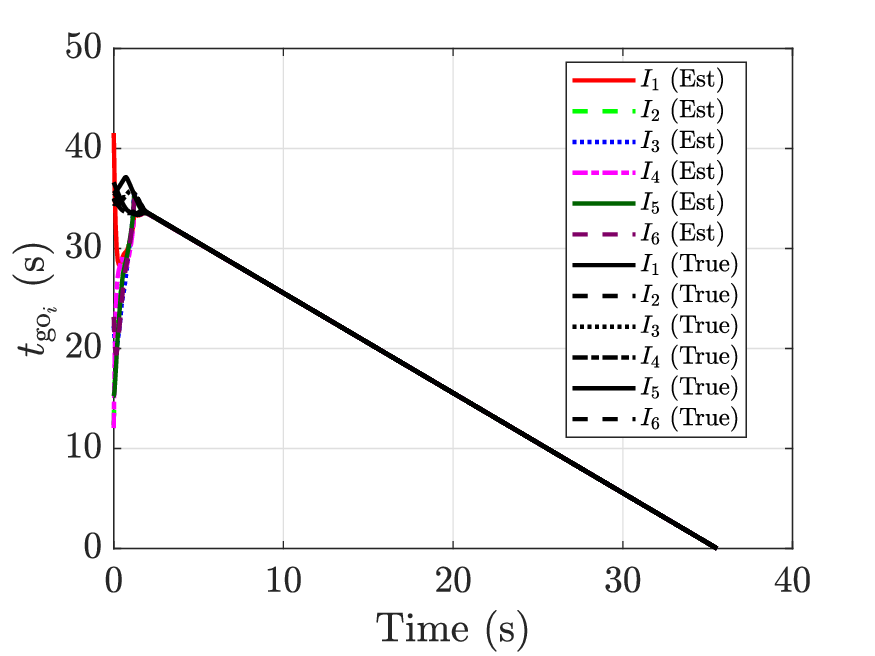}
        \includegraphics[width=0.49\linewidth]{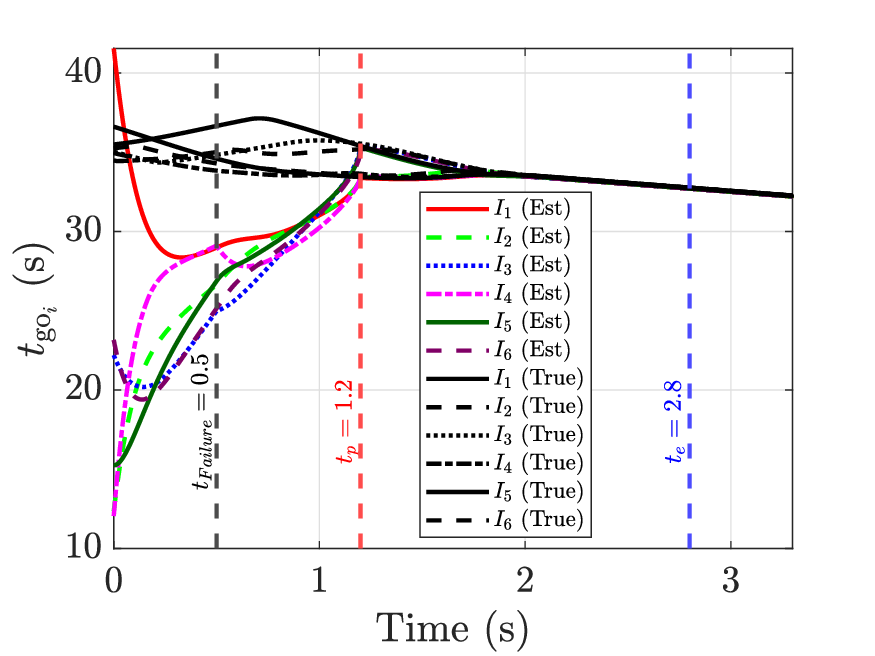}
        \caption{True vs Estimated Time-to-go (left) and Zoomed view (right).}
        \label{fig:truevsestim_combined5}
    \end{subfigure}
    
    \begin{subfigure}[t]{0.45\textwidth}
        \centering
        \includegraphics[width=\linewidth]{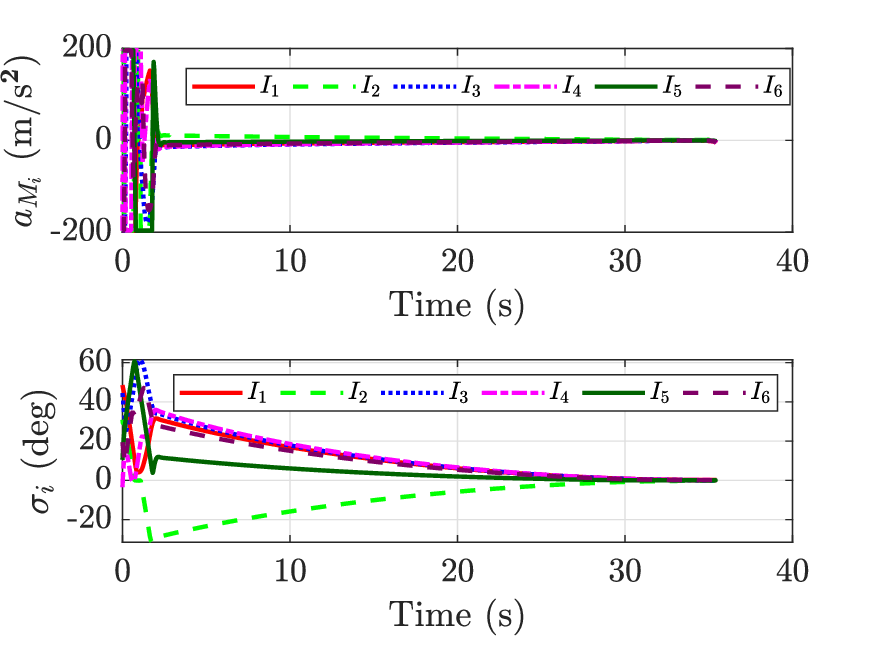}
        \caption{Interceptors' lateral acceleration and lead angle profiles.}
        \label{fig:acceleration5}
    \end{subfigure}
    \begin{subfigure}[t]{0.45\textwidth}
        \centering
        \includegraphics[width=\linewidth]{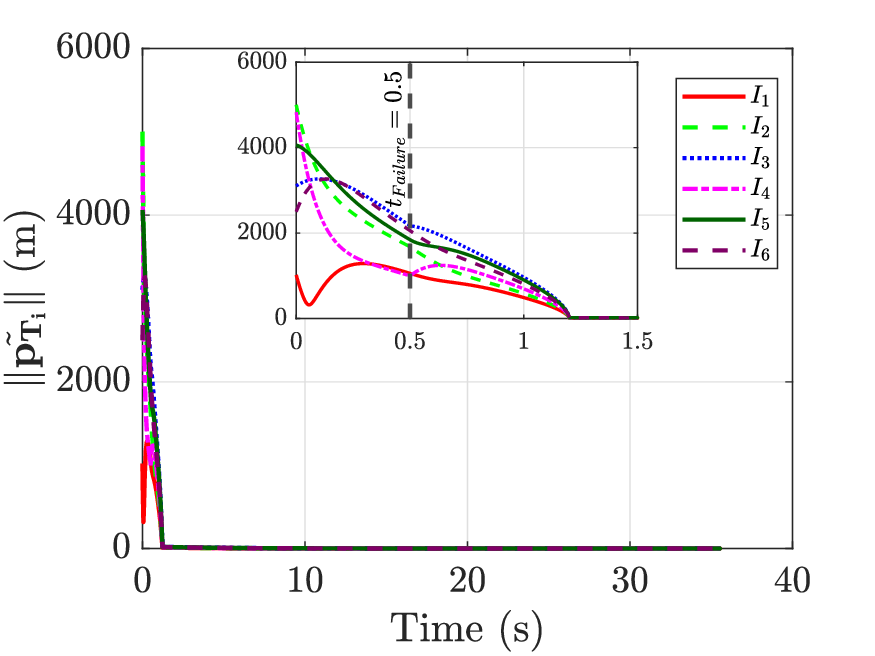}
        \caption{Target's state estimation error.}
        \label{fig:Observer_error5}
    \end{subfigure}
    
    \caption{Performance evaluation of the proposed strategy subjected to seeker and link failures. }
    \label{fig:stationary5}
\end{figure}

Consider a scenario in which the seeker of interceptor~4 fails at \( t = 0.5~\mathrm{s} \) due to factors such as loss of line-of-sight, adverse environmental-induced sensor saturation, or an onboard hardware malfunction, while the interceptor itself remains operational. Additionally, consider the link failures illustrated by the dotted lines in \Cref{fig:graphs_fail2}, which may result from intermittent communication outages, environmental disturbances, or temporary channel degradation. The engagement conditions remain the same, except for an increased observer convergence time of \( t_p = 1.2~\mathrm{s} \). As shown in \Cref{fig:stationary5},  despite seeker and link failures, all agents achieve cooperative simultaneous interception at approximately \( 35.55~\mathrm{s} \), despite the loss of one seeker-equipped interceptor. Moreover, the observer errors of all agents converge to zero as presented in \Cref{fig:Observer_error5}, demonstrating the effectiveness of the proposed framework. These results highlight the merits of the proposed framework. Cooperative state estimation remains reliable provided the target maintains a spanning tree in the sensing graph \(\mathscr{S}\) and the actuation graph \(\mathscr{A}\) stays strongly connected, even in the presence of seeker and link failures to guarantee a cooperative simultaneous interception.

It is worth noting that one of the advantages of the proposed strategy lies in its inherent robustness to the failure of seeker-equipped agents. In contrast to the leader–follower topology proposed in \cite{sp29}, where a single leader interceptor is the only source of target state information and its failure can compromise the entire mission, the proposed framework distributes sensing capability across multiple interceptors. This redundancy ensures that even if one or more seeker-equipped agents fail, the remaining interceptors can still reconstruct the target's state through cooperative estimation. Consequently, the interception task continues uninterrupted, demonstrating the strategy’s resilience to failures by seeker-equipped agents.

\subsection{Autopilot Performance with Test Inputs}
\begin{figure}[h!]
    \centering
    \begin{subfigure}[t]{0.45\textwidth}
        \centering
        \includegraphics[width=\textwidth]{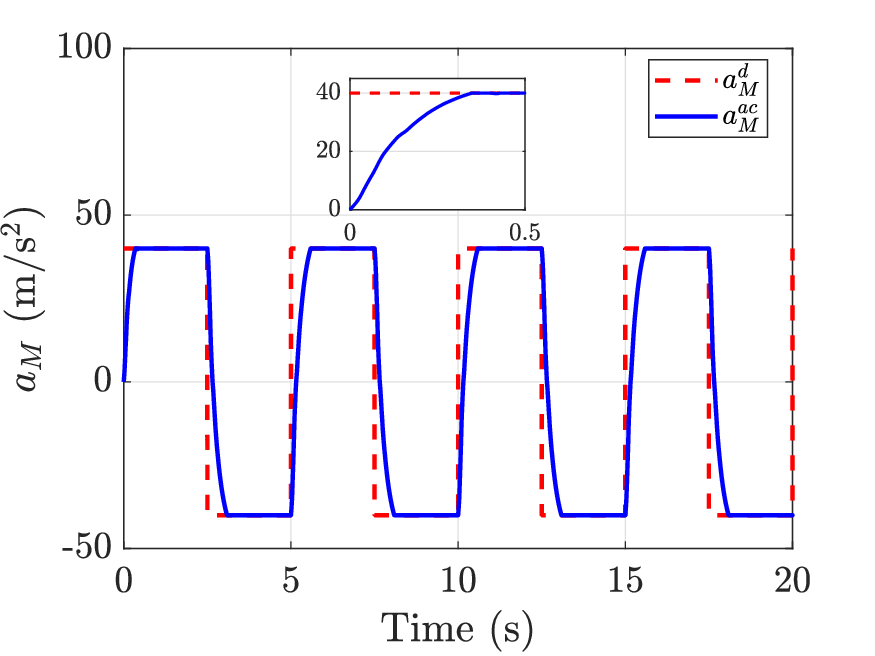}
        \caption{Square wave test input.}
        \label{fig:square_trajec}
    \end{subfigure}
    \hfill
    \begin{subfigure}[t]{0.45\textwidth}
        \centering
        \includegraphics[width=\textwidth]{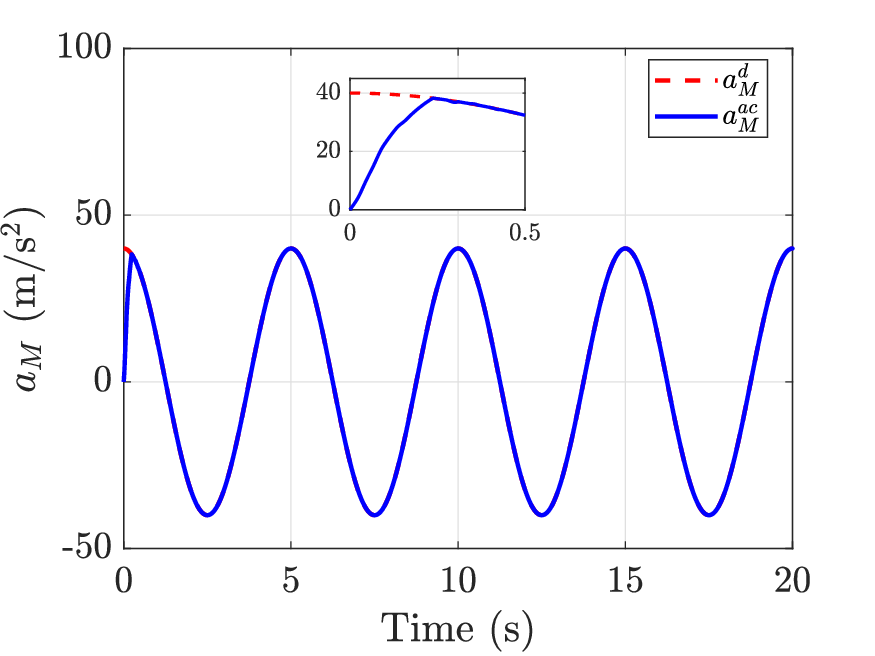}
        \caption{Sinusoidal test input.}
        \label{fig:cosine_trajec}
    \end{subfigure}

    \begin{subfigure}[t]{0.45\textwidth}
        \centering
        \includegraphics[width=\textwidth]{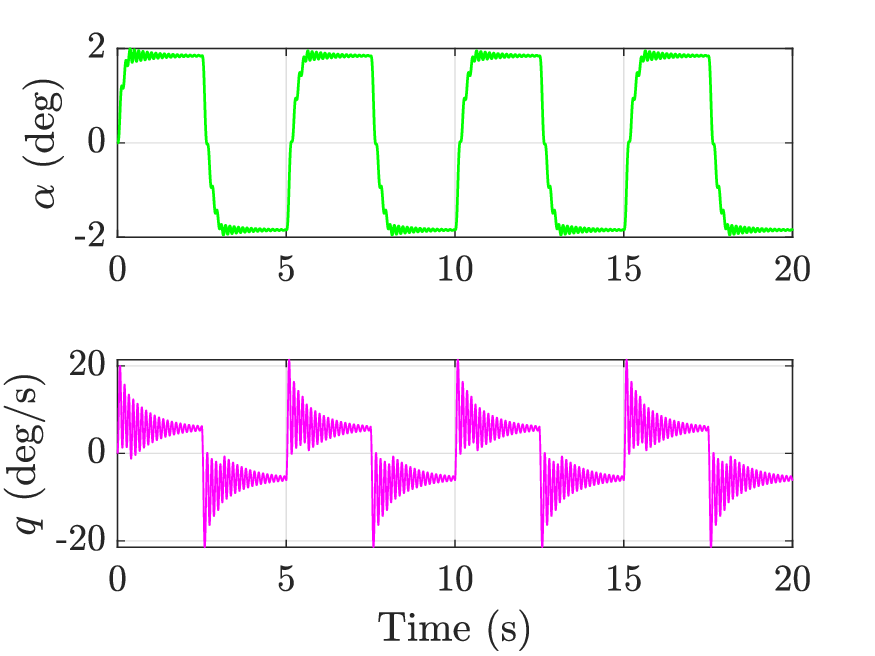}
        \caption{Angle of attack and pitch rate (square input).}
        \label{fig:square_aoa_q}
    \end{subfigure}
    \hfill
    \begin{subfigure}[t]{0.45\textwidth}
        \centering
        \includegraphics[width=\textwidth]{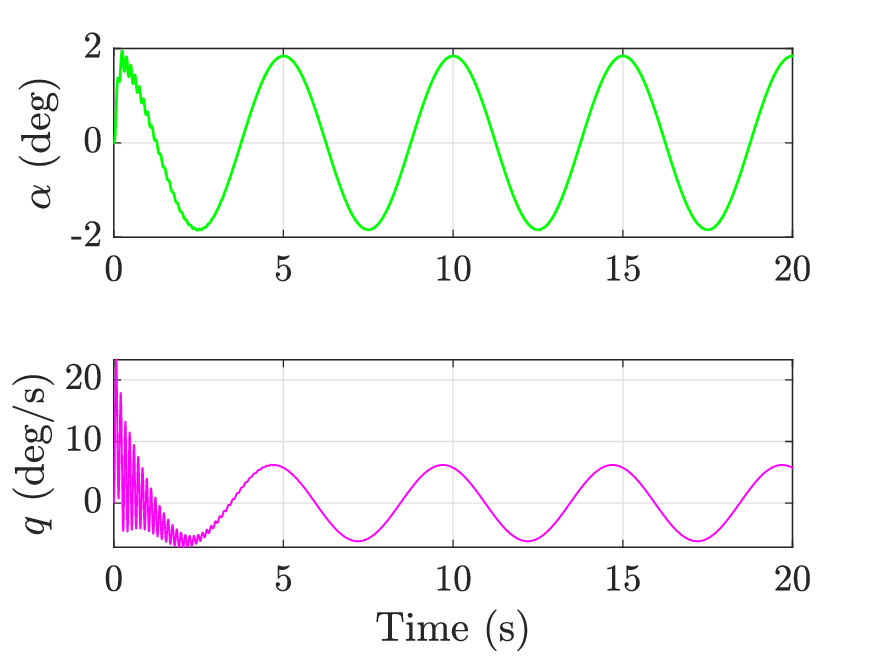}
        \caption{Angle of attack and pitch rate (sinusoidal input).}
        \label{fig:cosine_aoa_q}
    \end{subfigure}

    \begin{subfigure}[t]{0.45\textwidth}
        \centering
        \includegraphics[width=\textwidth]{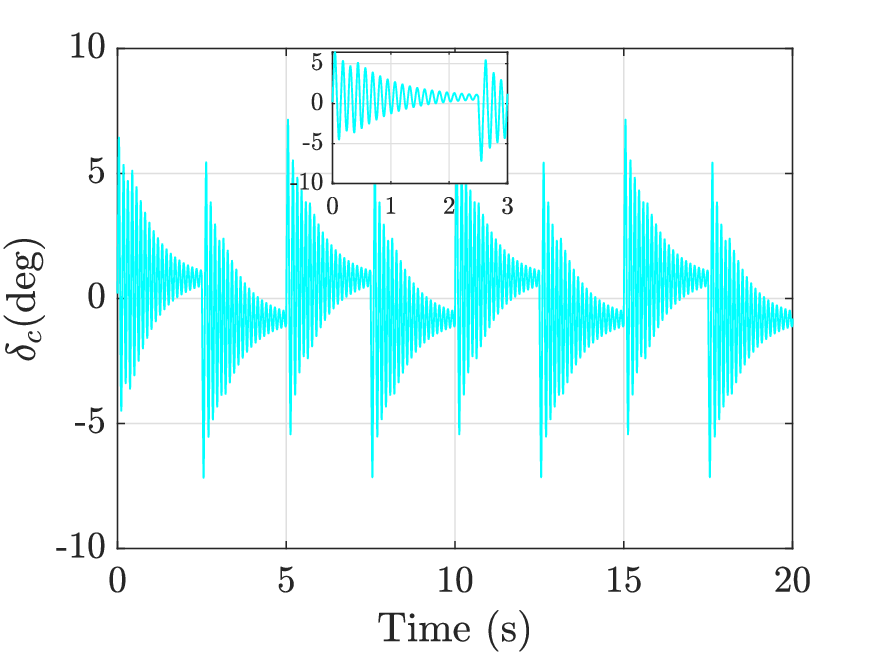}
        \caption{Canard deflection (square input)}
        \label{fig:square_canard}
    \end{subfigure}
    \hfill
    \begin{subfigure}[t]{0.45\textwidth}
        \centering
        \includegraphics[width=\textwidth]{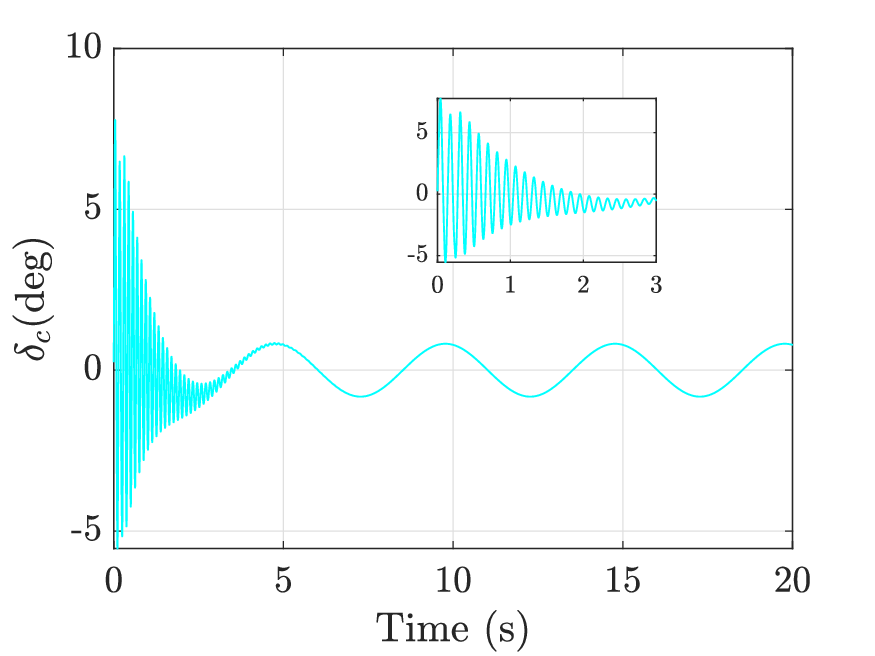}
        \caption{Canard deflection (sinusoidal input).}
        \label{fig:cosine_canard}
    \end{subfigure}

    \caption{Performance of interceptor autopilot for square and sinusoidal test inputs.}
    \label{fig:autopilot_test_inputs}
\end{figure}

The performance of the interceptors' autopilot for various test inputs is shown in \Cref{fig:autopilot_test_inputs}. The predefined settling time $t_a$ is chosen as $1 ~\mathrm{s}$. In the first case, a square waveform with a time period of 5~s is provided as the commanded input to the autopilot, followed by a cosine waveform of identical period. The proposed autopilot architecture ensures the accurate tracking of both test signals, achieving precise convergence within 0.5 s, i.e., almost instantaneously, thereby confirming the predefined-time convergence capability of the considered controller.

The corresponding variations of the states, angle of attack, pitch rate, and control surface deflections for both test inputs are illustrated in  \Cref{fig:autopilot_test_inputs}. From \Cref{fig:square_canard} and \Cref{fig:cosine_canard}, it can be observed that the canard deflections exhibit initial high-frequency oscillations whose amplitude gradually diminishes over time. For the sinusoidal input, these oscillations decay smoothly within $t=2.5~\mathrm{s}$, whereas for the square-wave input, the high-frequency components reappear periodically, consistent with the signal's discontinuities (sudden changes in the value of desired acceleration, in this case $a_{M}^d$). These high-frequency transients arise primarily from the rapid corrective action of the autopilot law during the initial response. In addition, part of these oscillations can be attributed to numerical errors introduced by the ODE solver, particularly when integrating fast-varying dynamics.

\begin{figure}[h!]
    \centering
    
    \begin{subfigure}[t]{0.45\textwidth}
        \centering
        \includegraphics[width=\linewidth]{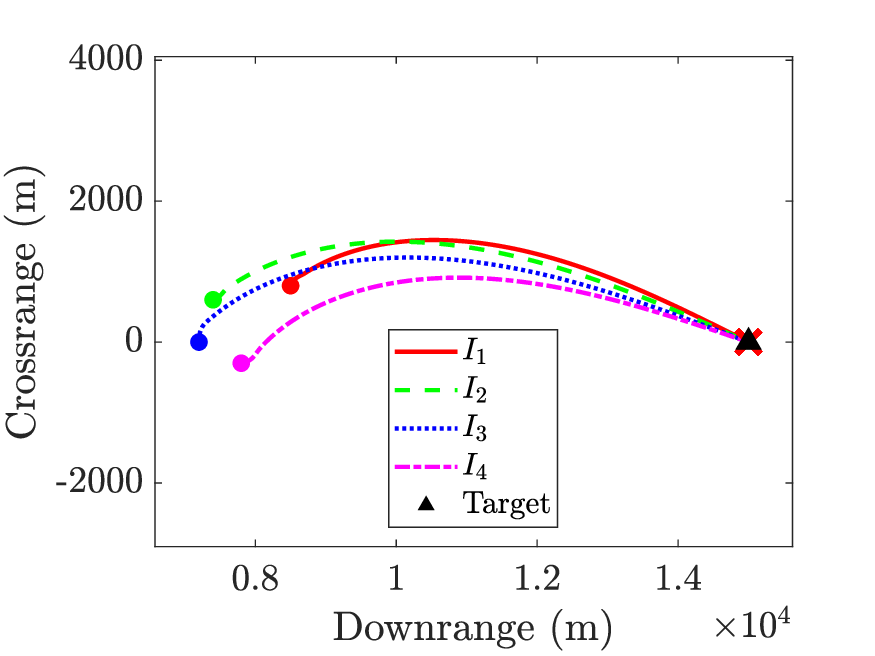}
        \caption{Interceptors' trajectories.}
        \label{fig:Trajectoryauto}
    \end{subfigure}
    \hfill
    \begin{subfigure}[t]{0.45\textwidth}
        \centering
        \includegraphics[width=\linewidth]{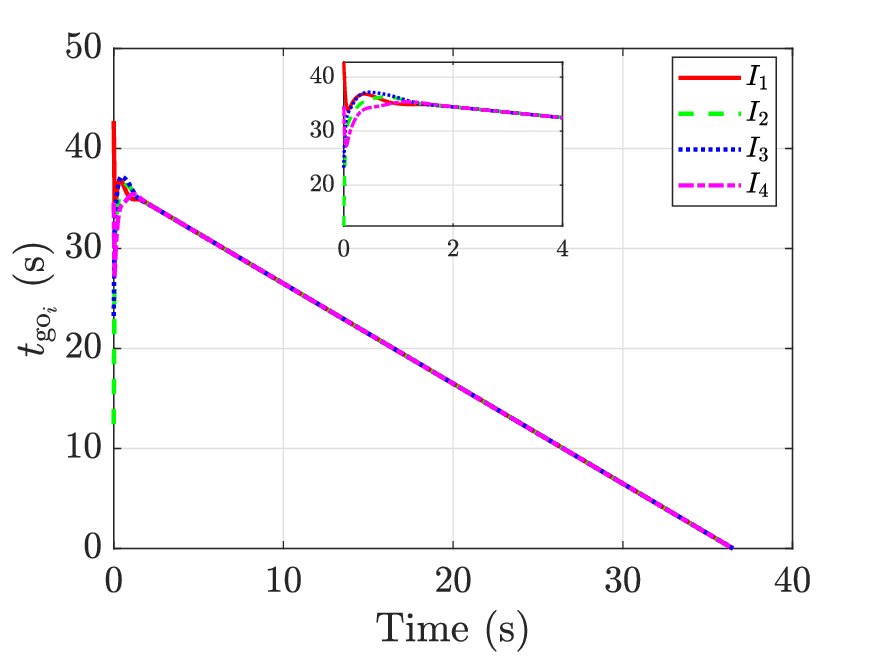}
        \caption{Interceptors' time-to-go values.}
        \label{fig:Time_to_goauto}
    \end{subfigure}

    \begin{subfigure}[t]{0.45\textwidth}
        \centering
        \includegraphics[width=\linewidth]{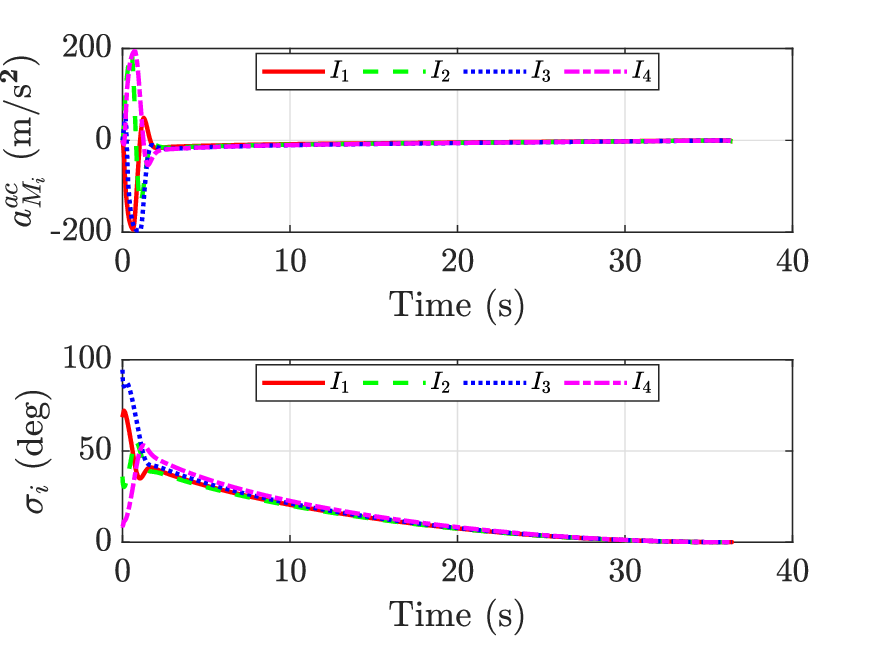}
        \caption{Interceptors' lateral acceleration and lead angle profiles.}
        \label{fig:Accelerationauto}
    \end{subfigure}
    \hfill
    \begin{subfigure}[t]{0.45\textwidth}
        \centering
        \includegraphics[width=\linewidth]{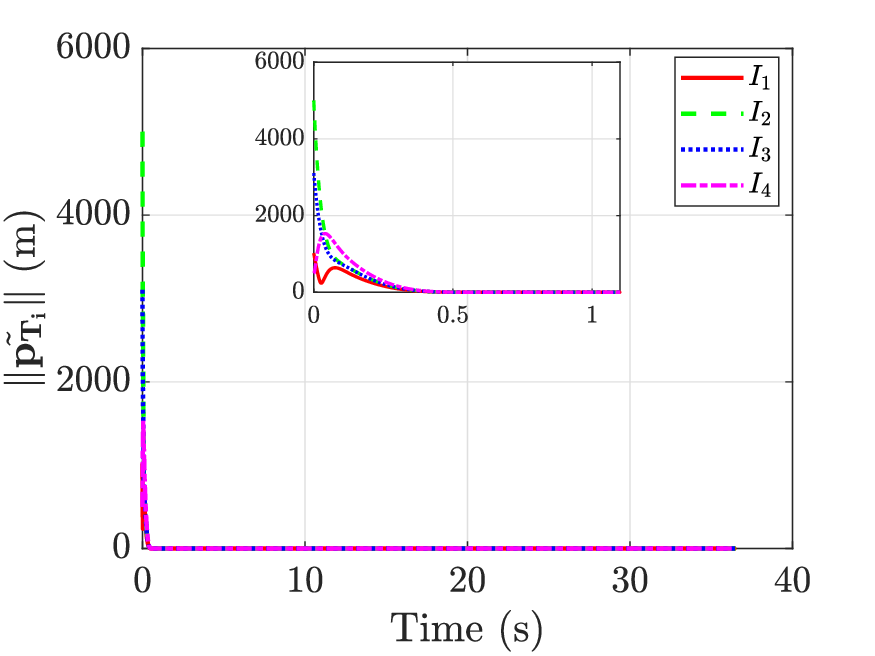}
        \caption{Target's state estimation error.}
        \label{fig:Observerauto}
    \end{subfigure}
    
    \begin{subfigure}[t]{0.45\textwidth}
        \centering
        \includegraphics[width=\linewidth]{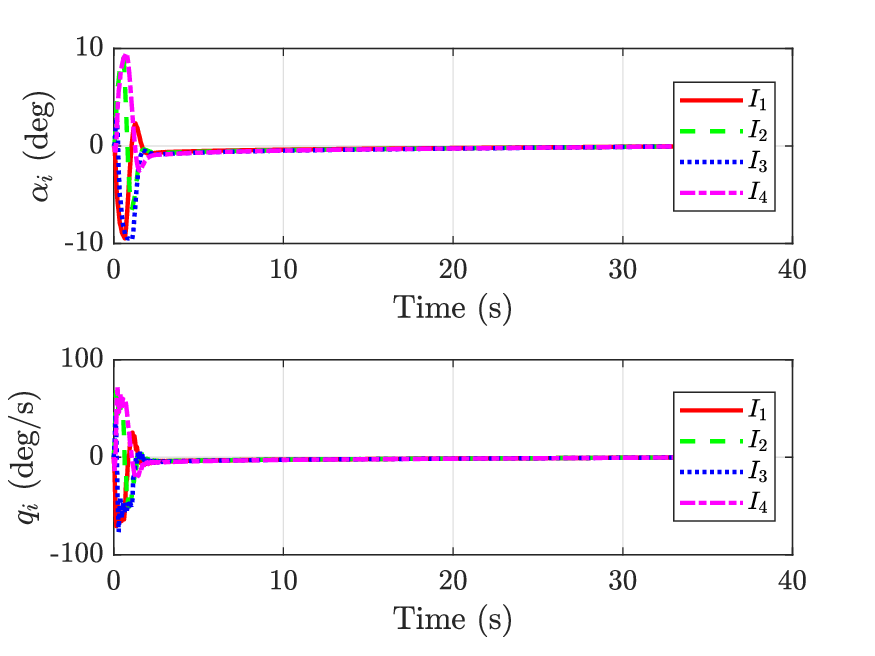}
        \caption{Interceptors' pitch rate and angle of attack.}
        \label{fig:alpha_q}
    \end{subfigure}
    \hfill
    \begin{subfigure}[t]{0.45\textwidth}
        \centering
        \includegraphics[width=\linewidth]{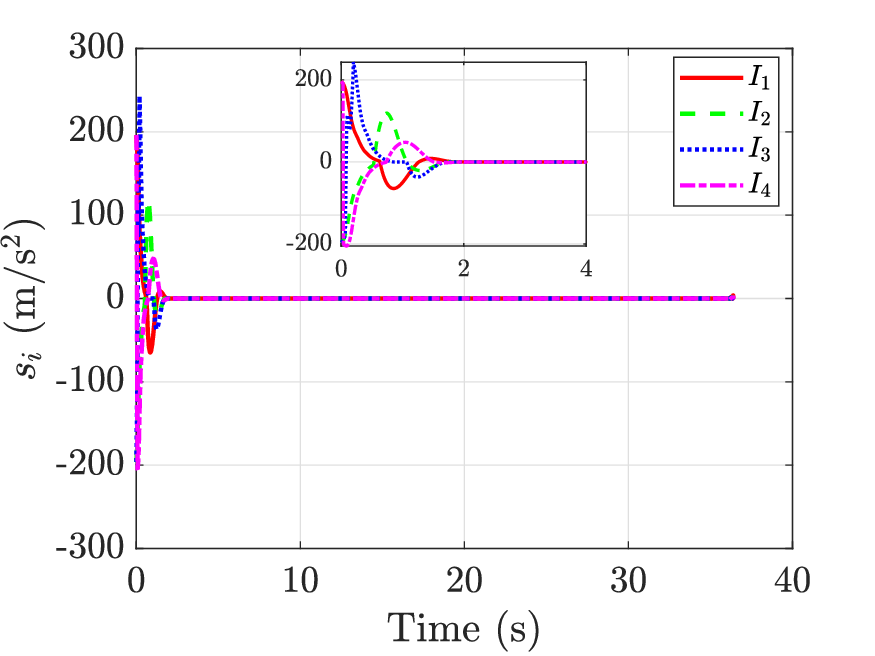}
        \caption{Interceptors' tracking error.}
        \label{fig:si}
    \end{subfigure}
    \caption{Performance evaluation of the proposed combined estimation–guidance–control architecture.}
    \label{fig:auto_full}
    
\end{figure}

\begin{figure}[h!]
    \ContinuedFloat
\centering
    \begin{subfigure}[t]{0.45\textwidth}
        \centering
        \includegraphics[width=\linewidth]{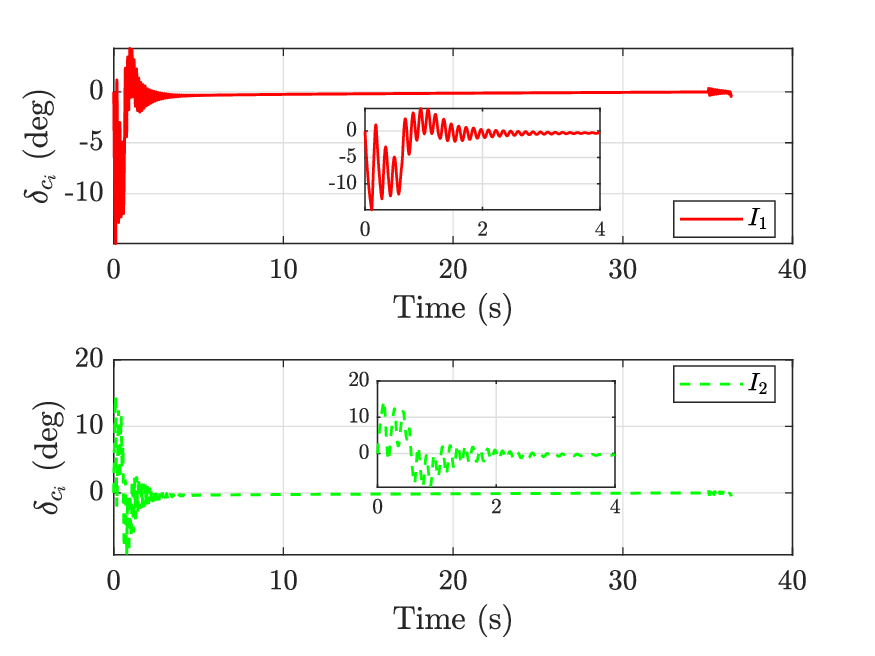}
        \caption{Canard deflections ($I_1$ and $I_2$). }
        \label{fig:delta_1_2}
    \end{subfigure}
    \hfill
    \begin{subfigure}[t]{0.45\textwidth}
        \centering
        \includegraphics[width=\linewidth]{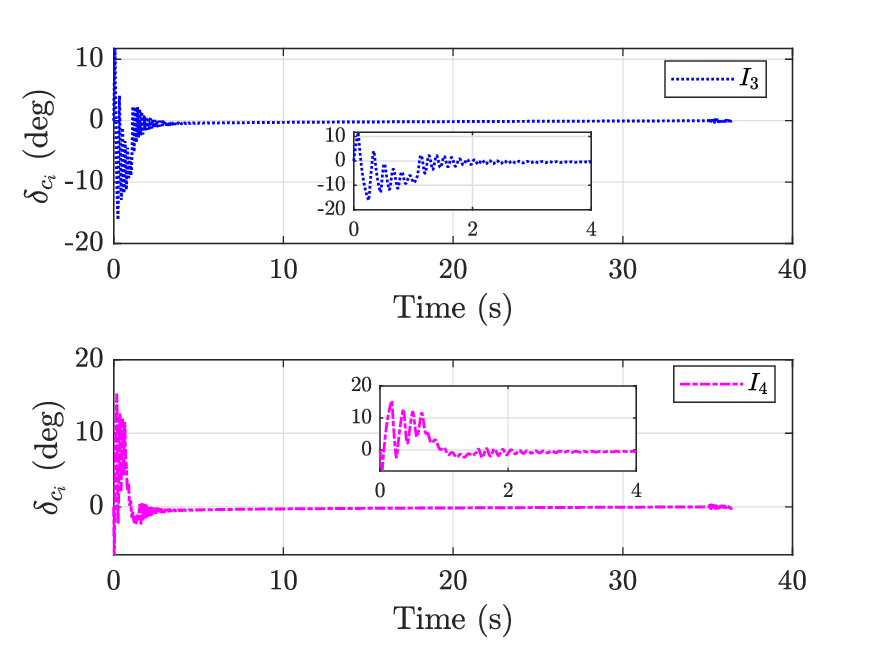}
        \caption{Canard deflections ($I_3$ and $I_4$).}
        \label{fig:delta_3_4}
    \end{subfigure}
    \captionsetup{list=off}
    \caption*{Figure.~\thefigure\ (continued).}
\end{figure}
\subsection{Autopilot Performance with Actual Commanded Inputs}
Thus far, the effectiveness of the proposed estimation–guidance framework has been established. To further substantiate its practical applicability, we now incorporate the autopilot dynamics of the interceptors and evaluate the performance of the complete estimation–guidance–control architecture for the case in which all interceptors are launched from a common initial position. It is assumed that the functions $k_{m_i}$, for all $m = 1, \ldots, 4$, are considered as hyperbolic tangent functions with an appropriate boundary layer. The lift and moment coefficients are adopted from \cite{sp27}. 

Due to physical limitations, the maximum canard deflection is constrained to $20^{\circ}$. The canard actuator time constant is set as $\tau_{c_i} = 0.1$ s for all interceptors. The observer convergence time is predefined as $t_p = 0.6~\mathrm{s}$, while the consensus controller is designed to achieve convergence within $t_e = 4.5   ~\mathrm{s}$. Moreover, the autopilot dynamics are assumed to settle within $t_a=2~\mathrm{s}$, thereby ensuring coherent operation across all control layers.

In this case, the performance of the proposed joint estimation–guidance design is validated, which is illustrated in \Cref{fig:auto_full}. Interceptors are launched from the same location with different initial heading angles, achieving cooperative simultaneous interception at $36.48~\mathrm{s}$. The corresponding interceptor trajectories are shown in \Cref{fig:Trajectoryauto}, while the evolution of time-to-go estimates reaching consensus within $t_e = 2.5~\mathrm{s}$ is depicted in \Cref{fig:Time_to_goauto}. The achieved lateral accelerations as a result of canard deflections and corresponding lead angles are presented in \Cref{fig:Accelerationauto}, and observer error convergence to zero within $t_p = 0.6~\mathrm{s}$ is illustrated in \Cref{fig:Observerauto}. The lateral accelerations and lead angles exhibit behavior similar to that observed in previous cases.  The corresponding variations of the angle of attack and pitch rate are illustrated in \Cref{fig:alpha_q}. The corresponding canard deflections of the interceptors are shown in \Cref{fig:delta_1_2} and \Cref{fig:delta_3_4}. It can be observed that the control inputs exhibit initial oscillations, whose amplitude diminishes significantly by approximately \( t = 2~\mathrm{s} \), which lies well within the user-specified \( t_e \). 

The effectiveness of the proposed autopilot design is evaluated through a comparative study with the finite-time autopilot reported in \cite{sp27}. It should be noted that the case considered corresponds to worst-case engagement conditions, characterized by large initial heading angle errors and significant uncertainties in the true target position, thereby ensuring a stringent and meaningful validation. Such adverse initial conditions place severe demands on the autopilot performance and robustness, thereby providing a rigorous assessment of its convergence and stability properties. It is well known that, by nature, one of the key advantages of predefined-time control over finite-time control lies in its convergence behavior. In particular, the convergence time of finite-time autopilots depends on the initial conditions and typically requires frequent tuning of controller parameters to achieve the desired performance. However, the proposed one-shot strategy embeds the predefined-time convergence and guarantees convergence of the necessary variables within a user-specified time, irrespective of the initial conditions. Consequently, the autopilot tracks the commanded input accurately, resulting in predictable and reliable tracking behavior. Therefore, to ensure a fair and meaningful comparison, the finite-time autopilot presented in \cite{sp27} is tuned such that its convergence time is set to \( t_a = 2~\mathrm{s} \). Moreover, the convergence times corresponding to the estimation and guidance subsystems are maintained identical to those used in the previous case to ensure a consistent comparison.

 \begin{figure}[h!]
    \centering
    
    \begin{subfigure}[t]{0.45\textwidth}
        \centering
        \includegraphics[width=\linewidth]{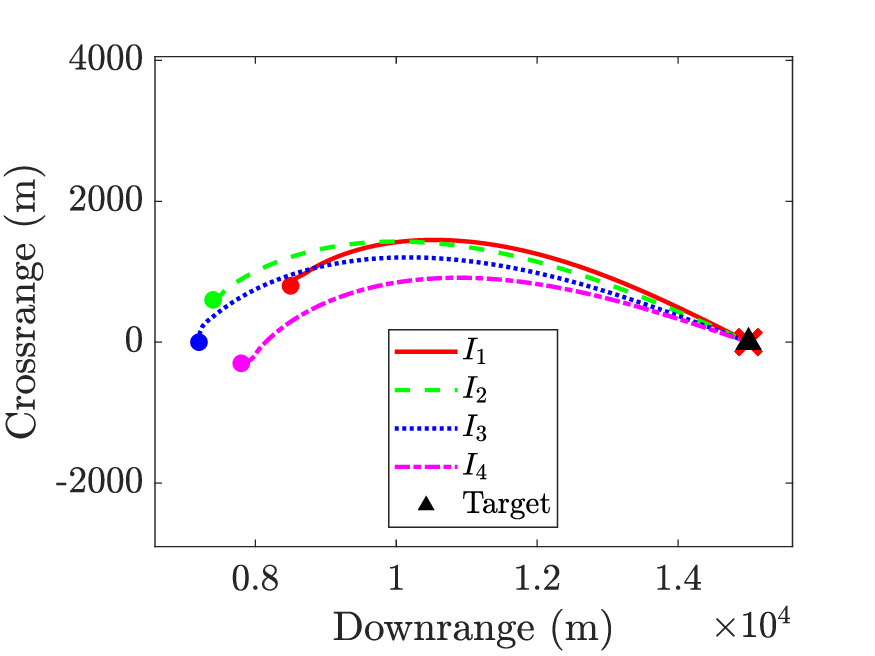}
        \caption{Interceptors' trajectories.}
        \label{fig:Trajectoryauto2}
    \end{subfigure}
    \hfill
    \begin{subfigure}[t]{0.45\textwidth}
        \centering
        \includegraphics[width=\linewidth]{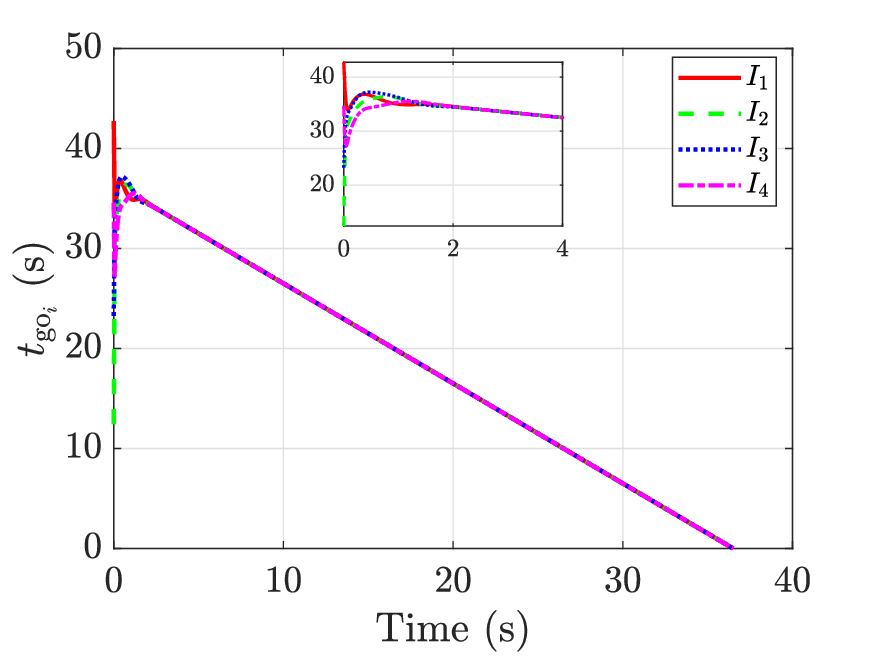}
        \caption{Interceptors' time-to-go values.}
        \label{fig:Time_to_goauto2}
    \end{subfigure}

    \begin{subfigure}[t]{0.45\textwidth}
        \centering
        \includegraphics[width=\linewidth]{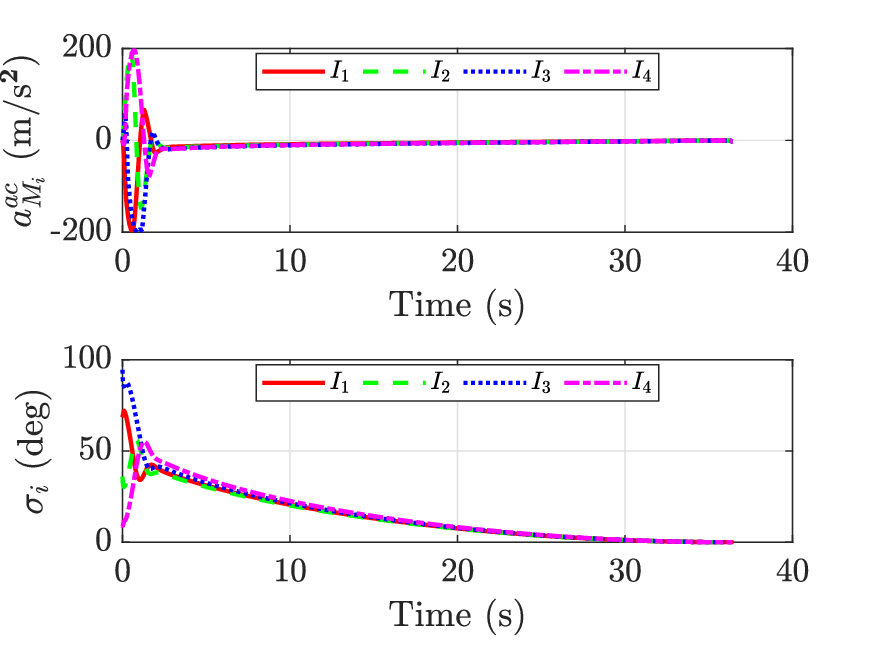}
        \caption{Interceptors' lateral acceleration and lead angle profiles.}
        \label{fig:Accelerationauto2}
    \end{subfigure}
    \hfill
    \begin{subfigure}[t]{0.45\textwidth}
        \centering
        \includegraphics[width=\linewidth]{Images/Observer_auto.eps}
        \caption{Target's state estimation error.}
        \label{fig:Observerauto2}
    \end{subfigure}
    
    \begin{subfigure}[t]{0.45\textwidth}
        \centering
        \includegraphics[width=\linewidth]{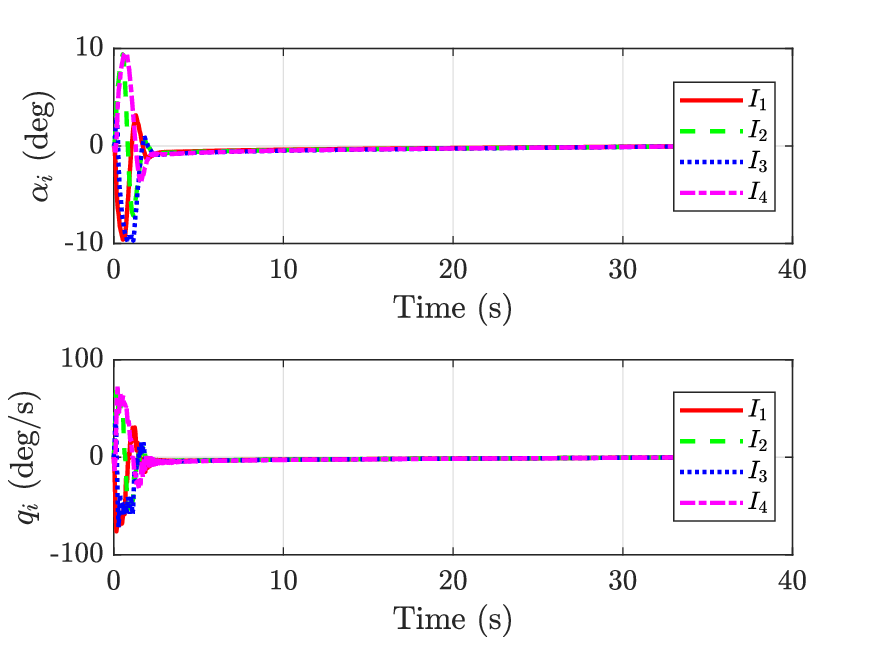}
        \caption{Interceptors' pitch rate and angle of attack.}
        \label{fig:alpha_q2}
    \end{subfigure}
    \hfill
    \begin{subfigure}[t]{0.45\textwidth}
        \centering
        \includegraphics[width=\linewidth]{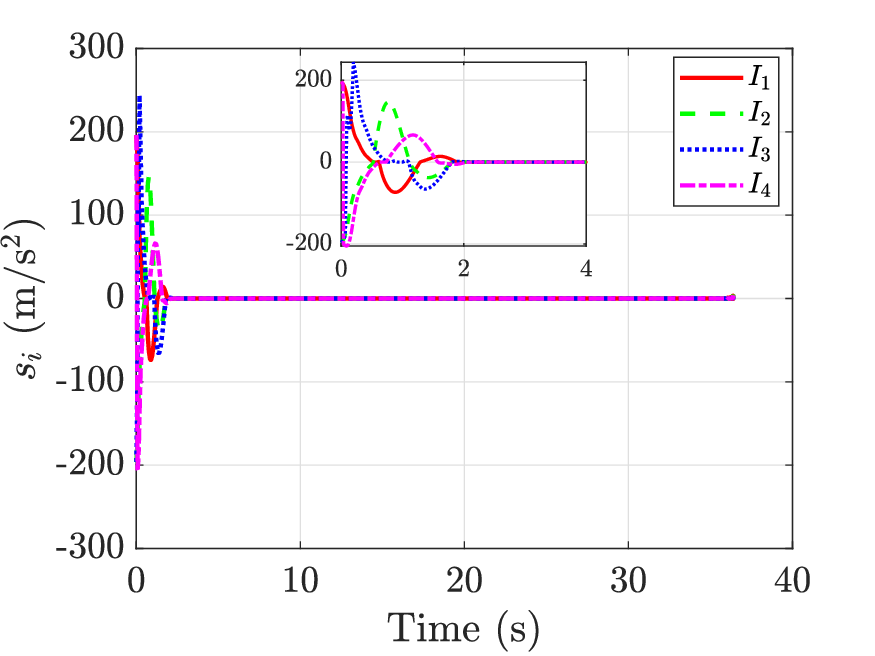}
        \caption{Interceptors' tracking error.}
        \label{fig:si2}
    \end{subfigure}
    \caption{Performance comparison of the proposed autopilot with the finite-time autopilot in \cite{sp27}.}
    \label{fig:auto_full2}
    
\end{figure}

\begin{figure}[h!]
    \ContinuedFloat
\centering
    \begin{subfigure}[t]{0.45\textwidth}
        \centering
        \includegraphics[width=\linewidth]{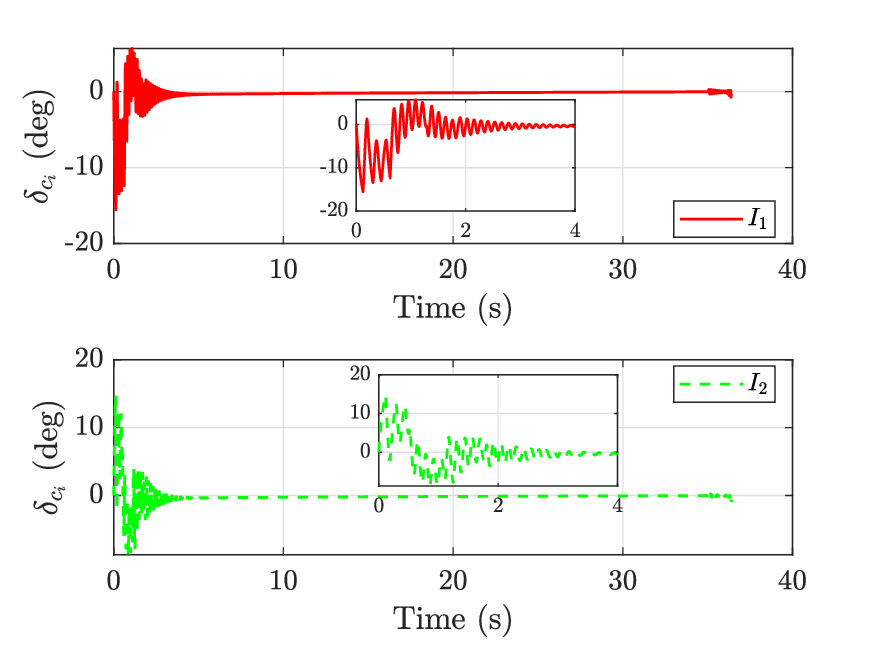}
        \caption{Canard deflections ($I_1$ and $I_2$). }
        \label{fig:delta_1_22}
    \end{subfigure}
    \hfill
    \begin{subfigure}[t]{0.45\textwidth}
        \centering
        \includegraphics[width=\linewidth]{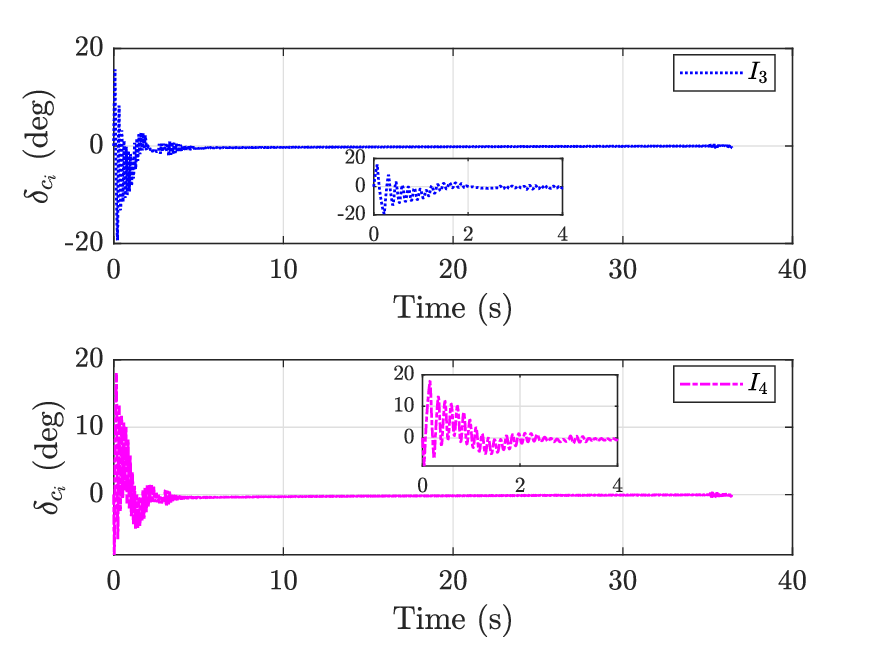}
        \caption{Canard deflections ($I_3$ and $I_4$).}
        \label{fig:delta_3_42}
    \end{subfigure}
    \captionsetup{list=off}
    \caption{Performance comparison of the proposed autopilot with the finite-time autopilot in \cite{sp27} (continued).}
\end{figure}

 \Cref{fig:auto_full2} illustrates the comparison scenario in which all interceptors achieve simultaneous interception at $t = 36.51~\mathrm{s}$. It should be noted that the interception time is identical to that of the proposed strategy, as the convergence time of the finite-time autopilot is tuned to match the predefined convergence time of the proposed approach. The corresponding interceptors' trajectories are depicted in \Cref{fig:Trajectoryauto2}. The evolution of the time-to-go estimates, which reach consensus within $t = 2.8~\mathrm{s}$, is shown in \Cref{fig:Time_to_goauto2}. The achieved lateral acceleration commands and lead angle histories are presented in \Cref{fig:Accelerationauto2}, exhibiting trends consistent with those observed in the earlier scenarios. The convergence of the observer estimation errors to zero within $t_p = 0.6~\mathrm{s}$ is illustrated in \Cref{fig:Observerauto2}. In addition, the time histories of the angle of attack and pitch rate are shown in \Cref{fig:alpha_q2}, while the corresponding canard deflection profiles for the interceptors are provided in \Cref{fig:delta_1_22,fig:delta_3_42}. With the autopilot convergence time tuned to  \(t_a = 2~\mathrm{s}\), the canard deflection exhibits a comparable transient response. In particular, the control input shows initial oscillatory behavior, with the oscillation amplitude decaying and settling by approximately \(t = 2.2~\mathrm{s}\), which lies within the convergence time \(t_e\). Furthermore, since the convergence time is enforced uniformly across all control layers, the performance in both approaches appears similar. This further motivates a quantitative assessment for a meaningful comparative evaluation, which may involve comparing the control-effort-based performance metrics.

\begin{table}[h!]
\centering
\caption{Comparison of control effort (deg$^2$ s) between the proposed strategy the one in \cite{sp27}.}
\label{tab:control_effort}
\begin{tabular}{c cc}
\hline
\textbf{Interceptor} 
& \textbf{Proposed Strategy} 
& \textbf{T.~Shima \emph{et~al.}~\cite{sp27}} \\
\hline
$J_{1}$ & 54.4090 & 64.8340 \\
$J_{2}$ & 54.1757 & 61.2017 \\
$J_{3}$ & 75.9851 & 94.7799 \\
$J_{4}$ & 54.7839 & 67.9214 \\
\hline
$J = \sum_{i=1}^{N} J_i$ 
& \textbf{239.3537} 
& \textbf{288.7370} \\
\hline
\end{tabular}
\end{table}
\begin{figure}[h!]
\centering
    \begin{subfigure}[t]{0.45\textwidth}
        \centering
        \includegraphics[width=\linewidth]{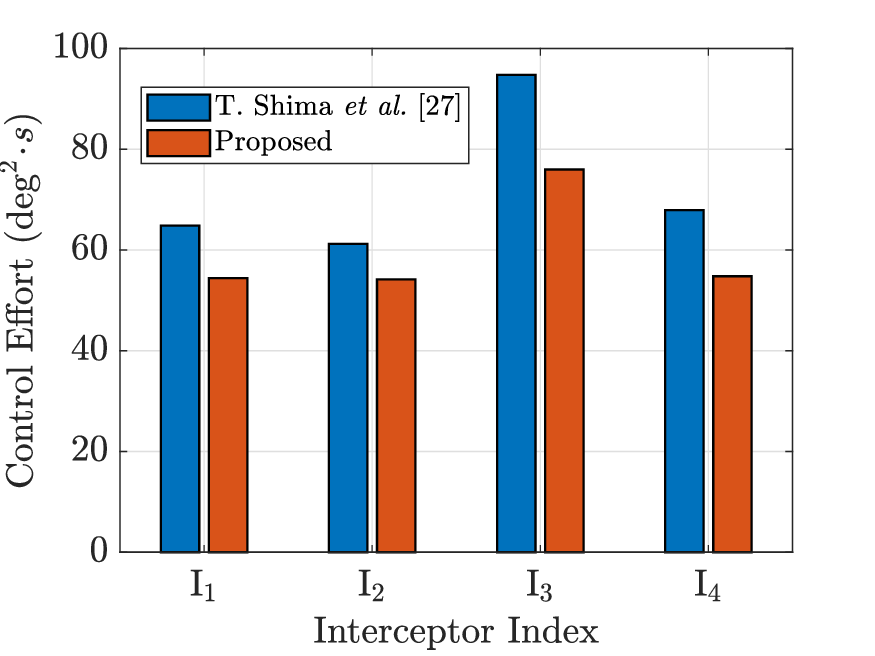}
        \caption{Absolute control effort comparison.}
        \label{fig:ce_absolute}
    \end{subfigure}
    \hfill
    \begin{subfigure}[t]{0.45\textwidth}
        \centering
        \includegraphics[width=\linewidth]{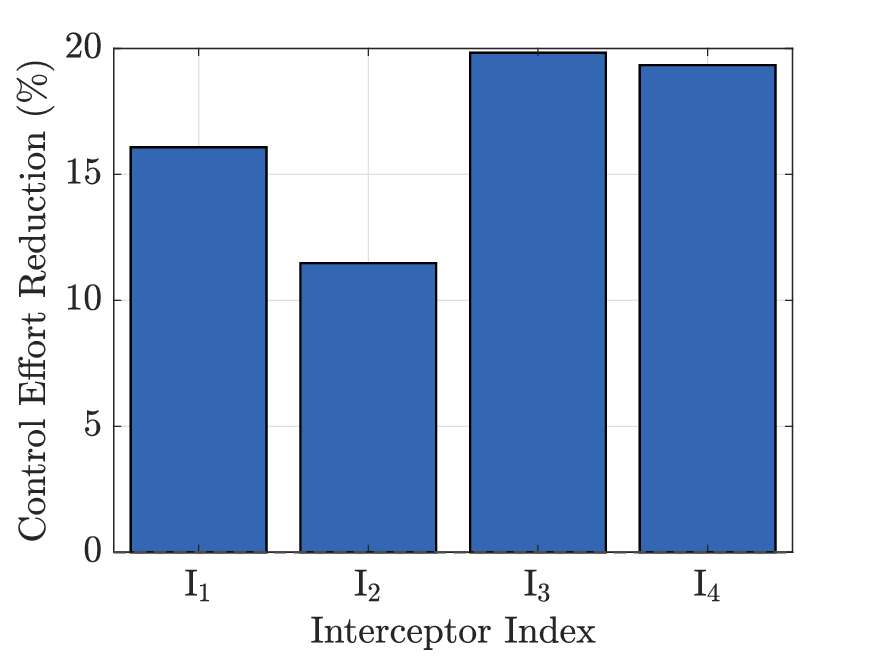}
        \caption{Percentage reduction in control effort.}
        \label{fig:ce_reduction}
    \end{subfigure}
    \caption{Comparison of control effort for the proposed strategy and the one in \cite{sp27}}
\end{figure}

The control effort for each interceptor is evaluated as 
$J_i = \int_{0}^{t_f} \delta_{c_i}^2(t)\,dt$,
where $t_f$ denotes the interception time. Moreover, we also compare the joint control effort defined as
$J = \sum_{i=1}^{N} J_i$,
which offers a more comprehensive and representative measure for evaluating the overall effectiveness.


The corresponding individual and joint control effort values are summarized in \Cref{tab:control_effort}, while \Cref{fig:ce_absolute} provides a clear visual comparison of the control effort across individual interceptors. From \Cref{tab:control_effort}, it can be observed that the proposed autopilot achieves approximately a 17\% reduction in joint control effort as compared to the finite-time autopilot presented in \cite{sp27}. Furthermore, \Cref{fig:ce_reduction} illustrates the corresponding percentage reduction in individual interceptor control effort, which is seen to consistently lie within the range of 10\,\%--20\,\% across all interceptors. Such a reduction in both individual and joint control effort is particularly advantageous in cooperative engagement scenarios, as it leads to lower actuator utilization, improved energy efficiency, and enhanced robustness of the overall control architecture. This demonstrates that the proposed autopilot is capable of achieving faster and reliable tracking within the predefined time while requiring comparatively smaller control effort in the absence of full state information.


Therefore, it can be inferred from the extensive simulation studies that, despite the inherent limitation of having only a subset of interceptors equipped with onboard seekers, the proposed one-shot predefined-time estimation–guidance–control architecture achieves performance comparable to strategies reported in the literature that assume continuous availability of full target information even under worst-case engagement conditions, including large initial heading angle errors and significant initial target position estimation errors. This advantage stems from embedding the estimator, guidance law, and autopilot into a single coherent framework with predefined-time convergence, which overcomes the limitations of conventional designs that treat these components independently and often fail to explicitly capture their mutual interactions, closed-loop stability properties, and impact on terminal accuracy. In addition, the observed reduction in both individual and joint control effort reflects improved control efficiency, which directly translates into lower actuator usage, reduced energy consumption, and favorable resource and cost implications. Moreover, the incorporation of a distributed observer enhances operational robustness by enabling simultaneous interception even in the event of seeker failures through inter-agent information sharing. This is unlike many existing strategies in the literature, which rely heavily on continuous access to true target information and consequently fail to guarantee cooperative interception when seeker failures occur. From a practical standpoint, this capability is beneficial in contested or degraded sensing environments such as electronic countermeasure scenarios or partial sensor outages where maintaining cooperative mission success with limited sensing resources is essential.

Therefore, the proposed one-shot framework is well-suited for resource-constrained cooperative interception applications, such as unmanned aerial vehicle swarms or distributed defense platforms, where limitations in sensing availability, onboard power, and actuator endurance necessitate efficient, robust, and scalable guidance and control solutions. The ability of the proposed framework to maintain cooperative interception performance under these practical constraints highlights its suitability for real-world deployments that demand cost,energy-efficient, resilient, and scalable cooperative guidance and control architectures

\section{Concluding Remarks}\label{sec:conclusions}
In this paper, a unified one-shot nonlinear estimation–guidance–control framework was developed to achieve cooperative simultaneous interception of a stationary target under heterogeneous sensing conditions. By explicitly accommodating a mixed team of seeker-equipped and seeker-less interceptors, the proposed approach addressed the challenges arising from partial observability using a predefined-time distributed observer that guarantees accurate target state reconstruction for uninformed agents through directed information exchange. The improved time-to-go estimation, integrated with a predefined-time consensus protocol, ensures that all interceptors synchronize their impact times within a designer-specified bound, enabling precise cooperative interception even under wide launch envelopes. Furthermore, the canard-based autopilot employing a predefined-time sliding mode control law realizes accurate tracking of the distributed guidance commands while ensuring non-singularity and enforcing strict convergence constraints. Our results across diverse engagement geometries validated the theoretical guarantees, demonstrating high estimation accuracy, coordinated interception performance, and fast autopilot response. Moreover, the distributed and predefined-time convergent design demonstrated robustness to agent failures, communication link failures, and seeker failures, allowing the cooperative interception mission to remain achievable even under degraded sensing or connectivity conditions. Overall, the proposed framework provides a reliable, scalable, and resilient solution for cooperative guidance under heterogeneous sensing conditions. 

\bibliographystyle{IEEEtran}
\bibliography{references}
\end{document}